\documentclass[a4paper,11pt]{article}
\usepackage{jheppub}
\usepackage{lineno}
\usepackage{amsmath,mathrsfs,multicol,multirow,array}
\usepackage{tikz-feynman}
\usepackage{booktabs}
\usepackage{tabularx}
\usepackage{adjustbox}
\usepackage{comment}
\usepackage{mathrsfs}
\usepackage[utf8]{inputenc}
\nolinenumbers

\allowdisplaybreaks

\title{\boldmath Exploring $Z/\gamma$-mediated heavy FCNCs at the FCC-ee}

\author[a]{Abhik Sarkar,}
\author[a]{ Subhajit Kala,}
\author[b,c]{ Amir Subba,}
\author[b,c]{ and Yu Shi}
\affiliation[a]{Department of Physics, Indian Institute of Technology Guwahati, North Guwahati, 781039, India}
\affiliation[b]{Wilczek Quantum Center, Shanghai Institute for Advanced Studies, Shanghai 201315, China}
\affiliation[c]{University of Science and Technology of China, Hefei 230026, China}
\emailAdd{sarkar.abhik@iitg.ac.in}
\emailAdd{s.kala@iitg.ac.in}
\emailAdd{amirsubba@ustc.edu.cn}
\emailAdd{yu\_shi@ustc.edu.cn}

\abstract{The flavor structure of the Standard Model (SM) remains one of the most compelling questions in particle physics, with the third generation being particularly intriguing due to its significantly larger masses and comparatively less precisely measured properties. These features make third-generation flavor transitions particularly interesting in context of search for physics beyond the SM. In this work, we investigate flavor-violating transitions between the third and the first two generations, mediated by the neutral gauge bosons ($Z/\gamma$), within the framework of the SM Effective Field Theory (SMEFT), using dipole and Higgs-current operators. We determine the optimal sensitivities using the optimal observable technique (OOT) at different center-of-mass energies of the upcoming Future Circular Collider in the $e^+e^-$ mode (FCC-ee). We further derive complementary constraints on the relevant SMEFT operators from low-energy flavor-violating observables and heavy fermion decay channels. Our analysis also reveals characteristic interference patterns among the dipole contributions, which depend on the underlying flavor transition and exhibit distinct behavior between the $Z$ pole and higher-energy FCC-ee stages. The FCC-ee provides a complementary and direct probe of flavor-violating interactions at the electroweak scale, with the projections showing improved sensitivity for several interactions and comparable sensitivity to existing flavor constraints for several others. This highlights the importance of a systematic assessment across the different FCC-ee energy stages, which provides a comprehensive picture of its potential to explore flavor-violating phenomena and its complementarity with low-energy flavor experiments.}

\begin{document}
\maketitle
\flushbottom

\section{Introduction}
\label{sec:intro}
The Standard Model (SM)~\cite{Weinberg:1967tq} has successfully described the interactions of elementary particles with remarkable precision. Nevertheless, several fundamental questions remain unanswered within its current framework. In particular, the SM does not explain the observed pattern of fermion masses, flavor mixing, or the hierarchy of Yukawa couplings spanning several orders of magnitude. Consequently, the underlying mechanism responsible for the flavor structure remains unknown, suggesting the possibility of physics beyond the SM (BSM). The study of flavor physics has therefore long served as one of the most important avenues for probing New Physics (NP), making both indirect precision measurements and direct searches for flavor-violating effects at high-energy colliders a central components of the contemporary particle physics program.

The third generation of fermions is of particular interest because of its comparatively large Yukawa couplings, resulting significantly larger masses than those in the first two generations. These large Yukawa couplings also imply stronger interactions with the SM Higgs boson, suggesting a possible connection to the mechanism of electroweak symmetry breaking. Consequently, many scenarios of NP predict enhanced interactions involving third-generation fermions~\cite{delaTorreTrishaFarooque:2022vqc}. From an experimental perspective, the properties and interactions of third-generation fermions are generally measured less precisely than those of the first and second-generation fermions ~\cite{Lannon:2012fp}. A notable exception is the bottom quark, whose properties have been studied with high precision at dedicated $B$-factory experiments~\cite{BaBar:2014omp, Belle-II:2018jsg}. This difference in experimental precision is particularly evident in the lepton sector. For instance, the anomalous magnetic moments of the $e$ and $\mu$ leptons are measured with relative precisions at the $\mathcal{O}(10^{-13})$~\cite{Fan:2022eto} and $\mathcal{O}(10^{-7})$~\cite{Muong-2:2026qnz} levels, respectively, whereas the corresponding precision for the $\tau$ lepton is only at the $\mathcal{O}(10^{-3})$ level~\cite{ATLAS:2025oiy}.

This leaves greater room for possible deviations in the third generation from the SM predictions. Consequently, flavor-violating transitions between the third and the first two generations, i.e. $3 \to 1,2$ processes, are expected to provide more sensitive probes of NP than transitions exclusively among the first two generations, i.e. $2 \to 1$ processes~\cite{Aguilar-Saavedra:2004mfd, Altmannshofer:2023bfk, Altmannshofer:2025lun, Sarkar:2025bgo, Bhattacharya:2025xwv, Bhattacharya:2025mlg, Bhattacharya:2023beo, Kala:2026xzo, Biswas:2022lhu}.

In the current work, we investigate neutral gauge boson ($Z/\gamma$)-mediated flavor-changing neutral currents (FCNCs) in the lepton and quark sectors at the multi-stage $e^{+}e^{-}$ Future Circular Collider (FCC-ee)~\cite{Agapov:2022bhm,FCC:2025lpp,FCC:2025uan}. Within the SM, neutral gauge bosons mediate only flavor-diagonal interactions, and any observation of FCNC would therefore constitute an unambiguous signature of NP. Since the FCC-ee is designed to operate at center-of-mass (CM) energies around the electroweak scale, the underlying NP degrees of freedom are expected to remain inaccessible as on-shell states. Their effects can instead be systematically parameterized through an Effective Field Theory (EFT) framework. For collider phenomenology, the Standard Model Effective Field Theory (SMEFT)~\cite{Buchmuller:1985jz} provides the most widely adopted model-independent framework, in which the effects of heavy NP are encoded in higher-dimensional operators constructed from SM fields. This enables a systematic and model-independent study of flavor-violating interactions at the FCC-ee. To estimate the sensitivities, we employ the Optimal Observable Technique (OOT)~\cite{Atwood:1991ka,Davier:1992nw,Diehl:1993br, Gunion:1996vv}, which has been widely used in collider phenomenology to obtain statistically optimal sensitivity projections for collider experiments~\cite{Jahedi:2024kvi, Bhattacharya:2023beo, Bhattacharya:2023mjr, Bhattacharya:2021ltd}.

FCNC and charged lepton flavor violation (LFV) are among the most sensitive probes of NP. Low-energy observables, including rare meson decays~\cite{LHCb:2021vsc, CMS:2022mgd,
LHCb:2022qnv, Belle:2014sac,
Belle-II:2024tru,
Belle:2017hum}, neutral meson mixing~\cite{HFLAV:2022esi}, and charged lepton flavor-violating decays~\cite{Belle:2021ysv,BaBar:2009hkt,Hayasaka:2010np}, have placed stringent constraints on flavor-violating interactions. Complementary constraints on the quark sector are also obtained from collider observables, such as top-quark FCNC decays~\cite{CMS:2023bjm,ATLAS:2023qzr}, which provide direct probes of flavor-violating interactions involving the top quark. Several phenomenological studies have explored FCNCs~\cite{Bhattacharya:2023beo, Sarkar:2025bgo,Altmannshofer:2023bfk, Altmannshofer:2025lun, Bhattacharya:2025mlg, Kala:2026xzo} and charged LFV~\cite{Jahedi:2024kvi,Crivellin:2013hpa, Ali:2023kua, Covone:2025lee} in the context of current and future collider experiments, providing projected sensitivities to a variety of flavor-violating interactions. The clean lepton collider environment, excellent detector performance, and high luminosity reach envisaged for the FCC-ee make it ideal for such searches. While several studies have investigated flavor-violating interactions within the SMEFT framework~\cite{Greljo:2024ytg,Altmannshofer:2023tsa, Denizli:2017cfx, Allwicher:2023shf}, a comprehensive study of $Z/\gamma$-mediated heavy (third generation) flavor-violating interactions across all proposed FCC-ee energy stages, together with a systematic comparison with existing low-energy constraints, is still lacking. Although some low-energy observables impose constraints that are beyond the projected sensitivity of collider experiments, assessing the corresponding reach of the FCC-ee remains essential for quantifying its capability to directly probe flavor-violating interactions and for benchmarking the flavor physics potential of future lepton colliders. In this work, we address this gap by providing a unified assessment of the FCC-ee sensitivity to such flavor-violating interactions in both the lepton and quark sectors. An important feature of this FCC-ee sensitivity study is the interference structure of the effective interactions. In particular, the dipole operators receive both photon and $Z$-mediated contributions, whose interference depends on the flavor structure and CM energy. We find that the resulting interference patterns can differ between different flavor transitions and, notably, between the $Z$-pole and higher-energy FCC-ee stages, with the up-type quark transitions exhibiting an opposite interference behavior compared to the leptonic and down-type quark sectors. This provides additional sensitivity to the underlying operator structure and highlights the complementarity of the different FCC-ee energy stages.

The remainder of this paper is organized as follows. In Sec.~\ref{sec:eft}, we introduce the effective framework used to describe flavor changing neutral currents and charged lepton flavor violating interactions. The existing constraints on the EFT parameters are presented in Sec.~\ref{sec:constraints}. In Sec.~\ref{sec:oot}, we outline the OOT, followed by the collider simulation framework and the projected sensitivities in Sec.~\ref{sec:collider}. Finally, we summarize our findings and conclude in Sec.~\ref{sec:conc}.

\section{SMEFT framework for FCNC}
\label{sec:eft}
Within the Standard Model, the FCNCs are highly suppressed and any observation of FCNC would be an unambiguous signature of BSM physics. In scenarios where no new resonant states are directly observed, EFT offers a systematic and model-independent framework to parameterize deviations from SM predictions through higher-dimensional operators, organized as an expansion in inverse powers of the heavy NP scale $\Lambda$. In this work, we consider the SMEFT framework, where the Higgs field transforms as an $\text{SU}(2)_{L}$ doublet and electroweak symmetry breaking (EWSB) proceeds via the standard non-zero vacuum expectation value (VEV) of the Higgs field. The SMEFT Lagrangian can be written as~\cite{Buchmuller:1985jz}
\begin{equation}
\mathcal{L}_{\mathrm{SMEFT}} = \mathcal{L}_{\mathrm{SM}} +
\sum_i \frac{c_i^{(5)}\mathcal{O}_i^{(5)}}{\Lambda} +
\sum_i \frac{c_i^{(6)}\mathcal{O}_i^{(6)}}{\Lambda^2} +\cdots,,
\end{equation}
where $\mathcal{L}_{\mathrm{SM}}$ denotes the renormalizable dimension-4 SM Lagrangian, $\mathcal{O}_i^{(d)}$ are higher dimensional operators of mass dimension $d>4$ that respect the SM gauge symmetry ($ \text{SU}(3)_C\times \text{SU}(2)_L \times \text{U}(1)_Y$), and $c_i^{(d)}$ are the corresponding dimensionless Wilson coefficients (WCs) encoding the low-energy effects of heavy BSM physics at the scale $\Lambda$. At dimension-5, there exists a unique operator~\cite{Weinberg:1979sa} commonly referred to as the Weinberg operator
\begin{equation}
[\mathcal{O}^{(5)}]_{pr}= \left(\overline{\ell^{\,c}_{p}}\,\widetilde{H}^{*}\right)
\left(\widetilde{H}^{\dagger}\ell_{r}\right),
\end{equation}
where $\ell$  and $H$ are lepton and Higgs doublet which transforms as $(1,2,-1/2)$ under the SM gauge symmetry. This operator is responsible for generating Majorana masses of neutrinos after EWSB and does not directly induce flavor-violating interactions among charged leptons. Since our analysis focuses on flavor-violating interactions involving charged leptons and quarks, the Weinberg operator is therefore not relevant to the processes considered here. The complete and non-redundant set of independent dimension-6 operators has been systematically classified in Ref.~\cite{Grzadkowski:2010es}, and is commonly referred to as the Warsaw basis. Since collider observables depend only on the combination $c_i^{(6)}/\Lambda^2$, we redefine
\begin{equation}
C_i \equiv \frac{c_i^{(6)}}{\Lambda^2}\,,
\end{equation}
thereby absorbing the explicit scale dependence into the effective WCs $C_i$. These rescaled WCs have mass dimension $[M^{-2}]$ (i.e. TeV$^{-2}$). SMEFT therefore provides a consistent and model-independent framework for probing BSM effects through precision measurements and global analyses of WCs using collider data. A subset of the dimension-6 operators gives rise to flavor-violating interactions. The relevant operators mediating flavor violation through $Z/\gamma$ exchange are listed in Tab.~\ref{tab:dim6}.
\begin{table}[htb!]
    \centering
    \renewcommand{\arraystretch}{1.25}{
    \begin{tabular}{|>{\centering\arraybackslash}p{0.100\textwidth}|>{\centering\arraybackslash}p{0.325\textwidth}|>{\centering\arraybackslash}p{0.100\textwidth}|>{\centering\arraybackslash}p{0.325\textwidth}|}
    \hline\hline
     \multicolumn{2}{|c|}{$\boldsymbol{\Psi^2 X H}$} & \multicolumn{2}{c|}{$\boldsymbol{\Psi^2H^2D}$}  \\
    \hline\hline
     $[\mathcal{O}_{eW}]_{pr}$ & $(\bar{\ell}_p\sigma^{\mu\nu}e_r)\tau^I H W^I_{\mu\nu}$ & $[\mathcal{O}_{H \ell }^{(1)}]_{pr}$ & $ (H^\dagger i \overleftrightarrow{D}_\mu H)(\bar{\ell}_p\gamma^\mu \ell_r)$\\
   $[\mathcal{O}_{eB}]_{pr} $ & $ (\bar{\ell}_p\sigma^{\mu\nu}e_r)H B_{\mu\nu}$&$[\mathcal{O}_{H \ell}^{(3)}]_{pr} $&$ (H^\dagger i \overleftrightarrow{D}_\mu^I H)(\bar{\ell}_p \tau^I \gamma^\mu \ell_r)$ \\
    $[\mathcal{O}_{uW}]_{pr} $ & $ (\bar{q}_p\sigma^{\mu\nu}u_r)\tau^I\widetilde{H}W^I_{\mu\nu}$& $[\mathcal{O}_{H e}]_{pr} $&$ (H^\dagger i \overleftrightarrow{D}_\mu H)(\bar{e}_p\gamma^\mu e_r)$\\
    $[\mathcal{O}_{uB}]_{pr} $ & $ (\bar{q}_p\sigma^{\mu\nu}u_r)\widetilde{H}B_{\mu\nu}$ & $[\mathcal{O}_{H q}^{(1)}]_{pr} $&$ (H^\dagger i \overleftrightarrow{D}_\mu H)(\bar{q}_p\gamma^\mu q_r)$\\
    $[\mathcal{O}_{dW}]_{pr} $&$ (\bar{q}_p\sigma^{\mu\nu}d_r)\tau^I H W^I_{\mu\nu}$ & $[\mathcal{O}_{H q}^{(3)}]_{pr} $ & $ (H^\dagger i \overleftrightarrow{D}_\mu^IH)(\bar{q}_p\tau^I\gamma^\mu q_r)$\\
    $[\mathcal{O}_{dB}]_{pr} $&$ (\bar{q}_p\sigma^{\mu\nu}d_r)H B_{\mu\nu}$ & $[\mathcal{O}_{H u}]_{pr} $ & $ (H^\dagger i \overleftrightarrow{D}_\mu H)(\bar{u}_p\gamma^\mu u_r)$\\
    & & $[\mathcal{O}_{H d}]_{pr} $&$ (H^\dagger i \overleftrightarrow{D}_\mu H)(\bar{d}_p\gamma^\mu d_r)$\\
    \hline\hline
    \end{tabular}}
    \caption{\label{tab:dim6}Dimension 6 SMEFT operators of dipole ($\Psi^2 X H$) and Higgs-current ($\Psi^2 H^2 D$) classes in the Warsaw basis, contributing to $Z/\gamma$-mediated flavor violation. Flavor violation via non-diagonal indices: $p\neq r$.}
\end{table}

It is worth mentioning that the Warsaw basis is defined in the flavor (interaction) basis. However, to connect the SMEFT operators with physical observables at colliders, a flavor rotation to the mass basis is required. In this work, we adopt \textit{up-aligned} rotation. The corresponding transformations between the flavor and mass eigenstate bases are,
\begin{align}
    u^{\prime}_{L(R)}&=U^u_{L(R)} u_{L(R)}\,,& d^{\prime}_{L(R)}&=U^d_{L(R)} d_{L(R)}\,, & q_L^{\prime}&=\begin{pmatrix}
        u_L \\ V_{\rm CKM} d_L 
    \end{pmatrix}, & V_{\rm CKM}=(U^u_L)^{\dagger} U_L^d\,.
\end{align}
In the \textit{up-aligned} basis, we choose
   $ U^u_L = \mathbf{1}\,,~ 
    ~U^{(u,d)}_R= \mathbf{1}, 
    ~U^d_L = V_{\rm CKM},$
while the charged-lepton sector is taken to be already in the mass basis. Here, the unprimed (primed) fields denote the mass (flavor) eigenstates, respectively. We absorb the effects of the flavor-rotation matrices into the definition of the unprimed SMEFT WCs.

In the context of charged LFV and FCNC processes, only the following combinations of the Higgs-current operators are phenomenologically relevant:
\begin{equation}
[C_{Hq}^{(\pm)}]_{pr} = [C_{Hq}^{(1)}]_{pr} \pm [C_{Hq}^{(3)}]_{pr},,
\qquad
[C_{H\ell}^{(\pm)}]_{pr} = [C_{H\ell}^{(1)}]_{pr} \pm [C_{H\ell}^{(3)}]_{pr},.
\end{equation}
After EWSB, the combinations $C_{Hq}^{(-)}$ and $C_{Hq}^{(+)}$ contribute to flavor-violating neutral currents involving up-type and down-type quarks, respectively, while $C_{H\ell}^{(-)}$ and $C_{H\ell}^{(+)}$ contribute to neutral currents involving neutrinos and charged leptons, respectively. For the dipole operators, both the $\text{SU}(2)_{L}$ and $\text{U}(1)_{Y}$ field-strength operators contribute to the $Z/\gamma$-mediated interactions after EWSB. Our analysis is restricted to a subset of $24$ operators relevant to the phenomenology under consideration, corresponding to the six flavor-violating transitions $\tau \to \mu$, $\tau \to e$, $b \to s$, $b \to d$, $t \to c$, and $t \to u$. The WCs associated with the dipole operators can, in general, be complex. In collider observables, the corresponding contributions are sensitive to the modulus of these complex couplings, and hence our sensitivity analysis is performed in terms of their magnitudes. The projected limits are therefore presented as bounds on the magnitudes of the WCs. For completeness, certain flavor observables provide comparable constraints on the real and imaginary parts of the WCs; these constraints are discussed separately in Appendix~\ref{Append:complex_fit}.

In addition to direct limits and projections associated with high-energy collider searches, the flavor-violating interactions considered here are also subject to constraints from low-energy observables. To consistently incorporate these indirect constraints, we first analyze the relevant flavor-changing processes below the electroweak scale using the appropriate EFT framework. For the low-energy analysis, we employ the Weak Effective Theory (WET)~\cite{Jenkins:2017jig} description of flavor-changing processes below the electroweak scale $(\mu_{\rm EW}=M_Z)$. This basis is particularly convenient for incorporating the available theoretical calculations of the relevant hadronic matrix elements. The WCs defined in the SMEFT Warsaw basis are matched onto the corresponding WET coefficients at the electroweak scale  and subsequently evolved to the relevant low-energy scales using the renormalization group equations (RGEs). This provides a consistent connection between the high-energy SMEFT description and the low-energy observables included in our analysis.

\section{Experimental Constraints}
\label{sec:constraints}
In this section, we summarize the existing experimental constraints on the flavor-violating interactions considered in our analysis. We focus on the constraints arising from flavor-violating processes involving bottom and top quarks, as well as charged lepton flavor-violating decays. These constraints provide complementary information to the direct collider sensitivities projected at the FCC-ee and allow us to assess the parameter space accessible to future lepton colliders.
\subsection{Bottom FCNC Processes}
To constrain the parameter space of down-type dipole and Higgs current operators, bottom quark flavour-changing neutral currents offer a powerful diagnostic tool. Specifically, rare (leptonic and semileptonic) and radiative $B$ meson decays act as complementary probes of these interactions. Because these effective operators induce FCNC transitions directly at the tree level within the EFT, precise measurements of these $B$ decay channels allow us to extract stringent bounds on the associated NP couplings.
\subsubsection*{Rare Decays}
The general low-energy effective Hamiltonian describing the $b \to d_i \ell^+\ell^-$ transition is expressed as \cite{Buras:1995iy,Bobeth:1999mk},
\begin{equation} \label{eq:Heff_b2di}
	\mathcal{H}^{b\to d_i\ell\ell}_{\text{eff}} = - \frac{4\,G_F}{\sqrt{2}} V_{tb}V_{td_i}^\ast
	\left[ \sum_{k=1}^{6} \mathcal{C}_k (\mu) \mathcal{O}_k(\mu) + \sum_{k=7,8,9,10,P,S} \biggl(\mathcal{C}_k (\mu) \mathcal{O}_k + \mathcal{C}'_k (\mu) \mathcal{O}'_k\biggr)\right] \,,
\end{equation}
where $\mathcal{O}_1$ through $\mathcal{O}_6$ are the four-quark operators, $\mathcal{O}_7$ and $\mathcal{O}_8$ are the electromagnetic and chromomagnetic dipole operators, $\mathcal{O}_9$ and $\mathcal{O}_{10}$ are the vector and axial-vector operators, and $\mathcal{O}_{S,P}$ and $\mathcal{O}_T$ denote the scalar, pseudoscalar, and tensor operators, respectively. The complete operator basis can be found in Ref.~\cite{Buras:1995iy}. However, in our analysis, the chosen set of effective operators, as tabulated in Tab.~\ref{tab:dim6}, only affects the dipole and vector (axial-vector) couplings. Therefore, the corresponding operators considered in this work are restricted to $\mathcal{O}_{7}^{(\prime)}$ and $\mathcal{O}_{9, 10}^{(\prime)}$.

\begin{subequations}
    \begin{align}
    \mathcal{O}_{7} &= \frac{e}{16 \pi^2} m_b
	(\bar{d_j} \sigma_{\mu \nu} P_R b) F^{\mu \nu} ,&
	\mathcal{O}_{7}^\prime &= \frac{e}{16 \pi^2} m_b
	(\bar{d_j} \sigma_{\mu \nu} P_L b) F^{\mu \nu} , \\
	\mathcal{O}_{9} &= \frac{e^2}{16 \pi^2} (\bar{d}_j \gamma_{\mu} P_L b)(\bar{\ell} \gamma^\mu \ell) \,,&
	\mathcal{O}_{9}^\prime &= \frac{e^2}{16 \pi^2} (\bar{d}_j \gamma_{\mu} P_R b)(\bar{\ell} \gamma^\mu \ell) \,, \\
	\mathcal{O}_{10} &=\frac{e^2}{16 \pi^2} (\bar{d}_j \gamma_{\mu} P_L b)( \bar{\ell} \gamma^\mu \gamma_5 \ell) \,,&
	\mathcal{O}_{10}^\prime &=\frac{e^2}{16 \pi^2} (\bar{d}_j \gamma_{\mu} P_R b)( \bar{\ell} \gamma^\mu \gamma_5 \ell) \,.
\end{align} 
\end{subequations}
In the SM, the operators $\mathcal{O}_7$, $\mathcal{O}_9$, and $\mathcal{O}_{10}$ primarily govern these rare radiative and dileptonic FCNC decays. The SM contributions from the chirality-flipped primed operators ($\mathcal{O}'_i$) are highly suppressed by the light quark masses and effectively vanish. However, in our framework, we obtain additional NP contributions to these operators. To disentangle the NP effects from the SM contributions, we redefine the low-energy WCs by explicitly separating the two contributions:
\begin{equation} \label{eq:WC_split}
	\mathcal{C}_i(\mu) = \mathcal{C}_i^{\text{SM}}(\mu) + \mathcal{C}_i^{\text{NP}}(\mu) \,, \quad \text{and} \quad \mathcal{C}'_i(\mu) = \mathcal{C}_i^{\prime \text{NP}}(\mu) \,,
\end{equation}
where $\mathcal{C}_i^{\text{SM}}$ denotes the known SM predictions and $\mathcal{C}_i^{(\prime)\text{NP}}$ encapsulates the non-standard NP modifications. The following matching relations in terms of SMEFT WCs can be written as, by evaluating the operators at the electroweak matching scale $\mu_{\text{EW}}$.
\begin{subequations}
\begin{align}\label{eq:b2dill_SMEFT}
   \lambda_2~ \mathcal{C}_9^{\rm NP} &= -\frac{g_Z^2 v^2}{2 M_Z^2}\left(\frac{1}{2}+2 ~q_e \sin^2\theta_W\right)\left[{C}_{H q}^{(+)}\right]_{d_i b}\,, \quad & \lambda_2~\mathcal{C}_{10}^{\rm NP} &=+\frac{g_Z^2 v^2}{4 M_Z^2}\left[{C}_{H q}^{(+)}\right]_{d_i b} \,, \\
    \lambda_2~\mathcal{C}_9^{\prime \rm NP} &= -\frac{g_Z^2 v^2}{2 M_Z^2}\left(\frac{1}{2}+2 ~q_e \sin^2\theta_W\right) \left[{C}_{H d}\right]_{d_i b}\,, \quad
    & \lambda_2~ \mathcal{C}_{10}^{\prime \rm NP}&=  +\frac{g_Z^2 v^2}{4 M_Z^2}\left[{C}_{H d}\right]_{d_i b} \,.
\end{align}
\end{subequations}
where 
\begin{align}
    \lambda_1&\equiv -\frac{4\cal{G}_F}{\sqrt{2}} \frac{m_b~e}{16\pi^2}V_{tb}V_{td_i}^*  \,,&\quad \lambda_2&\equiv -\frac{8\cal{G}_F}{\sqrt{2}} \frac{e^2}{16\pi^2}V_{tb}V_{td_i}^*\,.
\end{align}
To study the decay processes involving $B$-mesons, the WCs must be evaluated at the bottom-quark mass scale, $\mu_b = m_b$. In our analysis, we have incorporated the proper renormalisation group evolution (RGE) by utilising the anomalous dimension matrices (ADMs) of both the SMEFT and LEFT frameworks. For the numerical analysis, we have considered a comprehensive set of low-energy $B$-physics observables spanning the following leptonic and semileptonic modes:
\paragraph{Leptonic decays:} We include the branching fractions of the purely leptonic decays $B_s^0 \to \mu^+\mu^-$ and $B^0 \to \mu^+\mu^-$, within our framework, which are primarily sensitive to the vector and axial-vector couplings. The model-independent expression of the branching ratios can be written as,
\begin{equation}
\begin{split} \label{eq:BR_Bqll}
\mathcal{B}(B_q \rightarrow \mu^+ \mu^-) &= \tau_{B_q} f_{B_q}^2  m_{B_q}^3 \frac{G_F^2 \alpha^2}{64 \pi^3} |V^*_{tq}V_{tb}|^2 \beta_{\mu}(m_{B_q}^2) \left[   \frac{m_{B_q}^2}{m_b^2} |C_s - C'_s|^2 \left(1-\frac{4m_{\mu}^2}{m_{B_q}^2}\right) \right.\\ & \left.  + \bigg|\frac{m_{B_q}}{m_b}(C_p - C'_p) + 2\frac{m_{\mu}}{m_{B_q}} (C_{10} - C'_{10})\bigg|^2 \right]\,,
\end{split}
\end{equation}
where $\beta_{\mu}(q^2) = \sqrt{1 - \frac{4 m_{\mu}^2}{q^2}}$ is the kinematic phase-space factor, $f_{B_q}$ is the $B_q$ meson decay constant for $q \in \{d,s\}$, and, unless otherwise specified, all numerical input parameters are taken from the PDG~\cite{ParticleDataGroup:2024cfk}. The most recent experimental inputs are,
\begin{align}
    \mathcal{B}(B^{0} \to \mu^{+} \mu^{-}) &= \begin{cases}
          \left(1.2^{+0.8}_{-0.7}\pm0.1\right)\times10^{-10} \text{\cite{LHCb:2021vsc}} \\
         <  1.5 \times 10^{-10}\text{\cite{CMS:2022mgd}}
    \end{cases}
     ,\nonumber\\
    \mathcal{B}(B^{0}_{s} \to \mu^{+} \mu^{-}) &= \left(3.83 ^{+0.38 \, \, +0.24}_{-0.36\,\, -0.21}  \right) \times 10^{-9} \text{\cite{CMS:2022mgd}}\,.\nonumber
\end{align}

\paragraph{Semi Leptonic decays:} For the $b \to s \ell^+\ell^-$ transitions, we consider a comprehensive set of observables associated with the $B \to K^{(*)}\mu^+\mu^-$ and $B_s \to \phi\mu^+\mu^-$ decay channels. This includes differential branching fractions, $CP$ asymmetries, and angular observables measured by the LHCb, Belle, BaBar, CDF, ATLAS, and CMS collaborations~\cite{LHCb:2013lvw,LHCb:2014cxe,LHCb:2014mit,LHCb:2014vgu,LHCb:2015svh,LHCb:2020gog,LHCb:2021zwz,LHCb:2022qnv,Belle:2009zue,Belle:2016fev,BaBar:2012mrf,CDF:2011tds,ATLAS:2018gqc,CMS:2017rzx}. We also incorporate the lepton flavor universality (LFU) ratios $R_K$ and $R_{K^*}$ using the recent LHCb measurements~\cite{LHCb:2022qnv}. The corresponding theoretical expressions for these observables follow the frameworks established in Refs.\cite{Bobeth:2007dw,Altmannshofer:2008dz,Becirevic:2012fy,Descotes-Genon:2013vna,Capdevila:2016ivx,Biswas:2020uaq,Biswas:2022lhu,Kala:2025srq}. 

For the $b \to d \ell^+\ell^-$ transitions, we study the $B^+ \to \pi^+ \mu^+\mu^-$ decay channel. In particular, we incorporate the recent binned differential branching fraction and $CP$ asymmetry measurements reported by the LHCb collaboration~\cite{LHCb:2026huw}.
\subsubsection*{Radiative Decays}
The most general effective Hamiltonian describing radiative decays ($b \to d_i \gamma$) can be written as (in the WET basis):
\begin{equation}\label{eq:Heff_rad}
\mathcal{H}_{\rm eff}^{b \to d_i \gamma} = -\frac{4G_F}{\sqrt{2}} V_{tb} V_{td_i}^* \sum_{i=7,8} \left( \mathcal{C}_i(\mu) \mathcal{O}_i + \mathcal{C}_i'(\mu) \mathcal{O}_i' \right) ,
\end{equation}
where $d_i = b, s$ are the down-type quarks involved in the radiative transition. The NP contributions in terms of WET couplings can be written as,
\begin{align}
   \lambda_1 \mathcal{C}_7^{\rm NP}&=\frac{v}{\sqrt{2}}\left(\cos\theta_W \left[{C}_{dB}\right]_{d_i b} - \sin\theta_W\left[{C}_{dW}\right]_{d_i b}\right)\,,\nonumber\\
   \lambda_1 \mathcal{C}_7^{\prime \rm NP}&=\frac{v}{\sqrt{2}}\left(\cos\theta_W\left[{C}_{dB}^*\right]_{b d_i} -\sin\theta_W\left[{C}_{dW}^*\right]_{b d_i} \right)\,.
\end{align}

In our numerical analysis, we account the measurements of the branching fractions, as well as direct and mixing-induced CP asymmetries for both inclusive and exclusive radiative $B$-meson decays. All the relevant experimental measurements for these observables are tabulated in our previous work~\cite{Kala:2025srq}, which compiles data reported by the experimental collaborations~\cite{Belle:2014sac,Belle-II:2024tru,Belle:2017hum, HeavyFlavorAveragingGroupHFLAV:2024ctg}.

For the inclusive radiative decay $B \to X_s\gamma$, the branching fraction incorporating the NP WCs at the $\mu_{\rm EW}$ can be parametrised as:
\begin{equation}
	\mathcal{B}(B \to X_{s} \gamma) \times 10^{4} = (3.40 \pm 0.17) - 8.25\, \mathcal{C}_{7}^{\rm NP}(\mu_{\rm EW}) - 2.10\, \mathcal{C}_{8}^{\rm NP}(\mu_{\rm EW}) \,. 
\end{equation}
Similarly, the branching fraction for the exclusive radiative transitions $B_q \to V\gamma$, where $V$ represents a vector meson ($K^*$, $\rho$, or $\phi$), is given by:

\begin{equation}
\begin{split}
    \mathcal{B}(B_q \to V\gamma) & =  \tau_{B_q} \frac{G_F^2 \alpha_{\text{em}} m_{B_q}^3 m_b^2}{32 \pi^4} 
	\left( 1 - \frac{m_V^2}{m_{B_q}^2} \right)^3 
	|\lambda_t|^2  \left( |\mathcal{C}_7(\mu_b)|^2 + |\mathcal{C}_7'(\mu_b)|^2 \right) T_1(0),
\end{split}
\end{equation}
where $\lambda_t = V_{tb}V_{tq}^*$, and $T_1(0)$ is the tensor form factor evaluated at maximum recoil.

 The direct CP asymmetry for the inclusive decay is expressed as:
\begin{align}\label{eq:acpbsgamma}
&A_{\text{CP}}(b \to s\gamma) \equiv \frac{\Gamma(B \to X_{\bar{s}}\gamma) - \Gamma(\bar{B} \to X_s\gamma)}{\Gamma(B \to X_{\bar{s}}\gamma) + \Gamma(\bar{B} \to X_s\gamma)} \notag \\
&\simeq - \frac{1}{|\mathcal{C}_7(\mu_b)|^2 + |\mathcal{C}_7^\prime(\mu_b)|^2} \Big( 1.23 \, \text{Im}[\mathcal{C}_2(\mu_b) \mathcal{C}_7^*(\mu_b)] - 9.52 \, \text{Im}[\mathcal{C}_8(\mu_b) \mathcal{C}_7^*(\mu_b) + \mathcal{C}_8^\prime(\mu_b) \mathcal{C}_7^{\prime *}(\mu_b)] \notag \\
&+ 0.10 \, \text{Im}[\mathcal{C}_2(\mu_b) \mathcal{C}_8^*(\mu_b)] \Big) 
 - 0.5 \quad (\text{in } \%) , 
\end{align}
For the exclusive radiative decays $B\to V\gamma$, we use the expressions for the direct CP asymmetry $A_{\text{CP}}$ derived in Ref.~\cite{Bosch:2001gv}. 

Another important observable is the mixing-induced CP asymmetry in the neutral decay $B^0\to K^{*0}\gamma$, which is particularly sensitive to the interference between the left- and right-handed electromagnetic dipole operators. At leading order, it is given by:
\begin{equation}
\label{eq:Sbsgamma}
S_{K^{*}\gamma}
\simeq
\frac{2\,\mathrm{Im}\!\left(e^{-i\phi_d}\,\mathcal{C}_7(\mu_b)\,\mathcal{C}_7'(\mu_b)\right)}
{|\mathcal{C}_7(\mu_b)|^2+|\mathcal{C}_7'(\mu_b)|^2}\,,
\end{equation}
where $\phi_d$ denotes the weak phase associated with $B_d$--$\bar B_d$ mixing. In our numerical analysis, we use the current PDG average~\cite{ParticleDataGroup:2024cfk}, $\sin\phi_d = 0.710 \pm 0.011$, while the highly suppressed SM expectation is $S_{K^{*}\gamma}^{\rm SM}=(-2.3\pm1.6)\%$~\cite{Ball:2006eu}.

\subsubsection*{Invisible Decays}
The most general effective Hamiltonian describing the $b\to d_i\nu\bar{\nu}$ transitions is given by:
\begin{align}
    \mathcal{H}^{b\to d_i\nu\bar{\nu}}=\frac{4 G_F}{\sqrt{2}}V_{tb} V_{td_i}^*\sum_i C_i^{\nu} O_i^{\nu}
\end{align}
where the relevant operators are defined as:
\begin{align}
    \mathcal{O}_{L}^{\nu} = \frac{e^2}{16 \pi^2}  \left( \bar{d}_{j} \gamma_{\mu} P_{L}b \right) \left(\bar{\nu} \gamma^{\mu}P_{L}\nu\right) , & \ \ \ \ &  \mathcal{O}_{R}^{\nu}=  \frac{e^2}{16 \pi^2}  \left( \bar{d}_{j} \gamma_{\mu} P_{R}b \right) \left(\bar{\nu} \gamma^{\mu}P_{L}\nu\right)\,,
\end{align}
In the SM, only the left-handed vector current contributes. However, in our analysis, we obtain contributions from both chiralities of the vector operators. The corresponding matching relations with the WET couplings are as follows:
\begin{align}
    \lambda_2\,C_L^{\nu\,\rm NP}&=-\frac{g_Z^2 v^2}{2 M_Z^2}\big[C_{Hq}^{(+)}\big]_{d_i b}\,,& \quad \lambda_2\,C_R^{\nu\, \rm NP}=-\frac{g_Z^2 v^2}{2 M_Z^2}\big[C_{Hd}\big]_{d_i b}\,.
\end{align}
In our numerical analysis, we incorporate experimental inputs from the $B^+ \to K^+\nu\bar{\nu}$ and $B \to K^*\nu\bar{\nu}$ decay channels. 
\begin{align}
    \mathcal{B}(B^+\to K^+\nu\bar{\nu}) &= (2.3\pm 0.5^{+0.5}_{-0.4})\times 10^{-5}\text{\cite{Belle-II:2023esi}}\,, \\ 
    \mathcal{B}(B^0\to K^{*\,0}\nu\bar{\nu}) &= 
    \begin{cases}
        (3.8^{+2.9}_{-2.6})\times 10^{-5}\text{\cite{BaBar:2013npw}} \\ 
        < 1.8 \times 10^{-5}\text{\cite{Belle:2017oht}}
    \end{cases}\nonumber
\end{align}
The corresponding differential decay rates are given in Eq.~\eqref{eq:decay_inv}.
\begin{subequations}\label{eq:decay_inv}
\begin{align}
    \begin{split}
        \frac{d \mathcal{B}(B^+\to K^+\nu\bar{\nu})}{dq^2}&=\tau_B \frac{G_F^2 \alpha^2}{256 \pi^5}\frac{\lambda(q^2,m_B^2,m_K^2)^{3/2}}{m_B^3}\left|V_{tb}V_{ts}^*\right|^2\bigg[f_+(q^2)\bigg]^2\left|C_L^{\nu}+C_R^{\nu}\right|^2\,,
    \end{split}
\end{align}

\begin{align}
    \frac{d \mathcal{B}(B\to K^*\nu\bar{\nu})}{dq^2}&=\tau_B \frac{G_F^2 \alpha^2}{128 \pi^5}\frac{\lambda(q^2,m_B^2,m_{K^*}^2)^{1/2}\,q^2}{m_B^3}(m_B+m_{K^*})^2\left|V_{tb}V_{ts}^*\right|^2 \nonumber\\
    &\times \left\{\left(\bigg[A_1(q^2)\bigg]^2+ \frac{32 m_B^2 m_{K^*}^2}{q^2 (m_B+m_{K^*})^2}\bigg[A_{12}(q^2)\bigg]^2\right)\left|C_L^{\nu}-C_R^{\nu}\right|^2\right.\\& \left. +\frac{\lambda(q^2,m_B^2,m_{K^*}^2)}{(m_B+m_{K^*})^4}\bigg[V(q^2)\bigg]^2\left|C_L^{\nu}+C_R^{\nu}\right|^2\right\}\nonumber
\end{align}
\end{subequations}
where $\lambda$ is the K\"all\'en function and $q^2$ is the dineutrino invariant mass squared. The quantity $f_+(q^2)$ denotes the $B \to K$ transition form factor, whereas $A_1(q^2)$, $A_{12}(q^2)$, and $V(q^2)$ are the form factors associated with the $B \to K^*$ transition. We use the decay constants $f_K = (155.7 \pm 0.3)\,\text{MeV}$ and $f_{K^*} = (205 \pm 6)\,\text{MeV}$~\cite{FlavourLatticeAveragingGroupFLAG:2024oxs}. All form factors relevant to the $B \to K$ and $B \to K^*$ transitions are adopted from Ref.~\cite{Becirevic:2023aov}.
\subsubsection*{Meson Mixing}

Neutral meson mixing is described by the most general $\Delta F=2$ effective Hamiltonian. In our analysis, we adopt the BMU basis \cite{Buras:2000if} to define the complete set of four-quark operators, which is written as
\begin{align}
\label{eq:mixing_delF_2}
\mathcal{H}_{\rm eff}^{\Delta F=2} 
&= C_1^{\text{VLL}}(\mu)\, Q_1^{\text{VLL}} + \sum_{i=1}^{2} C_i^{\text{LR}}(\mu)\, Q_i^{\text{LR}} + \sum_{i=1}^{2} C_i^{\text{SLL}}(\mu)\, Q_i^{\text{SLL}} \nonumber \\
&\quad + C_1^{\text{VRR}}(\mu)\, Q_1^{\text{VRR}} + \sum_{i=1}^{2} C_i^{\text{SRR}}(\mu)\, Q_i^{\text{SRR}} + \text{h.c.}\,,
\end{align}
where the operators $Q_i^{a}$ correspond to the VLL, LR, and SLL sectors, and their chirality-flipped counterparts belong to the VRR and SRR sectors.

In the present analysis, the vector four-fermion operators $(Q_1^{\rm VLL}, Q_1^{\rm VRR})$ and tensor-type operators $(Q_2^{\rm SLL}, Q_2^{\rm SRR})$ receive new-physics contributions. The relevant operators are given by
\begin{align}\label{eq:meson_mixing_basis}
Q_1^{\text{VLL}} &= (\bar{q}^\alpha \gamma_\mu P_L b^\alpha)(\bar{q}^\beta \gamma^\mu P_L b^\beta)\,, &
Q_1^{\text{VRR}} &= (\bar{q}^\alpha \gamma_\mu P_R b^\alpha)(\bar{q}^\beta \gamma^\mu P_R b^\beta)\,, \nonumber\\[0.8em]
Q_2^{\text{SLL}} &= (\bar{q}^\alpha \sigma_{\mu\nu} P_L b^\alpha)(\bar{q}^\beta \sigma^{\mu\nu} P_L b^\beta)\,, &
Q_2^{\text{SRR}} &= (\bar{q}^\alpha \sigma_{\mu\nu} P_R b^\alpha)(\bar{q}^\beta \sigma^{\mu\nu} P_R b^\beta)\,,
\end{align}

Here, $\alpha$ and $\beta$ are the color indices, and $q \in \{d, s\}$ denotes the light down-type quark. This general notation allows the operators to simultaneously describe both $B_d^0-\bar{B}_d^0$ and $B_s^0-\bar{B}_s^0$ mixing systems. The corresponding contributions to the SMEFT WCs are
\begin{align}
C_1^{\rm VLL}
&=
-\frac{g_Z^2 v^4}{4 M_Z^2}
\left|
C_{Hq}^{(+)}
\right|_{d_i b}^{\,2}\,,
&
C_1^{\rm VRR}
&=
-\frac{g_Z^2 v^4}{4 M_Z^2}
\left|
C_{Hd}
\right|_{d_i b}^{\,2}\,,\\
C_2^{\rm SLL}&= \frac{v^2}{4}\left|\sin\theta_W~ \left[C_{dW}^*\right]_{d_i b}-\cos\theta_W~\left[C_{dB}^*\right]_{d_i b}\right|^2,\nonumber\\
C_2^{\rm SRR}&= \frac{v^2}{4}\left|\sin\theta_W~ \left[C_{dW}\right]_{d_i b}-\cos\theta_W~\left[C_{dB}\right]_{d_i b}\right|^2  \,.\nonumber
\end{align}
The experimental measurements for the neutral $B$-meson mixing observables are given by:
\begin{align}
    \Delta M_d &= (0.5062 \pm 0.0019)\,\mathrm{ps^{-1}}~\text{\cite{HFLAV:2022esi}}\,,&\qquad
    \Delta M_s &= (17.765 \pm 0.006)\,\mathrm{ps^{-1}}~\text{\cite{HFLAV:2022esi}}\,.
\end{align}

To connect the effective Hamiltonian to the mass difference $\Delta M_q=2\,|M_{12}| = |\mathcal{M}|/{m_{B_q}}$, we evaluate the hadronic matrix elements between the $B_q^0$ and $\bar{B}_q^0$ states. We utilise lattice inputs from \cite{FlavourLatticeAveragingGroupFLAG:2024oxs, FermilabLattice:2016ipl}, applying Fierz transformations to map the conventionally reported color-mixed basis onto the BMU basis. Due to QCD parity conservation, the matrix elements of chirality-flipped operators are identical to their left-handed counterparts. 
\begin{align}
\langle \bar{B}_q^0 | Q_1^{\text{VRR}} | B_q^0 \rangle &= \langle \bar{B}_q^0 | Q_1^{\text{VLL}} | B_q^0 \rangle = \frac{2}{3} m_{B_q}^2 f_{B_q}^2 B_{B_q}(\mu)\,, \nonumber \\[0.8em]
\langle \bar{B}_q^0 | Q_2^{\text{SRR}} | B_q^0 \rangle &= \langle \bar{B}_q^0 | Q_2^{\text{SLL}} | B_q^0 \rangle = - m_{B_q}^2 f_{B_q}^2 R_q(\mu) B_{B_q}(\mu)\,,
\end{align}
where $R_q(\mu) = \left[ m_{B_q} / (m_b(\mu) + m_q(\mu)) \right]^2$ is the chiral enhancement factor evaluated at the scale $\mu=\mu_b$, and $B_{B_q}$ is the corresponding bag parameter, which we set to $B_{B_q}^{(i)}=1$. For simplicity, we assume the NP couplings do not introduce any additional phase, allowing the total mass difference to be written as a linear sum $\Delta M_q=\Delta M_q^{\rm SM} + \Delta M_q^{\rm NP}$. 



\subsubsection*{Fit Results}
We perform a global $\chi^2$ analysis using the theoretical and experimental constraints from the various down-type FCNC processes to determine the allowed ranges of the relevant SMEFT WCs. The best-fit values and corresponding $1\sigma$ confidence intervals are presented in Tab.~\ref{tab:bottombound}.
\begin{table}[h!]
    \centering
    \begin{tabular}{|
    >{\centering\arraybackslash}p{0.10\textwidth}|
    >{\centering\arraybackslash}p{0.35\textwidth}|
    >{\centering\arraybackslash}p{0.10\textwidth}|
    >{\centering\arraybackslash}p{0.35\textwidth}|
    }
    \hline \hline
    \textbf{WCs} & \textbf{Values $\boldsymbol{(\mu_{\rm EW})\,[\mathrm{TeV}^{-2}]}$} & \textbf{WCs} & \textbf{Values $\boldsymbol{(\mu_{\rm EW})\,[\mathrm{TeV}^{-2}]}$}  \\
    \hline \hline
    $[C_{dW}]_{bs}$ & $(-0.67 \pm 4.50)\times 10^{-7}$ & $[C_{dW}]_{bd}$ & $(9.82 \pm 0.66)\times 10^{-6}$  \\
    $[C_{dB}]_{bs}$ & $(0.37 \pm 3.62)\times 10^{-7}$ & $[C_{dB}]_{bd}$ & $(-5.39 \pm 0.36)\times 10^{-6}$ \\
    $[C^{(+)}_{Hq}]_{bs}$ & $(0.99 \pm 0.16)\times 10^{-3}$ & $[C^{(+)}_{Hq}]_{bd}$ & $(1.59 \pm 1.19)\times 10^{-4}$ \\
    $[C_{Hd}]_{bs}$ & $(2.49 \pm 1.23)\times 10^{-4}$ & $[C_{Hd}]_{bd}$ & $(1.62 \pm 1.19)\times 10^{-4}$ \\
    \hline \hline
    \end{tabular}
    \caption{Best-fit values and corresponding allowed ranges of the real down-type FCNC SMEFT WCs at the electroweak scale.}
    \label{tab:bottombound}
\end{table}
In this primary analysis, all the extracted WCs are assumed to be strictly real. Within our framework, the dipole-type couplings ($C_{dW}$ and $C_{dB}$) are predominantly constrained by radiative decays at the $\mathcal{O}(10^{-7})$ order, whereas the current-type couplings ($C_{Hq}^{(+)}$ and $C_{Hd}$) are mostly bounded by rare leptonic and non-leptonic decays at the order $\mathcal{O}(10^{-4})$. However, since the dipole-type couplings are inherently non-Hermitian, they can generally possess CP-violating imaginary phases. To address this, we perform an additional, independent fit where these specific dipole couplings are treated as complex quantities. The best-fit values and bounds for this complex scenario are tabulated in Appendix~\ref{Append:complex_fit}.

\subsection{Top FCNC Processes}
To constrain the parameter space of top-quark FCNC interactions, we first utilise direct limits on various top FCNC branching fractions measured at the LHC. However, because the top quark is the heaviest elementary particle, its loop-level effects can be highly significant. In the current era of precision measurements, these virtual quantum contributions cannot be neglected. Therefore, we additionally investigate the indirect imprints of top FCNC interactions through their virtual contributions to low-energy flavour observables, as well as various electroweak precision observables (EWPOs) and Higgs sector measurements. Combining both direct high-$p_T$ measurements and indirect precision constraints allows us to comprehensively gauge the allowed parameter space for the top-quark FCNC couplings.

\subsubsection*{Direct Collider Bounds}
The current experimental upper bounds from the LHC on the top-quark FCNC branching fractions ($t \to q^{\prime} X$, where $q^{\prime} \in \{u, c\}$ and $X \in \{\gamma, Z\}$) utilised in our numerical analysis are summarised in Eq.~\eqref{eq:top_fcnc_exp}.

\begin{subequations}\label{eq:top_fcnc_exp}
\begin{align}
    &\mathcal{B}(t \to c\,\gamma)\quad \,< \quad 1.51 \times 10^{-5}~\text{\cite{CMS:2023bjm}}\,,\\
    &\mathcal{B}(t \to u\,\gamma)\quad \,< \quad 0.95 \times 10^{-5}~\text{\cite{CMS:2023bjm}}\,,\\
    &\mathcal{B}(t \to c\, Z)\quad < \quad
    1.3 \times 10^{-4}\;(\mathrm{LH})\,,\qquad
    1.2 \times 10^{-4}\;(\mathrm{RH})
    ~\text{\cite{ATLAS:2023qzr}}\,,\\
    &\mathcal{B}(t \to u\, Z)\quad < \quad
    6.2 \times 10^{-5}\;(\mathrm{LH})\,,\qquad
    6.6 \times 10^{-5}\;(\mathrm{RH})
    ~\text{\cite{ATLAS:2023qzr}}\,.
\end{align}
\end{subequations}
To translate these experimental limits into bounds on the NP parameter space, we express the theoretical branching fractions in terms of the relevant SMEFT WCs. The branching fractions for the respective $t \to qX$ processes can be written as:

\begin{align}\label{eq:collider_top_BR}
    \mathcal{B}\left(t \to q^{\prime} \gamma\right)
    &= \frac{v^2 m_t^3}{8\pi\,\Gamma_t}
    \Bigg[
    \cos^2\theta_W
    \left(
    \Big|[C_{uB}]_{tq^{\prime}}\Big|^2
    +
    \Big|[C_{uB}]^{*}_{tq^{\prime}}\Big|^2
    \right)
    +
    \sin^2\theta_W
    \nonumber\\
    &\quad \bigg(
    \Big|[C_{uW}]_{tq^{\prime}}\Big|^2
    +
    \Big|[C_{uW}]^{*}_{tq^{\prime}}\Big|^2
    \bigg) +
    2\cos\theta_W \sin\theta_W \\
    & \quad \Big(
    [C_{uB}]_{tq^{\prime}}[C_{uW}]_{tq^{\prime}}
    +
    [C_{uB}]^{*}_{tq^{\prime}}[C_{uW}]^{*}_{tq^{\prime}}
    \Big)
    \Bigg]\,,
    \nonumber
    \end{align}
    \begin{align}
    \mathcal{B}\left(t \to q^{\prime} Z\right)
    &= \frac{v^2 m_t^3}{32\pi\,\Gamma_t}
    \left(1-\frac{m_Z^2}{m_t^2}\right)^2
    \Bigg[
    \left(
    \frac{m_Z^2}{m_t^2}+2
    \right)
    \Bigg\{
    2\sin^2\theta_W
    \bigg(
    \Big|[C_{uB}]_{tq^{\prime}}\Big|^2
    \nonumber\\
    & +
    \Big|[C_{uB}]^{*}_{tq^{\prime}}\Big|^2
    \bigg) +
    2\cos^2\theta_W
    \left(
    \Big|[C_{uW}]_{tq^{\prime}}\Big|^2
    +
    \Big|[C_{uW}]^{*}_{tq^{\prime}}\Big|^2
    \right)\nonumber\\
    & -
    4\cos\theta_W \sin\theta_W
    \Big(
    [C_{uB}]_{tq^{\prime}}[C_{uW}]_{tq^{\prime}}
    +
    [C_{uB}]^{*}_{tq^{\prime}}[C_{uW}]^{*}_{tq^{\prime}}
    \Big)
    \Bigg\}
    \nonumber\\
    & +
    \left(
    2\frac{m_Z^2}{m_t^2}+1
    \right)
    \Bigg\{
    \Big|[C_{\phi q}^{(-)}]_{tq^{\prime}}\Big|^2
    +
    \Big|[C_{\phi u}]_{tq^{\prime}}\Big|^2
    \Bigg\} +
    6\sqrt{2}\frac{m_Z}{m_t}
    \\
    &\quad
    \Bigg\{
    \sin\theta_W
    \left(
    [C_{uB}]_{tq^{\prime}}[C_{\phi q}^{(-)}]_{tq^{\prime}}
    +
    [C_{uB}]^{*}_{tq^{\prime}}[C_{\phi u}]_{tq^{\prime}}
    \right)
     -
    \cos\theta_W \nonumber\\
    & \quad
    \left(
    [C_{uW}]_{tq^{\prime}}[C_{H q}^{(-)}]_{tq^{\prime}}
    +
    [C_{uW}]^{*}_{tq^{\prime}}[C_{H u}]_{tq^{\prime}}
    \right)
    \Bigg\}
    \Bigg]\,.\nonumber
\end{align}
Using the theoretical expressions derived in Eq.~\eqref{eq:collider_top_BR} alongside the current experimental upper limits given in Eq.~\eqref{eq:top_fcnc_exp}, we extract constraints on the individual SMEFT WCs. For this procedure, we adopt a single-parameter framework, assuming only one coupling to be non-zero at a time while setting the others to their SM values (which are identically zero for top FCNC couplings). The resulting single-parameter upper bounds on the relevant top FCNC WCs are presented in Tab.~\ref{tab:topbounds}.
\begin{table}[h!]
    \centering
    \begin{tabular}{|
    >{\centering\arraybackslash}p{0.10\textwidth}|
    >{\centering\arraybackslash}p{0.375\textwidth}|
    >{\centering\arraybackslash}p{0.375\textwidth}|}
    \hline \hline
    \textbf{WCs} & $\boldsymbol{t \to q^{\prime} \gamma}$ & $\boldsymbol{t \to q^{\prime} Z}$ \\
    \hline \hline
    $[C_{uW}]_{tc}$ & $<6.12 \times 10^{-2}$ & $<1.28 \times 10^{-1}$ \\
    $[C_{uB}]_{tc}$ & $<3.36 \times 10^{-2}$ & $<2.34 \times 10^{-1}$ \\
    $[C^{(-)}_{Hq}]_{tc}$ & $-$ & $<2.72 \times 10^{-1}$ \\
    $[C_{Hu}]_{tc}$ & $-$ &  $<2.72 \times 10^{-1}$ \\
    \hline \hline
    $[C_{uW}]_{tu}$ & $<4.85 \times 10^{-2}$ & $<0.89 \times 10^{-1}$ \\
    $[C_{uB}]_{tu}$ & $<2.66 \times 10^{-2}$ & $<1.61 \times 10^{-1}$ \\
    $[C^{(-)}_{Hq}]_{tu}$ & $-$ & $<1.88 \times 10^{-1}$ \\
    $[C_{Hu}]_{tu}$ & $-$ & $<1.88 \times 10^{-1}$ \\
    \hline \hline
    \end{tabular}
    \caption{Upper bounds on WCs ($\text{TeV}^{-2}$) from radiative top-FCNC decay modes.}
    \label{tab:topbounds}
\end{table}

\subsubsection*{Top FCNC Indirect Limits}
The complete set of observables providing indirect limits on top FCNC are detailed below:
\begin{itemize}
    \itemsep-0em
    \item \textbf{Low Energy FCNC:} \textit{$b\to sll$ observables}: Angular observables, isospin asymmetries, differential branchings $(dB/dq^2)$, LFUV ratios $(R_K, R_{K^*})$; \textit{Radiative decays}: $\mathcal{B}(B \rightarrow X_s\gamma)$, $\mathcal{B}(B_q \rightarrow V\gamma)$; \textit{Leptonic decays}: $\mathcal{B}(B_s^0 \rightarrow \mu^+\mu^-)$, $\mathcal{B}(B^0 \rightarrow \mu^+\mu^-)$, $\mathcal{B}(K_{L} \rightarrow \mu^+\mu^-)$; \textit{Invisible decays}: $\mathcal{B}(B^+ \rightarrow K^+\nu\bar{\nu})$, $\mathcal{B}(B^0 \rightarrow K^{*0}\nu\bar{\nu})$, $\mathcal{B}(K^+ \rightarrow \pi^+\nu\bar{\nu})$; \textit{Meson mixing}: $B^0-\bar{B}^0$  mixing amplitudes.
    \item \textbf{Low Energy FCCC:} \textit{Semi/leptonic decays}: Rates/branchings for $P \rightarrow Ml\nu_{l}$ and $P \rightarrow l\nu_l$, LFU ratios $R(D)$ and $R(D^*)$.
    \item \textbf{Anomalous $\boldsymbol{tWb}$}: \textit{Vertex modifications}: $|f_L V_{tb}|$, $V_R$, $g_L$, $g_R$.
    \item \textbf{EWPOs:} \textit{Oblique parameters}: $S$, $T$, $U$ parameters; \textit{$Z$ pole observables}: $\Gamma_Z$, $\sigma_{\rm had}$, $A_e$, $A_\mu$, $A_\tau$, $A_b$, $A_s$, $A_c$, $A_e^{FB}$, $A_\mu^{FB}$, $A_\tau^{FB}$, $A_b^{FB}$, $A_s^{FB}$, $A_c^{FB}$, $R_e$, $R_\mu$, $R_\tau$, $R_b$, $R_c$; \textit{$W$-pole observables}: $\Gamma_W$, $R_{Wc}$.
    \item \textbf{Higgs Observables}: \textit{Decay channels}: $h \rightarrow b\bar{b}$, $h \rightarrow WW^*$, $h \rightarrow \gamma\gamma$, $h \rightarrow Z\gamma$; \textit{Coupling modifiers}: $\kappa_\gamma$, $\kappa_{Z\gamma}$, $\kappa_g$.
    \item \textbf{Gauge and Top Quark Sector:} \textit{Trilinear gauge couplings (TGCs)}: $g_1^Z$, $\lambda_\gamma$, $\kappa_Z$, $\lambda_Z$, $g_4^Z$, $\tilde{\kappa}_Z$, $\tilde{\lambda}_Z$; \textit{Top quark dipole moments}: CMDM ($\mu_t$), CEDM ($d_t$).
    \item \textbf{Electric Dipole Moments (EDMs):} \textit{Nucleon EDM}: Neutron EDM ($d_n$).
\end{itemize}

\begin{table}[htb!]
    \centering
    \begin{tabular}{|
    >{\centering\arraybackslash}p{0.10\textwidth}|
    >{\centering\arraybackslash}p{0.35\textwidth}|
    >{\centering\arraybackslash}p{0.10\textwidth}|
    >{\centering\arraybackslash}p{0.35\textwidth}|
    }
    \hline \hline
    \textbf{WCs} & \textbf{Values $\boldsymbol{(\mu_{\rm EW})\,[\mathrm{TeV}^{-2}]}$} & \textbf{WCs} & \textbf{Values $\boldsymbol{(\mu_{\rm EW})\,[\mathrm{TeV}^{-2}]}$}  \\
    \hline \hline
    $[C_{uW}]_{tc}$ & $(-0.88\pm 2.48)\times 10^{-2}$ & $[C_{uW}]_{tu}$ & $(-0.15 \pm 0.92)\times 10^{-1}$ \\
    $[C_{uB}]_{tc}$ & $(-0.10\pm 1.45)\times 10^{-2}$ & $[C_{uB}]_{tu}$ & $(-0.15 \pm 5.79)\times 10^{-2}$ \\
    $[C^{(-)}_{Hq}]_{tc}$ & $(-0.58 \pm 2.02)\times 10^{-4}$ & $[C^{(-)}_{Hq}]_{tu}$ & $(+0.27 \pm 4.95)\times 10^{-4}$ \\
    $[C_{Hu}]_{tc}$ & $+0.79\pm 9.32$ & $[C_{Hu}]_{tu}$ & $-1.56 \pm 4.49$ \\
    \hline \hline
    \end{tabular}
    \caption{Fit results for top-quark FCNC processes from the global analysis using the observables listed above. Here, the WCs and all coefficients are evaluated at the scale $\mu = m_W$.}
    \label{tab:topbounds2}
\end{table}

Using this comprehensive dataset of low-energy, electroweak, Higgs, and top-quark observables (Tab.~\ref{tab:topbounds2}), we constrain the top-quark FCNC interactions for $t \to q^\prime \gamma$ and $t \to q^\prime Z$ ($q^\prime = u,c$). Comparing these indirect constraints with the direct collider measurements (Tab.~\ref{tab:topbounds}) reveals complementary sensitivities. For instance, indirect searches place tighter bounds on $C_{uB}$ and $C_{Hq}^{(-)}$, whereas $C_{Hu}$ remains loosely constrained and $C_{uW}$ limits are comparable across both approaches. These complementary sensitivities highlight the necessity of a combined global analysis that incorporates both high-$p_T$ collider data and low-energy precision inputs. Since the direct collider searches currently provide only upper bounds rather than central values, they can be statistically integrated into a combined $\chi^2$ fit by employing an appropriate half-Gaussian penalty function. A comprehensive discussion regarding this statistical treatment can be found in Ref.~\cite{Kala:2026xzo}.

\begin{figure}[h!]
    \centering
    \includegraphics[width=1\linewidth]{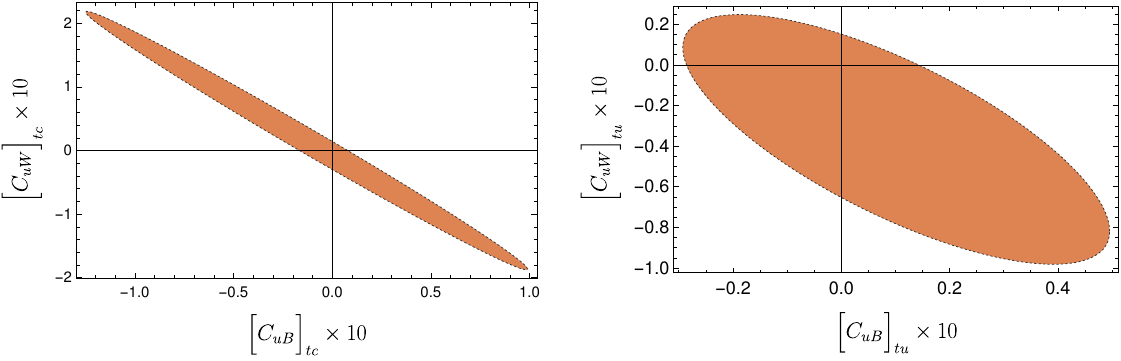}
    \caption{Correlations among the dipole type SMEFT WCs contributing to the top FCNC processes $t \to u_j \,\gamma(Z)$. }
    \label{fig:top_corr}
\end{figure}

In Fig.~\ref{fig:top_corr}, we present a correlation plot derived from the combined analysis, showing the interplay between the real WCs, $C_{uB}$ and $C_{uW}$. A strong anti-correlation is observed. This anti-correlation stems from the fact that both $C_{uB}$ and $C_{uW}$ are significantly more sensitive to radiative $B$-meson decays governed by the loop-induced $tu_j\gamma$ interaction than to the rare semileptonic $b \to s(d)\ell^+\ell^-$ processes mediated by the $tu_jZ$ interaction. In order to simultaneously satisfy the stringent experimental constraints from these radiative decays, the two couplings must dynamically compensate for one another, resulting in the observed negative correlation along a specific flat direction. Unlike the bottom FCNC dipole couplings, we also provided the complex fitted values of the top FCNC dipole WCs in Appendix~\ref{Append:complex_fit}.

\subsection{Charged LFV Modes}
The observation of neutrino oscillations~\cite{SuperKamiokande:1998kpq} establishes that lepton flavor is not an exact symmetry of nature. However, within the SM with massless neutrinos, charged LFV is absent, while its extension to accommodate neutrino masses generates charged LFV only at highly suppressed rates. Therefore, searches for lepton flavor-violating decays of charged leptons provide a sensitive probe of physics beyond the SM. In this section, we consider charged LFV processes involving the $\tau$ lepton via leptonic and  hadronic decay channels. In addition, we investigate the constraints arising from LFV $Z$-boson decays, which provide a complementary probe of the flavor-violating interactions at the electroweak scale.

To simplify the expressions, we absorb the Weinberg angle ($\theta_{W}$) dependence within the modified WCs. For the dipole operators, we define:
\begin{equation}
\begin{split}
    [C_{\gamma}]_{\tau l} =& - \sin{\theta_{W}} [C_{eW}]_{\tau l} + \cos{\theta_{W}} [C_{eB}]_{\tau l} \,, \\ 
    [C_{\gamma}]^{*}_{\tau l} =& - \sin{\theta_{W}} [C_{eW}]^{*}_{\tau l} + \cos{\theta_{W}} [C_{eB}]^{*}_{\tau l} \,,\\
    [C_{Z}]_{\tau l} =& -\sin{\theta_{W}} [C_{eB}]_{\tau l} - \cos{\theta_{W}} [C_{eW}]_{\tau l}\,, \\ 
    [C_{Z}]^{*}_{\tau l} =& - \sin{\theta_{W}} [C_{eB}]^{*}_{\tau l} - \cos{\theta_{W}} [C_{eW}]^{*}_{\tau l}\,.\\
\end{split}
\end{equation}
These combinations correspond to the effective dipole couplings entering the photon and $Z$-mediated LFV interactions. They allow the expressions for the relevant decay amplitudes and observables to be written directly in terms of the physical neutral gauge bosons. Similarly, for the Higgs-current operators, we define:
\begin{equation}
\begin{split}
    [C^{\ell}_{Z}]^{i}_{\tau l} = \alpha_{i} + \beta_{i} \sin^{2}{\theta_{W}} [C^{(+)}_{H\ell}]_{\tau l}\,, \qquad [C^{e'}_{Z}]^{i}_{\tau l} = \beta_{i} \sin^{2}{\theta_{W}} [C_{He}]_{\tau l}\,, \\
    [C^{e}_{Z}]^{i}_{\tau l} = \alpha_{i} + \beta_{i} \sin^{2}{\theta_{W}} [C_{He}]_{\tau l}\,, \qquad [C^{\ell'}_{Z}]^{i}_{\tau l} = \beta_{i} \sin^{2}{\theta_{W}} [C^{(+)}_{H\ell}]_{\tau l}\,,
\end{split}
\end{equation}
where the coefficients $\alpha_{i}$ are defined as follows: $\alpha_{\mu} = \alpha_{e} = -1$, $\alpha_{u} = 1$, and $\alpha_{d} = \alpha_{s} = -1$. Similarly, $\beta_{i}$ are defined as: $\beta_{\mu} = \beta_{e} = 2$, $\beta_{u} = -\frac{4}{3}$, and $\beta_{d} = \beta_{s} = \frac{2}{3}$. These coefficients account for the corresponding SM electroweak charges of the fermions appearing in the $Z$-mediated interactions. The expressions used throughout this section are adopted from Refs.~\cite{Altmannshofer:2023tsa,Altmannshofer:2025nbp}.

\subsubsection*{Leptonic Modes $\boldsymbol{(\tau \to l\gamma\,/\,3l)}$}
Radiative and three-body charged lepton flavor-violating $\tau$ decays provide stringent constraints on the dipole and Higgs-current operators. The current experimental bounds on the radiative decay modes are
\begin{equation}
\begin{aligned}
    \mathcal{B}(\tau^{\pm} \to \mu^{\pm} \gamma) &< 4.2 \times 10^{-8}~\text{\cite{Belle:2021ysv}}\,, &\qquad
    \mathcal{B}(\tau^{\pm} \to e^{\pm} \gamma) &< 3.3 \times 10^{-8}~\text{\cite{BaBar:2009hkt}}\,.
\end{aligned} 
\end{equation}
The charged LFV $\tau \to l\gamma$ branching ratios can be expressed in terms of the dipole WCs as
\begin{equation}
\begin{split}
\frac{\mathcal{B}(\tau^{\pm} \to \mu^{\pm} \gamma)}{\mathcal{B}(\tau^{\pm} \to \mu^{\pm} \nu_{\tau}{\nu}_{\mu})} &\simeq \frac{48 \pi^{2} v^{6}}{m^{2}_{\tau}} \Bigg(\Big|[C_{\gamma}]_{\tau\mu}\Big|^{2} + \Big|[C_{\gamma}]^{*}_{\tau\mu}\Big|^{2} \Bigg)\,,\\
\end{split}
\end{equation}
\begin{equation}
\begin{split}
\frac{\mathcal{B}(\tau^{\pm} \to e^{\pm} \gamma)}{\mathcal{B}(\tau^{\pm} \to e^{\pm} \nu_{\tau}{\nu}_{e})} &\simeq \frac{48 \pi^{2} v^{6}}{m^{2}_{\tau}} \Bigg(\Big|[C_{\gamma}]_{\tau e}\Big|^{2} + \Big|[C_{\gamma}]^{*}_{\tau e}\Big|^{2} \Bigg)\,,\\
\end{split}
\end{equation}
where, $\mathcal{B}(\tau^{\pm} \to \mu^{\pm} (e^{\pm}) \nu_{\tau}{\nu}_{\mu}) \simeq 17.37\%~(17.85\%)$. The three-body charged lepton flavor-violating decay modes provide additional sensitivity to both dipole and Higgs-current operators. The relevant experimental bounds are
\begin{equation}
\begin{aligned}
    \mathcal{B}(\tau^{\pm} \to \mu^{\pm}\mu^+\mu^-) &< 1.9 \times 10^{-8}~\text{\cite{Hayasaka:2010np}}\,, &\qquad
    \mathcal{B}(\tau^{\pm} \to e^{\pm}\mu^+\mu^-) &< 2.7 \times 10^{-8}~\text{\cite{Hayasaka:2010np}}\,, \\
    \mathcal{B}(\tau^{\pm} \to \mu^{\pm}e^+e^-) &< 1.8 \times 10^{-8}~\text{\cite{Hayasaka:2010np}}\,, &\qquad
    \mathcal{B}(\tau^{\pm} \to e^{\pm}e^+e^-) &< 2.7 \times 10^{-8}~\text{\cite{Hayasaka:2010np}}\,.
\end{aligned} 
\end{equation}
The $\tau \to 3l$ branching ratios associated with $\tau\mu$ flavor violation are as follows:
\begin{equation}
\begin{split}
\frac{\mathcal{B}(\tau^{\pm} \to \mu^{\pm} \mu^+ \mu^-)}{\mathcal{B}(\tau^{\pm} \to \mu^{\pm} \nu_{\tau}{\nu}_{\mu})} & \simeq \frac{v^{4}}{4} \Bigg[ 2 \Big|[C^{\ell}_{Z}]^{\mu}_{\tau\mu}  \Big|^{2} + 2 \Big|[C^{e'}_{Z}]^{\mu}_{\tau\mu}  \Big|^{2} + \Big|[C^{e}_{Z}]^{\mu}_{\tau\mu}  \Big|^{2} +  \Big|[C^{\ell'}_{Z}]^{\mu}_{\tau\mu}  \Big|^{2} \\ & - \frac{4\sqrt{2} ev}{m_{\tau}} \mathfrak{Re} \bigg\{[C_{\gamma}]_{\tau\mu} \Big(2[C^{\ell}_{Z}]^{\mu}_{\tau\mu} + [C^{\ell'}_{Z}]^{\mu}_{\tau\mu} \Big)^* + [C_{\gamma}]^{*}_{\tau\mu} \Big(2[C^{e}_{Z}]^{\mu}_{\tau\mu} \\ & + [C^{e'}_{Z}]^{\mu}_{\tau\mu} \Big)^* \bigg\} + \frac{16 e^{2} v^{2}}{m^{2}_{\tau}} \bigg([C_{\gamma}]_{\tau\mu} + [C_{\gamma}]^{*}_{\tau\mu}\bigg) \bigg(\log{\frac{m^{2}_{\tau}}{m^{2}_{e}}} - \frac{11}{4}\bigg) \Bigg]\,, \\
\end{split}
\end{equation}
\begin{equation}
\begin{split}
\frac{\mathcal{B}(\tau^{\pm} \to \mu^{\pm} e^+e^-)}{\mathcal{B}(\tau^{\pm} \to \mu^{\pm} \nu_{\tau}{\nu}_{\mu})} & \simeq \frac{v^{4}}{4} \Bigg[ \Big|[C^{\ell}_{Z}]^{e}_{\tau\mu}  \Big|^{2} + \Big|[C^{e'}_{Z}]^{e}_{\tau\mu}  \Big|^{2} + \Big|[C^{e}_{Z}]^{e}_{\tau\mu}  \Big|^{2} +  \Big|[C^{\ell'}_{Z}]^{e}_{\tau\mu}  \Big|^{2} \\ & - \frac{4\sqrt{2} ev}{m_{\tau}} \mathfrak{Re} \bigg\{[C_{\gamma}]_{\tau\mu} \Big(2[C^{\ell}_{Z}]^{e}_{\tau\mu} + [C^{\ell'}_{Z}]^{e}_{\tau\mu} \Big)^* + [C_{\gamma}]^{*}_{\tau\mu} \Big(2[C^{e}_{Z}]^{e}_{\tau\mu} \\ & + [C^{e'}_{Z}]^{e}_{\tau\mu} \Big)^* \bigg\} + \frac{16 e^{2} v^{2}}{m^{2}_{\tau}} \bigg([C_{\gamma}]_{\tau\mu} + [C_{\gamma}]^{*}_{\tau\mu}\bigg) \bigg(\log{\frac{m^{2}_{\tau}}{m^{2}_{e}}} - 3 \bigg) \Bigg]\,.
\end{split}
\end{equation}
Similarly, the $\tau \to 3l$ branching ratios associated with $\tau e$ flavor violation are as follows:
\begin{equation}
\begin{split}
\frac{\mathcal{B}(\tau^{\pm} \to e^{\pm}e^+e^-)}{\mathcal{B}(\tau^{\pm} \to \mu^{\pm} \nu_{\tau}{\nu}_{\mu})} & \simeq \frac{v^{4}}{4} \Bigg[ 2 \Big|[C^{\ell}_{Z}]^{e}_{\tau e}  \Big|^{2} + 2 \Big|[C^{e'}_{Z}]^{e}_{\tau e}  \Big|^{2} + \Big|[C^{e}_{Z}]^{e}_{\tau e}  \Big|^{2} +  \Big|[C^{\ell'}_{Z}]^{e}_{\tau e}  \Big|^{2} \\ & - \frac{4\sqrt{2} ev}{m_{\tau}} \mathfrak{Re} \bigg\{[C_{\gamma}]_{\tau e} \Big(2[C^{\ell}_{Z}]^{e}_{\tau e} + [C^{\ell'}_{Z}]^{e}_{\tau e} \Big)^* + [C_{\gamma}]^{*}_{\tau e} \Big(2[C^{e}_{Z}]^{e}_{\tau e} \\ & + [C^{e'}_{Z}]^{e}_{\tau e} \Big)^* \bigg\} + \frac{16 e^{2} v^{2}}{m^{2}_{\tau}} \bigg([C_{\gamma}]_{\tau e} + [C_{\gamma}]^{*}_{\tau e}\bigg) \bigg(\log{\frac{m^{2}_{\tau}}{m^{2}_{e}}} - \frac{11}{4}\bigg) \Bigg]\,, \\
\end{split}
\end{equation}
\begin{equation}
\begin{split}
\frac{\mathcal{B}(\tau^{\pm} \to e^{\pm}\mu^+\mu^-)}{\mathcal{B}(\tau^{\pm} \to \mu^{\pm} \nu_{\tau}{\nu}_{\mu})} & \simeq \frac{v^{4}}{4} \Bigg[ \Big|[C^{\ell}_{Z}]^{\mu}_{\tau e}  \Big|^{2} + \Big|[C^{e'}_{Z}]^{\mu}_{\tau e}  \Big|^{2} + \Big|[C^{e}_{Z}]^{\mu}_{\tau e}  \Big|^{2} +  \Big|[C^{\ell'}_{Z}]^{\mu}_{\tau e}  \Big|^{2} \\ & - \frac{4\sqrt{2} ev}{m_{\tau}} \mathfrak{Re} \bigg\{[C_{\gamma}]_{\tau e} \Big(2[C^{\ell}_{Z}]^{\mu}_{\tau e} + [C^{\ell'}_{Z}]^{\mu}_{\tau e} \Big)^* + [C_{\gamma}]^{*}_{\tau e} \Big(2[C^{e}_{Z}]^{\mu}_{\tau e} \\ & + [C^{e'}_{Z}]^{\mu}_{\tau e} \Big)^* \bigg\} + \frac{16 e^{2} v^{2}}{m^{2}_{\tau}} \bigg([C_{\gamma}]_{\tau e} + [C_{\gamma}]^{*}_{\tau e}\bigg) \bigg(\log{\frac{m^{2}_{\tau}}{m^{2}_{e}}} - 3 \bigg) \Bigg]\,.
\end{split}
\end{equation}

The resulting bounds on the WCs are summarized in Tab.~\ref{tab:taubounds1}. The radiative decay modes provide the strongest constraints on the dipole operators, with bounds at the $\mathcal{O}(10^{-6})~\text{TeV}^{-2}$ level. In comparison, the three-body decay modes provide weaker constraints on the dipole operators, while also constraining the Higgs-current operators at the $\mathcal{O}(10^{-2})~\text{TeV}^{-2}$ level. The bounds obtained from the $\tau\mu$ and $\tau e$ combinations are broadly comparable, with small differences arising from the corresponding phase-space and logarithmic contributions.

\begin{table}[htb!]
    \centering
    \begin{tabular}{|
    >{\centering\arraybackslash}p{0.10\textwidth}|
    >{\centering\arraybackslash}p{0.25\textwidth}|
    >{\centering\arraybackslash}p{0.25\textwidth}|
    >{\centering\arraybackslash}p{0.25\textwidth}|}
    \hline \hline
    \textbf{WCs} & $\boldsymbol{\tau^\pm \to l^\pm \gamma}$ & $\boldsymbol{\tau^{\pm} \to l^{\pm}\mu^+\mu^-}$ & $\boldsymbol{\tau^{\pm} \to l^{\pm}e^+e^-}$ \\
    \hline \hline
    $[C_{eW}]_{\tau\mu}$ & $< 3.96 \times 10^{-6}$ & $< 2.60 \times 10^{-5}$ & $< 2.56 \times 10^{-5}$ \\
    $[C_{eB}]_{\tau\mu}$ & $< 2.17 \times 10^{-6}$ & $< 1.43 \times 10^{-5}$ & $< 1.40 \times 10^{-5}$ \\
    $[C^{(+)}_{H\ell}]_{\tau\mu}$ & $-$ & $< 1.23 \times 10^{-2}$ & $< 1.50 \times 10^{-2}$ \\
    $[C_{He}]_{\tau\mu}$ & $-$ & $< 1.29 \times 10^{-2}$ & $< 1.50 \times 10^{-2}$ \\
    \hline \hline
    $[C_{eW}]_{\tau e}$ & $< 3.47 \times 10^{-6}$ & $< 3.13 \times 10^{-5}$ & $< 3.10 \times 10^{-5}$ \\
    $[C_{eB}]_{\tau e}$ & $< 1.90 \times 10^{-6}$ & $< 1.72 \times 10^{-5}$ & $< 1.70 \times 10^{-5}$ \\
    $[C^{(+)}_{H\ell}]_{\tau e}$ & $-$ & $< 1.84 \times 10^{-2}$ & $< 1.46 \times 10^{-2}$ \\
    $[C_{He}]_{\tau e}$ & $-$ & $< 1.84 \times 10^{-2}$ & $< 1.54 \times 10^{-2}$ \\
    \hline \hline
    \end{tabular}
    \caption{Upper bounds on WCs ($\text{TeV}^{-2}$) from leptonic $\tau$ decay modes.}
    \label{tab:taubounds1}
\end{table}

\subsubsection*{Hadronic Modes $\boldsymbol{(\tau \to l\pi^0/l\rho^{0}/l\phi)}$}
The $\tau$ lepton, being considerably heavier than the $e$ and $\mu$, has a substantial phase space for hadronic decays via weak charged-current interactions. This also makes hadronic decay modes particularly relevant for probing charged LFV, as the flavor-violating neutral currents induced by the SMEFT operators can mediate $\tau \to l$ transitions accompanied by neutral mesons. Here, we consider the three representative hadronic modes $\tau \to l\pi^0$, $\tau \to l\rho^0$, and $\tau \to l\phi$, which probe different combinations of the effective couplings to light quarks. The current experimental upper bounds on the corresponding branching ratios are
\begin{equation}
\begin{aligned}
    \mathcal{B}(\tau^{\pm} \to \mu^{\pm} \pi^0) &< 1.1 \times 10^{-7}~\text{\cite{BaBar:2006jhm}}\,, &\qquad
    \mathcal{B}(\tau^{\pm} \to e^{\pm} \pi^0) &< 8.0 \times 10^{-8}~\text{\cite{Belle:2007cio}}\,, \\
    \mathcal{B}(\tau^{\pm} \to \mu^{\pm} \rho^0) &< 1.7 \times 10^{-8}~\text{\cite{Belle:2023ziz}}\,, &\qquad
    \mathcal{B}(\tau^{\pm} \to e^{\pm} \rho^0) &< 2.2 \times 10^{-8}~\text{\cite{Belle:2023ziz}}\,, \\
    \mathcal{B}(\tau^{\pm} \to \mu^{\pm} \phi) &< 2.3 \times 10^{-8}~\text{\cite{Belle:2023ziz}}\,, &\qquad
    \mathcal{B}(\tau^{\pm} \to e^{\pm} \phi) &< 2.0 \times 10^{-8}~\text{\cite{Belle:2023ziz}}\,.
\end{aligned} 
\end{equation}
The charged lepton flavor-violating $\tau \to l\pi^0$ branching ratios receive contributions from the lepton flavor-violating $Z$ couplings to the light quarks. Normalizing to the corresponding charged-current decay $\tau^{\pm} \to \nu\,\pi^{\pm}$, with $\mathcal{B}(\tau^{\pm} \to \nu\,\pi^{\pm}) \simeq 10.82\%$, we obtain
\begin{align}
\frac{\mathcal{B}(\tau^{\pm} \to \mu^{\pm} \pi^{0})}{\mathcal{B}(\tau^{\pm} \to \nu\, \pi^{\pm})} & \simeq \frac{v^{4}}{8} \Bigg[ \Big| [C^{\ell}_{Z}]^{u}_{\tau\mu} - [C^{\ell}_{Z}]^{d}_{\tau\mu} - [C^{\ell'}_{Z}]^{u}_{\tau\mu} + [C^{\ell'}_{Z}]^{d}_{\tau\mu} \Big|^{2} \nonumber\\ & + \Big| [C^{e}_{Z}]^{u}_{\tau\mu} - [C^{e}_{Z}]^{d}_{\tau\mu} - [C^{e'}_{Z}]^{u}_{\tau\mu} + [C^{e'}_{Z}]^{d}_{\tau\mu} \Big|^{2} \Bigg],\\
\frac{\mathcal{B}(\tau^{\pm} \to e^{\pm} \pi^{0})}{\mathcal{B}(\tau^{\pm} \to \nu\, \pi^{\pm})} & \simeq \frac{v^{4}}{8} \Bigg[ \Big| [C^{\ell}_{Z}]^{u}_{\tau e} - [C^{\ell}_{Z}]^{d}_{\tau e} - [C^{\ell'}_{Z}]^{u}_{\tau e} + [C^{\ell'}_{Z}]^{d}_{\tau e} \Big|^{2}\nonumber \\ & + \Big| [C^{e}_{Z}]^{u}_{\tau e} - [C^{e}_{Z}]^{d}_{\tau e} - [C^{e'}_{Z}]^{u}_{\tau e} + [C^{e'}_{Z}]^{d}_{\tau e} \Big|^{2} \Bigg].
\end{align}
For the vector meson $\rho^0$, both photon and $Z$-mediated contributions are present. The corresponding branching ratios, normalized to $\mathcal{B}(\tau^{\pm} \to \nu\,\rho^{\pm}) \simeq 25.49\%$, are given by
\begin{equation}
\begin{split}
\frac{\mathcal{B}(\tau^{\pm} \to \mu^{\pm} \rho^{0})}{\mathcal{B}(\tau^{\pm} \to \nu\, \rho^{\pm})} & \simeq \frac{v^{4}}{8} \bigg(1 + \frac{2 m_{\rho}^{2}}{m_{\tau}^{2}} \bigg)^{-1} \Bigg[ \frac{8e^2 v^2}{m_{\rho}^{2}} \bigg(\Big|[C_{\gamma}]_{\tau\mu}\Big|^{2} + \Big|[C_{\gamma}]^{*}_{\tau\mu}\Big|^{2} \bigg) \\ & \quad \bigg(2 + \frac{m_{\rho}^{2}}{m_{\tau}^{2}} \bigg)  + \bigg( \Big| [C^{\ell}_{Z}]^{u}_{\tau\mu} - [C^{\ell}_{Z}]^{d}_{\tau\mu} + [C^{\ell'}_{Z}]^{u}_{\tau\mu} - [C^{\ell'}_{Z}]^{d}_{\tau\mu} \Big|^{2} \\ & + \Big| [C^{e}_{Z}]^{u}_{\tau\mu} - [C^{e}_{Z}]^{d}_{\tau\mu} + [C^{e'}_{Z}]^{u}_{\tau\mu} - [C^{e'}_{Z}]^{d}_{\tau\mu} \Big|^{2} \bigg) \bigg(1 + \frac{2 m_{\rho}^{2}}{m_{\tau}^{2}} \bigg) \\ & + \frac{12\sqrt{2}ev}{m_{\tau}} \mathfrak{Re} \bigg\{[C_{\gamma}]_{\tau\mu} \Big([C^{\ell}_{Z}]^{u}_{\tau\mu} - [C^{\ell}_{Z}]^{d}_{\tau\mu} + [C^{\ell'}_{Z}]^{u}_{\tau\mu} - [C^{\ell'}_{Z}]^{d}_{\tau\mu} \Big)^* \\ & + [C_{\gamma}]_{\tau\mu}^* \Big([C^{e}_{Z}]^{u}_{\tau\mu} - [C^{e}_{Z}]^{d}_{\tau\mu} + [C^{e'}_{Z}]^{u}_{\tau\mu} - [C^{e'}_{Z}]^{d}_{\tau\mu} \Big)^* \bigg\} \Bigg]\,, \\
\end{split}
\end{equation}
\begin{equation}
\begin{split}
\frac{\mathcal{B}(\tau^{\pm} \to e^{\pm} \rho^{0})}{\mathcal{B}(\tau^{\pm} \to \nu\, \rho^{\pm})} & \simeq \frac{v^{4}}{8} \bigg(1 + \frac{2 m_{\rho}^{2}}{m_{\tau}^{2}} \bigg)^{-1} \Bigg[ \frac{8e^2 v^2}{m_{\rho}^{2}} \bigg(\Big|[C_{\gamma}]_{\tau e}\Big|^{2} + \Big|[C_{\gamma}]^{*}_{\tau e}\Big|^{2} \bigg) \\ & \quad \bigg(2 + \frac{m_{\rho}^{2}}{m_{\tau}^{2}} \bigg)  + \bigg( \Big| [C^{\ell}_{Z}]^{u}_{\tau e} - [C^{\ell}_{Z}]^{d}_{\tau e} + [C^{\ell'}_{Z}]^{u}_{\tau e} - [C^{\ell'}_{Z}]^{d}_{\tau e} \Big|^{2} \\ & + \Big| [C^{e}_{Z}]^{u}_{\tau e} - [C^{e}_{Z}]^{d}_{\tau e} + [C^{e'}_{Z}]^{u}_{\tau e} - [C^{e'}_{Z}]^{d}_{\tau e} \Big|^{2} \bigg) \bigg(1 + \frac{2 m_{\rho}^{2}}{m_{\tau}^{2}} \bigg) \\ & + \frac{12\sqrt{2}ev}{m_{\tau}} \mathfrak{Re} \bigg\{[C_{\gamma}]_{\tau e} \Big([C^{\ell}_{Z}]^{u}_{\tau e} - [C^{\ell}_{Z}]^{d}_{\tau e} + [C^{\ell'}_{Z}]^{u}_{\tau e} - [C^{\ell'}_{Z}]^{d}_{\tau e} \Big)^* \\ & + [C_{\gamma}]_{\tau e}^* \Big([C^{e}_{Z}]^{u}_{\tau e} - [C^{e}_{Z}]^{d}_{\tau e} + [C^{e'}_{Z}]^{u}_{\tau e} - [C^{e'}_{Z}]^{d}_{\tau e} \Big)^* \bigg\} \Bigg]\,,
\end{split}
\end{equation}
The $\phi$ meson, being predominantly an $s\bar{s}$ state, provides sensitivity to the flavor-violating couplings involving strange quarks. Using $(f_{\phi}/f_{\rho}) \simeq 1.09$, the corresponding expressions are
\begin{equation}
\begin{split}
\frac{\mathcal{B}(\tau^{\pm} \to \mu^{\pm} \rho^{0})}{\mathcal{B}(\tau^{\pm} \to \nu\, \rho^{\pm})}  \simeq &\frac{v^{4}}{4} \frac{f^{2}_{\phi}}{f^{2}_{\rho}} \bigg(1 + \frac{2 m_{\rho}^{2}}{m_{\tau}^{2}} \bigg)^{-1} \bigg(1 - \frac{m_{\rho}^{2}}{m_{\tau}^{2}} \bigg)^{-2} \bigg(1 - \frac{m_{\phi}^{2}}{m_{\tau}^{2}} \bigg)^{2} \\ & \Bigg[ \frac{8e^2v^2}{9 m^{2}_{\phi}} \bigg(\Big|[C_{\gamma}]_{\tau\mu}\Big|^{2} + \Big|[C_{\gamma}]^{*}_{\tau\mu}\Big|^{2} \bigg) \bigg(2 + \frac{m_{\phi}^{2}}{m_{\tau}^{2}} \bigg)^{2} +  \bigg(1 + \frac{2 m_{\phi}^{2}}{m_{\tau}^{2}} \bigg) \\ & \bigg( \Big| [C^{\ell}_{Z}]^{s}_{\tau\mu} + [C^{\ell'}_{Z}]^{s}_{\tau\mu} \Big|^{2} + \Big| [C^{e}_{Z}]^{s}_{\tau\mu} + [C^{e'}_{Z}]^{s}_{\tau\mu} \Big|^{2} \bigg) - \frac{4\sqrt{2}ev}{m_{\tau}} \\ & \mathfrak{Re} \bigg\{[C_{\gamma}]_{\tau\mu} \Big([C^{\ell}_{Z}]^{s}_{\tau\mu} +[C^{\ell'}_{Z}]^{s}_{\tau\mu} \Big)^* + [C_{\gamma}]_{\tau\mu}^* \Big([C^{e}_{Z}]^{s}_{\tau\mu} + [C^{e'}_{Z}]^{s}_{\tau\mu} \Big)^* \bigg\} \Bigg]\,, \\
\end{split}
\end{equation}
\begin{equation}
\begin{split}
\frac{\mathcal{B}(\tau^{\pm} \to e^{\pm} \rho^{0})}{\mathcal{B}(\tau^{\pm} \to \nu\, \rho^{\pm})} \simeq & \frac{v^{4}}{4} \frac{f^{2}_{\phi}}{f^{2}_{\rho}} \bigg(1 + \frac{2 m_{\rho}^{2}}{m_{\tau}^{2}} \bigg)^{-1} \bigg(1 - \frac{m_{\rho}^{2}}{m_{\tau}^{2}} \bigg)^{-2} \bigg(1 - \frac{m_{\phi}^{2}}{m_{\tau}^{2}} \bigg)^{2} \\ & \Bigg[ \frac{8e^2v^2}{9 m^{2}_{\phi}} \bigg(\Big|[C_{\gamma}]_{\tau e}\Big|^{2} + \Big|[C_{\gamma}]^{*}_{\tau e}\Big|^{2} \bigg) \bigg(2 + \frac{m_{\phi}^{2}}{m_{\tau}^{2}} \bigg)^{2} +  \bigg(1 + \frac{2 m_{\phi}^{2}}{m_{\tau}^{2}} \bigg) \\ & \bigg( \Big| [C^{\ell}_{Z}]^{s}_{\tau e} + [C^{\ell'}_{Z}]^{s}_{\tau e} \Big|^{2} + \Big| [C^{e}_{Z}]^{s}_{\tau e} + [C^{e'}_{Z}]^{s}_{\tau e} \Big|^{2} \bigg) - \frac{4\sqrt{2}ev}{m_{\tau}} \\ & \mathfrak{Re} \bigg\{[C_{\gamma}]_{\tau e} \Big([C^{\ell}_{Z}]^{s}_{\tau e} +[C^{\ell'}_{Z}]^{s}_{\tau e} \Big)^* + [C_{\gamma}]_{\tau e}^* \Big([C^{e}_{Z}]^{s}_{\tau e} + [C^{e'}_{Z}]^{s}_{\tau e} \Big)^* \bigg\} \Bigg]\,.
\end{split}
\end{equation}

The resulting bounds on the WCs from the hadronic decay modes are summarized in Tab.~\ref{tab:taubounds2}. The $\rho^0$ channel provides the strongest constraints among the hadronic modes, with the dipole coefficients constrained at the $\mathcal{O}(10^{-5})\,\text{TeV}^{-2}$ level and the Higgs-current coefficients at the $\mathcal{O}(10^{-2})\,\text{TeV}^{-2}$ level. The bounds from the $\phi$ channel are comparatively weaker, particularly for the dipole operators, owing to its different kinematics and sensitivity to the strange-quark current. The $\pi^0$ mode constrains only the Higgs-current operators, with bounds at the $\mathcal{O}(10^{-2})\,\text{TeV}^{-2}$ level, comparable to those obtained from the $\rho^0$ and $\phi$ channels. The $\tau\mu$ and $\tau e$ flavor structures yield broadly comparable constraints across the three hadronic channels.

\begin{table}[htb!]
    \centering
    \begin{tabular}{|
    >{\centering\arraybackslash}p{0.10\textwidth}|
    >{\centering\arraybackslash}p{0.25\textwidth}|
    >{\centering\arraybackslash}p{0.25\textwidth}|
    >{\centering\arraybackslash}p{0.25\textwidth}|}
    \hline \hline
    \textbf{WCs} & $\boldsymbol{\tau^\pm \to l^\pm\pi^0}$ & $\boldsymbol{\tau^{\pm} \to l^{\pm}\rho^0}$ & $\boldsymbol{\tau^{\pm} \to l^{\pm}\phi}$ \\
    \hline \hline
    $[C_{eW}]_{\tau\mu}$ & $-$ & $<5.18 \times 10^{-5}$ & $<1.18 \times 10^{-4}$ \\
    $[C_{eB}]_{\tau\mu}$ & $-$ & $<2.84 \times 10^{-5}$ & $<6.49 \times 10^{-5}$ \\
    $[C^{(+)}_{H\ell}]_{\tau\mu}$ & $<2.36 \times 10^{-2}$ & $<1.12 \times 10^{-2}$ & $<1.45 \times 10^{-2}$ \\
    $[C_{He}]_{\tau\mu}$ & $<2.36 \times 10^{-2}$ & $<1.12 \times 10^{-2}$ & $<1.45 \times 10^{-2}$ \\
    \hline \hline
    $[C_{eW}]_{\tau e}$ & $-$ & $<5.90 \times 10^{-5}$ & $<1.10 \times 10^{-4}$ \\
    $[C_{eB}]_{\tau e}$ & $-$ & $<3.23 \times 10^{-5}$ & $<6.05 \times 10^{-5}$ \\
    $[C^{(+)}_{H\ell}]_{\tau e}$ & $<2.01 \times 10^{-2}$ & $<1.28 \times 10^{-2}$ & $<1.35 \times 10^{-2}$ \\
    $[C_{He}]_{\tau e}$ & $<2.01 \times 10^{-2}$ & $<1.28 \times 10^{-2}$ & $<1.35 \times 10^{-2}$ \\
    \hline \hline
    \end{tabular}
    \caption{Upper bounds on WCs ($\text{TeV}^{-2}$) from hadronic $\tau$ decay modes.}
    \label{tab:taubounds2}
\end{table}

\subsubsection*{LFV $\boldsymbol{Z}$ Decays $\boldsymbol{(Z \to l\tau)}$}
In addition to the lepton flavor-violating $\tau$ decays discussed above, we consider flavor-violating decays of the $Z$ boson, which provide a complementary probe of the same effective interactions at the electroweak scale. The current experimental upper bounds on the corresponding branching ratios are
\begin{equation}
\begin{aligned}
    \mathcal{B}(Z \to \mu^{\pm} \tau^{\mp}) &< 6.5 \times 10^{-6}~\text{\cite{ATLAS:2021bdj}}\,, &\qquad
    \mathcal{B}(Z \to e^{\pm} \tau^{\mp}) &< 5.0 \times 10^{-6}~\text{\cite{ATLAS:2021bdj}}\,.
\end{aligned} 
\end{equation}
The charged lepton flavor-violating $Z \to l^\pm \tau^\mp$ branching ratios can be connected to the corresponding WCs as follows. Taking $\mathcal{B}(Z \to l^{\pm}l^{\mp}) \simeq 3.3658\%$, we have
\begin{equation}
\begin{split}
\frac{\mathcal{B}(Z \to \mu^{\pm} \tau^{\mp})}{\mathcal{B}(Z \to l^{\pm}l^{\mp})} &\simeq \frac{2v^4}{1 - 4 \sin^2{\theta_{W}} + 8 \sin^4{\theta_{W}}}  \Bigg[\Big|[C_{Z}]_{\tau\mu}  \Big|^{2} \\ &+ \Big|[C_{Z}]^*_{\tau\mu}  \Big|^{2} + \Big|[C^{(+)}_{H\ell}]_{\tau\mu}  \Big|^{2} + \Big|[C_{He}]_{\tau\mu}  \Big|^{2} \Bigg]  \,,\\
\end{split}
\end{equation}
\begin{equation}
\begin{split}
\frac{\mathcal{B}(Z \to e^{\pm} \tau^{\mp})}{\mathcal{B}(Z \to l^{\pm}l^{\mp})} & \simeq \frac{2v^4}{1 - 4 \sin^2{\theta_{W}} + 8 \sin^4{\theta_{W}}}  \Bigg[\Big|[C_{Z}]_{\tau e}  \Big|^{2} \\ & + \Big|[C_{Z}]^*_{\tau e}  \Big|^{2} + \Big|[C^{(+)}_{H\ell}]_{\tau e}  \Big|^{2} + \Big|[C_{He}]_{\tau e}  \Big|^{2} \Bigg]  \,.\\
\end{split}
\end{equation}

The resulting constraints on the WCs are summarized in Tab.~\ref{tab:zbounds}. Compared with the constraints from charged LFV $\tau$ decays, the bounds from LFV $Z$ decays are weaker, with the WCs constrained at the $\mathcal{O}(10^{-1})\,\text{TeV}^{-2}$ level. Nevertheless, these limits are particularly relevant as they directly probe flavor-violating interactions at the electroweak scale. The bounds for the $\tau\mu$ and $\tau e$ flavor structures are also broadly comparable, with the latter yielding slightly stronger constraints. 

\begin{table}[htb!]
    \centering
    \begin{tabular}{|
    >{\centering\arraybackslash}p{0.10\textwidth}|
    >{\centering\arraybackslash}p{0.325\textwidth}|
    >{\centering\arraybackslash}p{0.10\textwidth}|
    >{\centering\arraybackslash}p{0.325\textwidth}|}
    \hline \hline
    \textbf{WCs} & $\boldsymbol{Z \to \mu^\pm\tau^\mp}$ & \textbf{WCs} & $\boldsymbol{Z \to e^\pm\tau^\mp}$ \\
    \hline \hline
    $[C_{eW}]_{\tau\mu}$ & $<2.9\times10^{-1}$ & $[C_{eW}]_{\tau e}$ & $<2.6\times10^{-1}$ \\
    $[C_{eB}]_{\tau\mu}$ & $<5.4\times10^{-1}$ & $[C_{eB}]_{\tau e}$ & $<4.7\times10^{-1}$ \\
    $[C^{(+)}_{H\ell}]_{\tau\mu}$ & $<3.6\times10^{-1}$ & $[C^{(+)}_{H\ell}]_{\tau e}$ & $<3.2\times10^{-1}$ \\
    $[C_{He}]_{\tau\mu}$ & $<3.6\times10^{-1}$ & $[C_{He}]_{\tau e}$ & $<3.2\times10^{-1}$ \\
    \hline \hline
    \end{tabular}
    \caption{Upper bounds on WCs ($\text{TeV}^{-2}$) from LFV $Z$ boson decay modes.}
    \label{tab:zbounds}
\end{table}

In addition, the electron and muon EDM measurements provide constraints on the imaginary components of the corresponding leptonic dipole operators, as discussed in Appendix~\ref{Append:complex_fit}.

\section{Optimal Observable Technique}
\label{sec:oot}
The OOT 
is a powerful method used to extract model parameters or couplings from collider observables with maximum statistical sensitivity. By constructing observables that are derived from the dependence of the differential cross-section on the parameters of interest, this approach ensures the most efficient use of the available kinematic information. It is particularly useful in precision measurements and EFT analyses, where small deviations from the SM predictions are expected. The differential cross section for any given collider process can be written in the form
\begin{equation}
    \begin{aligned}
    \mathscr{O}(\phi) &= \frac{d\sigma_{\rm}}{d\phi}\bigg|_{O} = \frac{d\sigma_{\rm}}{d\phi}\bigg|_{S} + \frac{d\sigma_{\rm}}{d\phi}\bigg|_{B} = \sum_{i}\; g_{i} f_{i}(\phi)\,,
    \end{aligned}
\end{equation}
where $\phi$ is a phase space variable, $g_{i}$ are the functions containing the NP couplings and $f_{i}$ are the functions of phase space variable, $\phi$. The subscripts $O$, $S$ and $B$ refers to observed, signal and background events, respectively. It should be noted that these differential cross sections are following kinematic cuts use to minimize background contributions. The optimal covariance matrix is defined as
\begin{equation}
    V_{ij} = \frac{M^{-1}_{ij}}{\mathfrak{L}_{\rm int}} = \frac{1}{\mathfrak{L}_{\rm int}}\int \frac{f_i(\phi)f_j(\phi)}{\mathcal{O}(\phi)}d\phi.
\end{equation}
And the optimal $(\chi^{2}_{\rm OOT})$ is given by
\begin{equation}
    \chi^{2}_{\rm OOT} = \sum_{i,j}\; (g_{i}-g^{0}_{i})\;V^{-1}_{ij}\;(g_{j}-g^{0}_{j})\; \bigg|_{g=g^{0}}
\end{equation}
where $g^{0}$ correspond to the seed values of the coefficients, $g$. The input seed values can come from different sources like previous measurements, predictions from a different experiment, etc. In case of sensitivity prediction of an unobserved NP scenario, the seed value is usually determined by setting the NP couplings to zero.

For our analysis, we construct the optimal observable using the cosine of the polar angle between selected final-state particles, denoted by $\cos\theta_{i}$. Since multiple signal channels are considered, the choice of the final-state $i$ depends on the process under investigation. In each case, the observable is chosen to maximize the sensitivity to the EFT contributions. To construct $\mathscr{O}(\cos\theta_{i})$, we adopt a semi-analytical approach that consistently incorporates the effects of the background processes. The observable is built from the kinematic distributions after applying all event-selection criteria to suppress background contamination. For the background processes, the coefficients $g_i$ are independent of the NP couplings and remain unchanged throughout the parameter space. We therefore parametrize the corresponding angular dependence by fitting the basis functions up to quadratic order, $f_i(\cos\theta_{i})=\left\{1\,,\cos\theta_{i}\,,\cos^2\theta_{i}\right\}$. For the signal processes, benchmark points distributed over the NP parameter space are generated to extract the dependence of the differential distributions on both the EFT parameters and the phase-space variables. The overall strategy closely follows that of Ref.~\cite{Bhattacharya:2023zln}. This semi-analytical parameterization enables sensitivity projections at different confidence levels without the need to regenerate Monte Carlo samples for every benchmark point.

\section{Collider Analysis \& Projections}
\label{sec:collider}
Collider experiments provide one of the most powerful avenues for probing NP by enabling direct searches for new particles and precision measurements of deviations from SM predictions. Even when the underlying NP scale is beyond the kinematic reach of a collider, its virtual effects can manifest as small modifications to production rates and kinematic distributions. In the present analysis, we consider the four planned operational stages of the FCC-ee~\cite{FCC:2025lpp,FCC:2025uan}: the $Z$ pole, the $W^{+}W^{-}$ threshold, the Higgs factory, and the $t\bar{t}$ threshold. Although the $Z$ pole run features an energy scan in the range $\sqrt{s} = [87.9, 94.3]$~GeV, we adopt the nominal on-peak energy of $\sqrt{s} = 91.2$~GeV to accurately model the Breit-Wigner resonance in our simulations. Similarly, while the $W^{+}W^{-}$ phase includes few scan points in the range $\sqrt{s} = [157.5, 162.5]$~GeV, we evaluate this stage at $\sqrt{s} = 161$~GeV to properly capture the on-shell threshold production kinematics. The Higgs factory and $t\bar{t}$ threshold stages are simulated at their standard benchmark energies of $\sqrt{s} = 240$~GeV and $\sqrt{s} = 365$~GeV, respectively. The integrated luminosities to be accumulated at each stage of FCC-ee are as follows, $Z$ pole: $205\text{ ab}^{-1}$, $W^{+}W^{-}$ threshold: $19.2\text{ ab}^{-1}$, Higgs factory: $10.8\text{ ab}^{-1}$, and $t\bar{t}$ threshold: $2.70\text{ ab}^{-1}$. The FCC-ee running scenarios considered in this analysis are summarized below:
\begin{itemize}
    \itemsep-0em
    \item \textbf{Stage 1}: $\sqrt{s} = 91.2\text{ GeV}\,, \mathfrak{L}_{\rm int} = 205\text{ ab}^{-1}$,
    \item \textbf{Stage 2}: $\sqrt{s} = 161\text{ GeV}\,, \mathfrak{L}_{\rm int} = 19.2\text{ ab}^{-1}$,
    \item \textbf{Stage 3}: $\sqrt{s} = 240\text{ GeV}\,, \mathfrak{L}_{\rm int} = 10.8\text{ ab}^{-1}$,
    \item \textbf{Stage 4}: $\sqrt{s} = 365\text{ GeV}\,, \mathfrak{L}_{\rm int} = 2.70\text{ ab}^{-1}$.
\end{itemize}

We investigate six flavor-violating production channels induced by 24 flavor-specific dimension-6 SMEFT operators introduced in Sec.~\ref{sec:eft}. Each process is parameterized by four independent WCs, corresponding to two dipole operators and two Higgs-current operators. The complete list of the processes considered, together with their associated EFT operators, is summarized in Tab.~\ref{tab:fcc-ee}. Throughout this work, only one flavor structure is considered at a time, allowing the sensitivity to the corresponding set of WCs to be determined independently.
\begin{table}[htb!]
    \centering
    \renewcommand{\arraystretch}{1.25}{
    \begin{tabular}{|
    >{\centering\arraybackslash}p{0.40\textwidth}|
    >{\centering\arraybackslash}p{0.10\textwidth}
    >{\centering\arraybackslash}p{0.10\textwidth}
    >{\centering\arraybackslash}p{0.10\textwidth}
    >{\centering\arraybackslash}p{0.10\textwidth}|
    }
    \hline\hline
    \textbf{Process} & \multicolumn{4}{c|}{\textbf{Operators}} \\
    \hline\hline
    $\boldsymbol{e^+e^- \to \tau^{\pm} \mu^{\mp}}$ & $[\mathcal{O}_{eW}]_{\tau \mu}$ & $[\mathcal{O}_{eB}]_{\tau \mu}$ & $[\mathcal{O}^{(+)}_{H \ell}]_{\tau \mu}$ & $[\mathcal{O}_{H e}]_{\tau \mu}$ \\ \hline
    $\boldsymbol{e^+e^- \to \tau^{\pm} e^{\mp}}$ & $[\mathcal{O}_{eW}]_{\tau e}$ & $[\mathcal{O}_{eB}]_{\tau e}$ & $[\mathcal{O}^{(+)}_{H \ell}]_{\tau e}$ & $[\mathcal{O}_{H e}]_{\tau e}$ \\ \hline \hline
    $\boldsymbol{e^+e^- \to b\overline{s}\,/\, \overline{b}s}$ & $[\mathcal{O}_{dW}]_{bs}$ & $[\mathcal{O}_{dB}]_{bs}$ & $[\mathcal{O}^{(+)}_{H q}]_{bs}$ & $[\mathcal{O}_{H d}]_{bs}$ \\ \hline
    $\boldsymbol{e^+e^- \to b\overline{d}\,/\, \overline{b}d}$ & $[\mathcal{O}_{dW}]_{bd}$ & $[\mathcal{O}_{dB}]_{bd}$ & $[\mathcal{O}^{(+)}_{H q}]_{bd}$ & $[\mathcal{O}_{H d}]_{bd}$ \\
    \hline \hline
    $\boldsymbol{e^+e^- \to t\overline{c}\,/\, \overline{t}c}$ & $[\mathcal{O}_{uW}]_{tc}$ & $[\mathcal{O}_{uB}]_{tc}$ & $[\mathcal{O}^{(-)}_{H q}]_{tc}$ & $[\mathcal{O}_{H u}]_{tc}$ \\ \hline
    $\boldsymbol{e^+e^- \to t\overline{u}\,/\, \overline{t}u}$ & $[\mathcal{O}_{uW}]_{tu}$ & $[\mathcal{O}_{uB}]_{tu}$ & $[\mathcal{O}^{(-)}_{H q}]_{tu}$ & $[\mathcal{O}_{H u}]_{tu}$ \\
    \hline\hline
    \end{tabular}}
    \caption{EFT coefficients associated with different flavor-violating processes in context of FCC-ee.}
    \label{tab:fcc-ee}
\end{table}

\begin{figure}[htb!]
    \begin{center}
    \begin{tikzpicture}[baseline={(current bounding box.center)},style={scale=0.8, transform shape}]
        \begin{feynman}
            \vertex (a);
            \vertex [above left = 1.25cm of a] (b) {$e^{-}$};
            \vertex [below left = 1.25cm of a] (c) {$e^{+}$};
            \vertex [dot, right = 1.25cm of a] (d){};
            \vertex [above right = 1.65cm of d] (e) {$\tau^{-}$};
            \vertex [below right = 1.65cm of d] (f) {$e^{+}/\mu^{+}$};
            
            \diagram* {
                (b) -- [thick, fermion, arrow size = 1 pt] (a),
                (a) -- [thick, fermion, arrow size = 1 pt] (c),
                (a) -- [thick, boson, edge label = $Z/\gamma$] (d),
                (d) -- [thick, fermion, arrow size = 1 pt] (e),
                (f) -- [thick, fermion, arrow size = 1 pt] (d)
            };
        \end{feynman}
    \end{tikzpicture}
    \begin{tikzpicture}[baseline={(current bounding box.center)},style={scale=0.8, transform shape}]
        \begin{feynman}
            \vertex [dot](a){};
            \vertex [left = 1.65cm of a] (b) {$e^{-}$};
            \vertex [right = 1.65cm of a] (c) {$\tau^{-}$};
            \vertex [below = 2.25cm of a] (d);
            \vertex [left = 1.25cm of d] (e) {$e^{+}$};
            \vertex [right = 1.25cm of d] (f) {$e^{+}$};
            
            \diagram* {
                (b) -- [thick, fermion, arrow size = 1 pt] (a),
                (a) -- [thick, fermion, arrow size = 1 pt] (c),
                (a) -- [thick, boson, edge label = $Z/\gamma$] (d),
                (d) -- [thick, fermion, arrow size = 1 pt] (e),
                (f) -- [thick, fermion, arrow size = 1 pt] (d)
            };
        \end{feynman}
    \end{tikzpicture}
    \begin{tikzpicture}[baseline={(current bounding box.center)},style={scale=0.8, transform shape}]
        \begin{feynman}
            \vertex (a);
            \vertex [above left = 1.25cm of a] (b) {$e^{-}$};
            \vertex [below left = 1.25cm of a] (c) {$e^{+}$};
            \vertex [dot, right = 1.25cm of a] (d){};
            \vertex [above right = 1.65cm of d] (e) {$b$};
            \vertex [below right = 1.65cm of d] (f) {$\overline{d}/\overline{s}$};
            
            \diagram* {
                (b) -- [thick, fermion, arrow size = 1 pt] (a),
                (a) -- [thick, fermion, arrow size = 1 pt] (c),
                (a) -- [thick, boson, edge label = $Z/\gamma$] (d),
                (d) -- [thick, fermion, arrow size = 1 pt] (e),
                (f) -- [thick, fermion, arrow size = 1 pt] (d)
            };
        \end{feynman}
    \end{tikzpicture}
    \begin{tikzpicture}[baseline={(current bounding box.center)},style={scale=0.8, transform shape}]
        \begin{feynman}
            \vertex (a);
            \vertex [above left = 1.25cm of a] (b) {$e^{-}$};
            \vertex [below left = 1.25cm of a] (c) {$e^{+}$};
            \vertex [dot, right = 1.25cm of a] (d){};
            \vertex [above right = 1.65cm of d] (e) {$t$};
            \vertex [below right = 1.65cm of d] (f) {$\overline{u}/\overline{c}$};
            
            \diagram* {
                (b) -- [thick, fermion, arrow size = 1 pt] (a),
                (a) -- [thick, fermion, arrow size = 1 pt] (c),
                (a) -- [thick, boson, edge label = $Z/\gamma$] (d),
                (d) -- [thick, fermion, arrow size = 1 pt] (e),
                (f) -- [thick, fermion, arrow size = 1 pt] (d)
            };
        \end{feynman}
    \end{tikzpicture}
    \caption{Feynman diagram corresponding to different flavor-violating FCC-ee processes. The dark blob represent the insertion of SMEFT operator.}
    \label{fig:1}
    \end{center}
\end{figure}
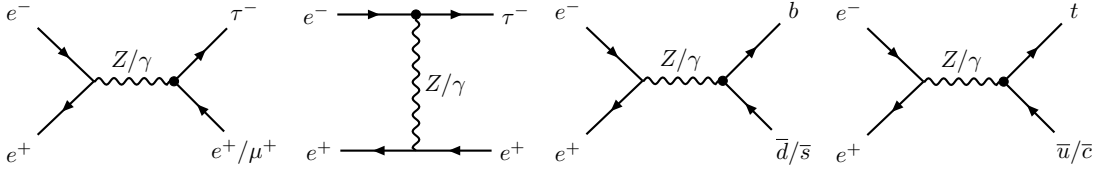

The leading-order Feynman diagrams for the six signal processes are shown in Fig.~\ref{fig:1}. The channels $\tau\mu$, $bs/bd$, and $tc/tu$ proceed exclusively through an $s$-channel exchange of a $Z/\gamma$. In contrast, the process $e^{+}e^{-}\to\tau^{\pm}e^{\mp}$ receives contributions from both $s$- and $t$-channel diagrams owing to the presence of electrons in both the initial and final states. Unlike the $s$-channel contribution, which is suppressed away from the $Z$ resonance, the $t$-channel contribution does not suffer from such propagator suppression and therefore becomes increasingly important at higher CM energies. Consequently, the $\tau e$ channel exhibits a qualitatively different energy dependence compared to the other flavor-violating processes considered in this work.

The collider simulation workflow employed in this analysis is as follows. The dimension-6 operators listed in Tab.~\ref{tab:dim6} are implemented in \textsc{FeynRules}~\cite{Alloul:2013bka} to generate a \textsc{UFO} model~\cite{Degrande:2011ua}, which is subsequently interfaced with the Monte Carlo event generator \textsc{MadGraph5\_aMC}~\cite{Alwall:2011uj} for parton-level event generation. The generated events are then passed to \textsc{Pythia8}~\cite{Bierlich:2022pfr} for parton showering and hadronization, followed by a fast detector simulation using \textsc{Delphes3}~\cite{deFavereau:2013fsa}. Jet reconstruction is performed within \textsc{Delphes3} using the anti-$k_T$ clustering algorithm and the default detector card (\texttt{delphes\_default\_card}). The same detector configuration is used to implement the $\tau$ and $b$-tagging algorithms employed throughout the analysis. In the following sections, we present the signal and background analyses for each flavor-violating channel and discuss the corresponding projected sensitivities to the EFT operators at the different CM energy stages of the FCC-ee.

\subsection{Channel: $e^{+}e^{-} \to \tau\mu\,/\,\tau e$}
\label{sec:lfv0}
The signal consists of one electron/muon ($N_{\mu}/N_{e} = 1$) and one hadronically decaying tau i.e. $\tau$-tagged jet,  ($N^{\tau}_{j}=1$) in the final state with no electrons/muons or photons. The major background processes are $\tau^+\tau^-$, and $W^+W^-$ (Stage 2 and beyond). For signal-background discrimination, we utilize two variables:
\begin{itemize}
    \itemsep-0em
    \item The invariant mass of the lepton ($l=\mu/e$) and the $\tau$-tagged jet, defined as
    \begin{equation}
        M_{l\tau} = \sqrt{\left(p_{l} + p_{\tau} \right)^{2}} = \sqrt{\left(E_{l} + E_{\tau} \right)^{2} - \left|\overrightarrow{p_{l}} + \overrightarrow{p_{\tau}} \right|^{2}}\,,
    \end{equation}
    where, $p_{l} \equiv \left(E_{l}, \overrightarrow{p_{l}}\right)$ and $p_{\tau} \equiv \left(E_{\tau}, \overrightarrow{p_{\tau}}\right)$ are the 4-momenta of the lepton and the $\tau$-tagged jet, respectively. Since the signal processes contain fewer invisible particles in the final state, a larger fraction of the collision energy is carried by the visible final-state objects. Consequently, the corresponding kinematic distributions are shifted toward higher energies compared to those of the background processes (see Fig.~\ref{fig:dists1}).
    \item The energy of the charged lepton ($l=\mu,e$), denoted by $E_l$, provides a powerful discriminating observable. For the signal, the charged lepton is produced directly in the underlying $2 \to 2$ hard scattering process. Consequently, for a fixed CM energy, its energy is kinematically constrained, resulting in a sharp peak around half of the CM energy. In contrast, for the background processes, the charged lepton typically originates from the decays of intermediate particles, leading to a much broader energy spectrum that extends to significantly lower energies. The corresponding distribution are shown in Fig.~\ref{fig:dists1}.
\end{itemize}
The kinematic cuts employed for the $\tau\mu/\tau e$ signal channels, together with the corresponding signal ($S$) and background ($B$) efficiencies, are presented below. Throughout this work, the quoted signal efficiencies are obtained by averaging over ten randomly sampled benchmarks in the four-dimensional EFT parameter space corresponding to each signal process.
\subsubsection*{Stage 1: $\boldsymbol{(\sqrt{s} = 91.2\;{\rm GeV}\,, \mathfrak{L}_{\rm int} = 205\;{\rm ab}^{-1})}$}
At this stage, the dominant SM background arises from the $\tau^{+}\tau^{-}$ process. To suppress this background, the following kinematic requirements are imposed:
\begin{equation}
M_{l\tau} > 75~\mathrm{GeV}\,, \qquad E_{l} > 42.5~\mathrm{GeV}\,.
\end{equation}
The resulting signal and background efficiencies are
\begin{equation}
\epsilon_{S}:
\begin{cases}
\tau\mu &= 0.20\,,\\
\tau e &= 0.17\,,
\end{cases}
\qquad
\epsilon_{B} :
\begin{cases}
\tau^+\tau^-~(\mu) &= 1.7\times10^{-4}\,,\\
\tau^+\tau^-~(e) &= 3.0\times10^{-4}\,.
\end{cases}
\end{equation}

\subsubsection*{Stage 2: $\boldsymbol{(\sqrt{s} = 161\;{\rm GeV}\,, \mathfrak{L}_{\rm int} = 19.2\;{\rm ab}^{-1})}$}
At this stage, $W^+W^-$ emerges as an additional background, alongside $\tau^{+}\tau^{-}$. The following kinematic cuts are imposed:
\begin{equation}
M_{l\tau} > 120~\mathrm{GeV}\,, \qquad E_{l} > 75~\mathrm{GeV}\,.
\end{equation}
The $W^+W^-$ background is completely wiped out by these cuts. The signal and remaining background efficiencies following the cuts are
\begin{equation}
\epsilon_{S}:
\begin{cases}
\tau\mu &= 0.25\,,\\
\tau e &= 0.19\,,
\end{cases}
\qquad
\epsilon_{B} :
\begin{cases}
\tau^+\tau^-~(\mu) &= 4.8\times10^{-4}\,,\\
\tau^+\tau^-~(e) &= 4.0\times10^{-4}\,.
\end{cases}
\end{equation}

\subsubsection*{Stage 3: $\boldsymbol{(\sqrt{s} = 240\;{\rm GeV}\,, \mathfrak{L}_{\rm int} = 10.8\;{\rm ab}^{-1})}$}
The backgrounds are same as Stage 3. We impose following kinematic cuts:
\begin{equation}
M_{l\tau} > 170~\mathrm{GeV}\,, \qquad E_{l} > 110~\mathrm{GeV}\,.
\end{equation}
The $W^{+}W^{-}$ background is wiped out completely like the previous stage. The cut efficiencies are
\begin{equation}
\epsilon_{S}:
\begin{cases}
\tau\mu &= 0.27\,,\\
\tau e &= 0.22\,,
\end{cases}
\qquad
\epsilon_{B} :
\begin{cases}
\tau^+\tau^-~(\mu) &= 6.9\times10^{-4}\,,\\
\tau^+\tau^-~(e) &= 5.8\times10^{-4}\,.
\end{cases}
\end{equation}

\subsubsection*{Stage 4: $\boldsymbol{(\sqrt{s} = 365\;{\rm GeV}\,, \mathfrak{L}_{\rm int} = 2.70\;{\rm ab}^{-1})}$}
The backgrounds are same as Stage 2 and 3. The following kinematic cuts are applied:
\begin{equation}
M_{l\tau} > 200~\mathrm{GeV}\,, \qquad E_{l} > 180~\mathrm{GeV}\,.
\end{equation}
Similar to Stage 2 and 3, the diboson background is eliminated. The signal and background efficiencies are
\begin{equation}
\epsilon_{S}:
\begin{cases}
\tau\mu &= 0.26\,,\\
\tau e &= 0.24\,,
\end{cases}
\qquad
\epsilon_{B} :
\begin{cases}
\tau^+\tau^-~(\mu) &= 1.1\times10^{-4}\,,\\
\tau^+\tau^-~(e) &= 1.0\times10^{-4}\,.
\end{cases}
\end{equation}

To estimate the optimal sensitivities, we employ the observable $\cos\theta_l$, where $\theta_l$ denotes the polar angle of the charged lepton ($l=\mu,e$) with respect to the beam axis. The corresponding test statistic, $\chi^2_{\rm OOT}$, is constructed following the prescription described in Sec.~\ref{sec:oot}. The projected sensitivity limits at the 95\% confidence level (C.L.) correspond to $\chi^2_{\rm OOT}=3.84$ and $\chi^2_{\rm OOT}=5.99$ for the one and two-parameter analyses, respectively. The resulting 95\% C.L. sensitivities to the WCs (in TeV$^{-2}$) for the processes $e^{+}e^{-}\to\tau\mu$ and $e^{+}e^{-}\to\tau e$ are presented in Tab.~\ref{tab:sensitivity1}.

\begin{table}[htb!]
    \centering
    \renewcommand{\arraystretch}{1.25}{
    \begin{tabular}{|
    >{\centering\arraybackslash}p{0.075\textwidth}|
    >{\centering\arraybackslash}p{0.195\textwidth}|
    >{\centering\arraybackslash}p{0.195\textwidth}|
    >{\centering\arraybackslash}p{0.195\textwidth}|
    >{\centering\arraybackslash}p{0.195\textwidth}|}
    \hline \hline
    \multirow{3}*{\textbf{WCs}} & \textbf{Stage 1} & \textbf{Stage 2} & \textbf{Stage 3} & \textbf{Stage 4} \\
     & $\boldsymbol{\sqrt{s}=91.2\,{\rm GeV}}$ & $\boldsymbol{\sqrt{s}=161\,{\rm GeV}}$ & $\boldsymbol{\sqrt{s}=240\,{\rm GeV}}$ & $\boldsymbol{\sqrt{s}=365\,{\rm GeV}}$ \\
     & $\boldsymbol{\mathfrak{L}_{\rm int}= 205\,{\rm ab}^{-1}}$ & $\boldsymbol{\mathfrak{L}_{\rm int}= 19.2\,{\rm ab}^{-1}}$ & $\boldsymbol{\mathfrak{L}_{\rm int}= 10.8\,{\rm ab}^{-1}}$ & $\boldsymbol{\mathfrak{L}_{\rm int}= 2.70\,{\rm ab}^{-1}}$ \\
    \hline \hline
    $[C_{eW}]_{\tau e}$ & $\pm7.83\times10^{-3}$ & $\pm1.46\times10^{-2}$ & $\pm1.30\times10^{-2}$ & $\pm7.43\times10^{-3}$ \\
    $[C_{eB}]_{\tau e}$ & $\pm1.34\times10^{-2}$ & $\pm9.84\times10^{-3}$ & $\pm9.04\times10^{-3}$ & $\pm5.42\times10^{-3}$ \\ \hline
    $[C^{(+)}_{H\ell}]_{\tau e}$ & $\pm5.02\times10^{-3}$ & $\pm5.42\times10^{-2}$ & $\pm4.92\times10^{-2}$ & $\pm3.07\times10^{-2}$ \\
    $[C_{He}]_{\tau e}$ & $\pm4.62\times10^{-3}$ & $\pm4.22\times10^{-2}$ & $\pm4.20\times10^{-2}$ & $\pm2.83\times10^{-2}$ \\
    \hline \hline
    $[C_{eW}]_{\tau\mu}$ & $\pm1.22\times10^{-2}$ & $\pm2.63\times10^{-2}$ & $\pm3.03\times10^{-2}$ & $\pm2.35\times10^{-2}$ \\
    $[C_{eB}]_{\tau\mu}$ & $\pm2.27\times10^{-2}$ & $\pm2.59\times10^{-2}$ & $\pm2.63\times10^{-2}$ & $\pm1.95\times10^{-2}$ \\ \hline
    $[C^{(+)}_{H\ell}]_{\tau\mu}$ & $\pm4.22\times10^{-3}$ & $\pm7.83\times10^{-2}$ & $\pm1.46\times10^{-1}$ & $\pm1.71\times10^{-1}$ \\
    $[C_{He}]_{\tau\mu}$ & $\pm3.82\times10^{-3}$ & $\pm7.03\times10^{-2}$ & $\pm1.30\times10^{-1}$ & $\pm1.63\times10^{-1}$ \\
    \hline \hline
    \end{tabular}}
    \caption{Projected 95\% C.L. sensitivities on WCs (in TeV$^{-2}$) from $e^{+}e^{-} \to \tau\mu\,/\,\tau e$ at the FCC-ee.}
    \label{tab:sensitivity1}
\end{table}

At Stage 1, both the dipole and Higgs-current operators exhibit excellent sensitivity owing to the resonant enhancement of the signal at the $Z$ pole. Furthermore, the $\tau e$ channel yields stronger constraints than the $\tau\mu$ channel because of the additional $t$-channel contribution, which supplements the $s$-channel amplitude. Finally, it should be noted that the four FCC-ee stages differ not only in center-of-mass energy but also in their projected integrated luminosities. Consequently, the quoted sensitivities should not be interpreted as illustrating a simple dependence on either energy or luminosity alone; rather, they represent the projected reach under the corresponding baseline experimental configuration of each FCC-ee stage.

\begin{figure}[htb!]
    \centering
    \includegraphics[width=0.475\linewidth]{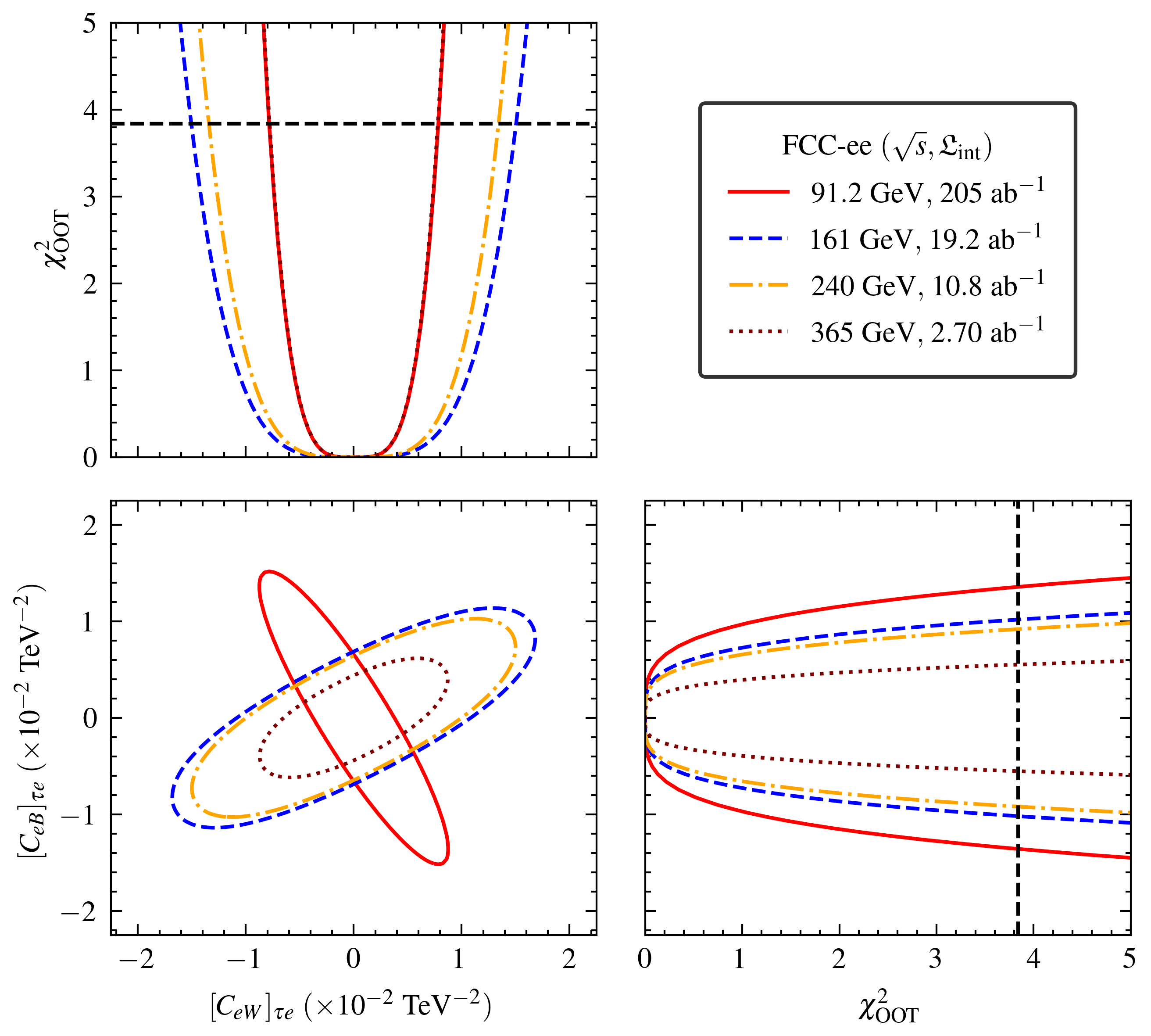}
    \includegraphics[width=0.475\linewidth]{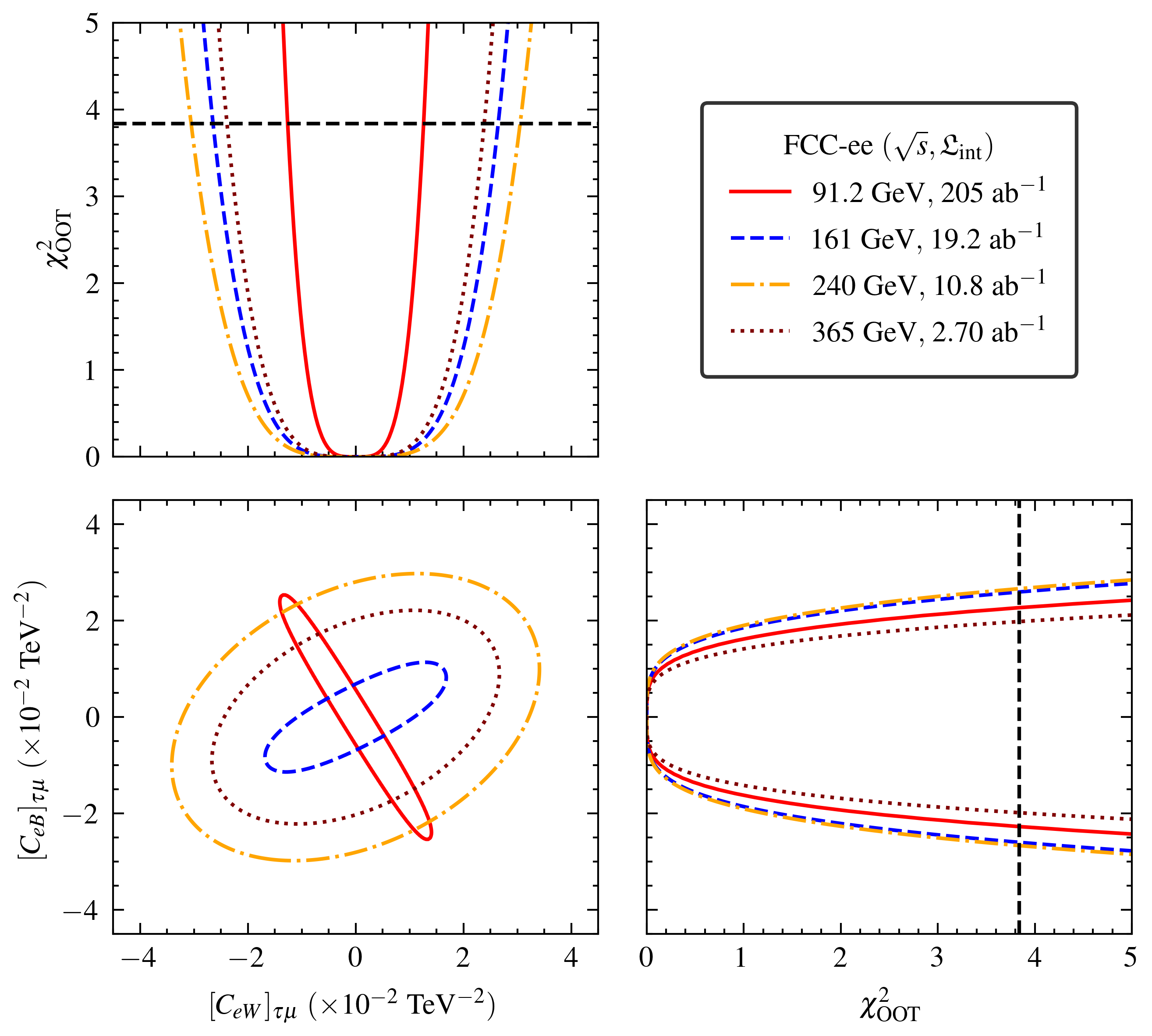} \\
    \includegraphics[width=0.475\linewidth]{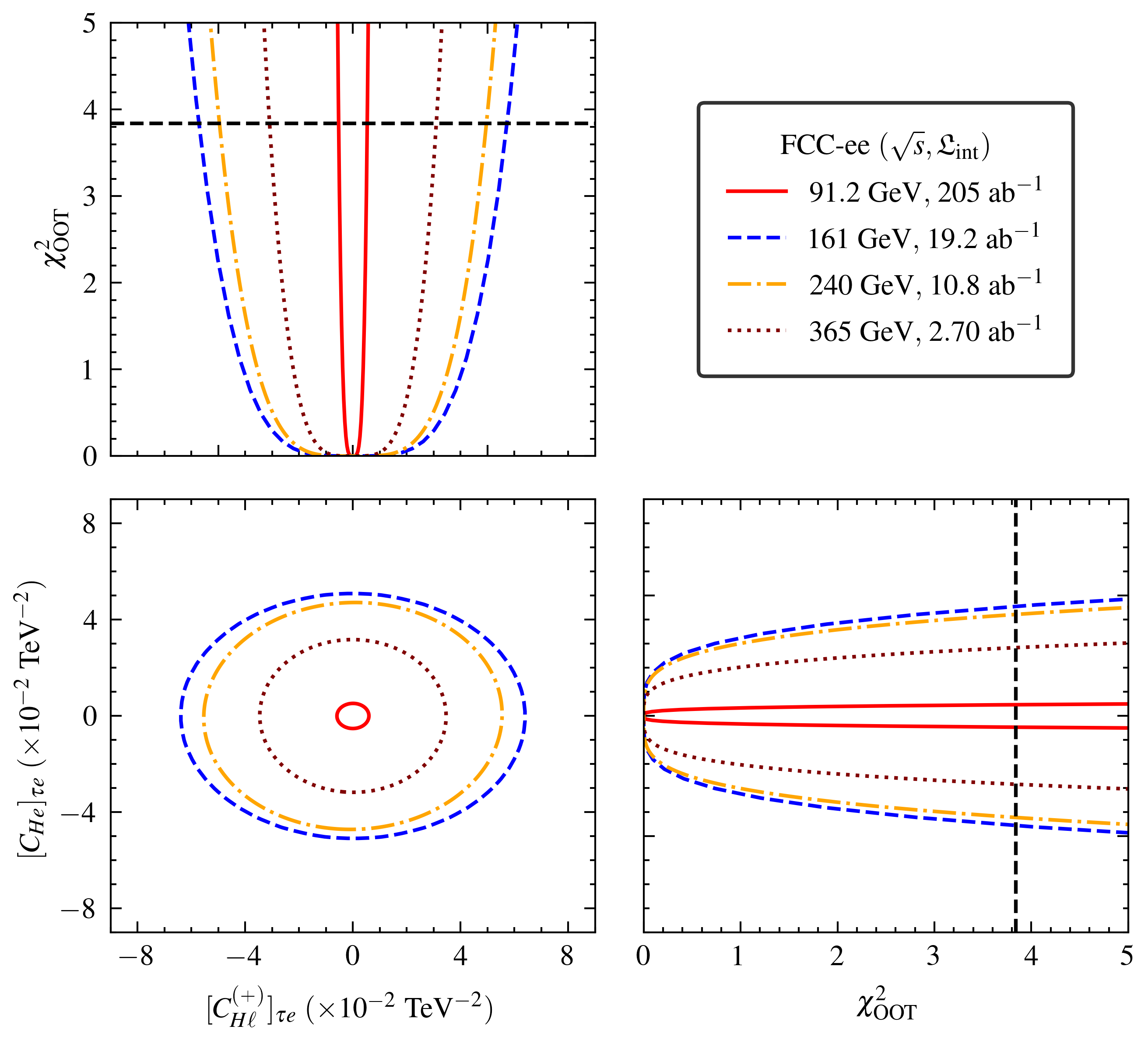}
    \includegraphics[width=0.475\linewidth]{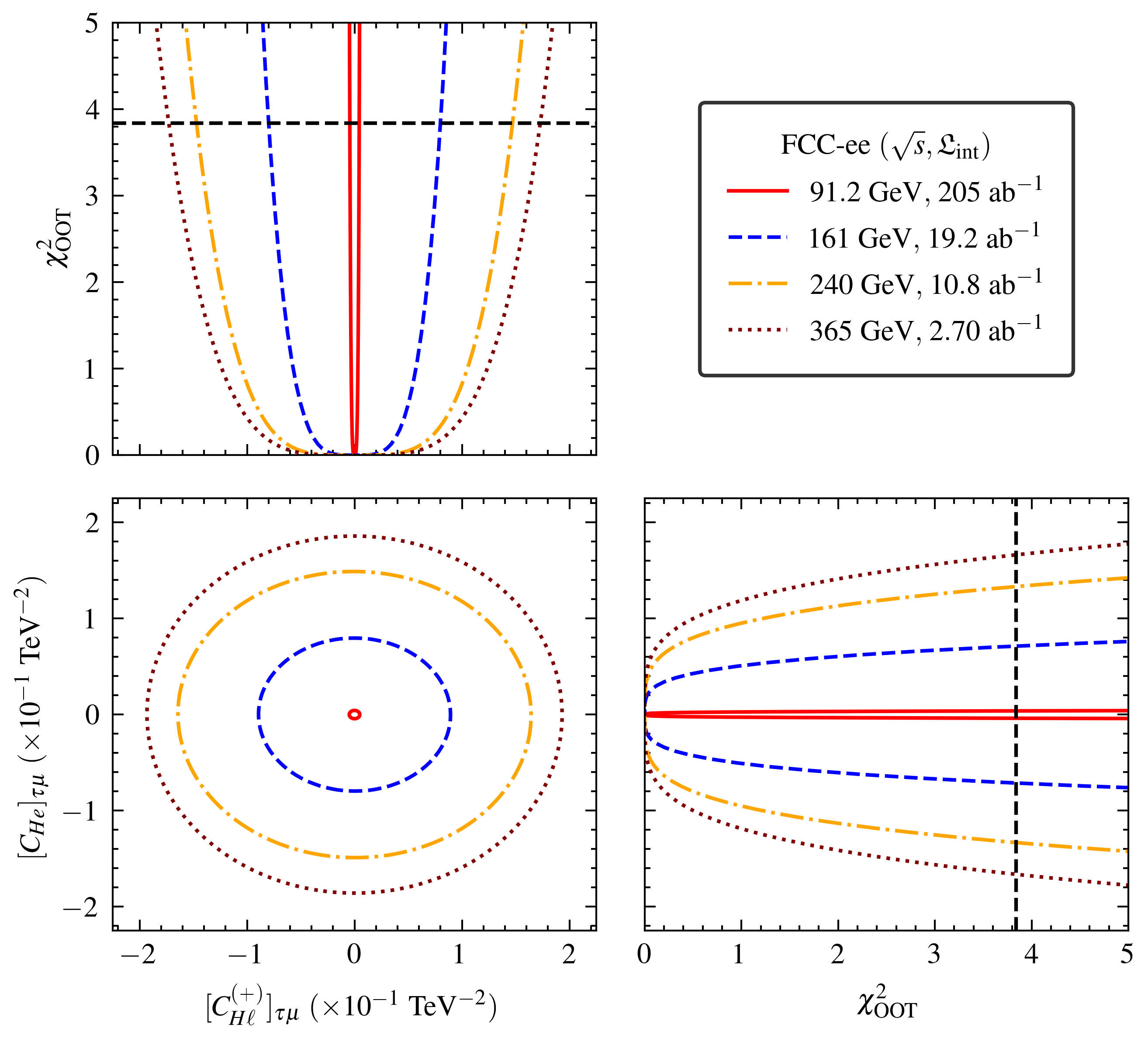}
    \caption{Projected 95\% C.L. plots/sensitivity contours for the WCs (in TeV$^{-2}$) obtained from the processes $e^{+}e^{-}\to\tau\mu/\tau e$ at the FCC-ee.}
    \label{fig:contour1}
\end{figure}

In Fig.~\ref{fig:contour1}, we present the projected 95\% C.L. sensitivity contours for the WCs (in TeV$^{-2}$) obtained from the processes $e^{+}e^{-}\to\tau\mu/\tau e$ at the different FCC-ee energy stages. Several noteworthy features can be observed. The two dipole operators couple to left- and right-handed leptons and therefore interfere with one another. This interference manifests itself through the diagonally oriented elliptical contours in the dipole parameter planes. In contrast, the Higgs-current operators $\mathcal{O}^{(+)}_{H\ell}$ and $\mathcal{O}_{He}$ involve left-handed and right-handed leptons, respectively, and hence do not interfere, resulting in contours that are nearly aligned with the coordinate axes and exhibit a more circular shape. Furthermore, for the dipole operators, the interference is constructive at the $Z$ pole, whereas it becomes destructive at higher CM energies. This behavior can be understood from the effective couplings entering the photon and $Z$-mediated amplitudes,
\begin{equation}
\begin{aligned}
C^e_\gamma &= -\sin\theta_W\,C_{eW}+\cos\theta_W\,C_{eB},,\\
C^e_Z &= -\cos\theta_W\,C_{eW}-\sin\theta_W\,C_{eB},.
\end{aligned}
\end{equation}
The relative signs in these combinations imply that the interference between $C_{eW}$ and $C_{eB}$ is destructive for the photon-mediated contribution, while it is constructive for the $Z$-mediated contribution. Around the $Z$ resonance, the $Z$-exchange amplitude dominates because of the resonant enhancement of the propagator, leading to an overall constructive interference. As the CM energy moves away from the $Z$ pole, the resonant enhancement is lost and the photon-mediated contribution becomes increasingly important. Consequently, the destructive interference associated with the photon exchange dominates, resulting in the observed change in the orientation of the sensitivity contours.

\subsection{Channel: $e^{+}e^{-} \to bs\,/\,bd$}
\label{sec:qfv1}
The signal is characterized by a two-jet final state consisting of one $b$-tagged jet ($N_j^b=1$) and one light jet\footnote{A light jet refers to a jet originating from a gluon ($g$) or a light quark ($u,d,s,c$).} ($N_j=1$). The dominant SM background processes are $e^{+}e^{-}\to\tau^{+}\tau^{-}$, $jj$ (including $b\bar{b}$), $W^{+}W^{-}$ (Stage 2 and beyond), and $ZZ$ (Stage 3 and beyond). Although the dijet background has a significantly larger production cross section than the signal, it can be efficiently suppressed by suitable event-selection criteria. To discriminate the signal from the SM backgrounds, we employ the following kinematic observables:
\begin{itemize}
\itemsep-0em
\item The invariant mass of the $b$-tagged jet and the light jet, defined as
\begin{equation}
M_{bj}=\sqrt{\left(p_b + p_j\right)^2} =\sqrt{\left(E_b + E_j\right)^2 \left|\vec{p_b}+\vec{p_j}\right|^2}\,,
\end{equation}
where $p_b\equiv(E_b,\vec{p_b})$ and $p_j\equiv(E_j,\vec{p_j})$ denote the four-momenta of the $b$-tagged jet and the light jet, respectively. Since the signal originates from a hard $2 \to 2$ scattering process, the visible final-state particles carry a larger fraction of the available center-of-mass energy. Consequently, the $M_{bj}$ distribution is shifted towards higher values compared to those of most background processes, with the exception of the dijet background (see Fig.~\ref{fig:dists2}).
\item The missing energy of an event, defined as
\begin{equation}
    E_{\rm miss}=\sqrt{s} - \sum_{\rm visible}E\,,
\end{equation}
where the summation runs over the energies of all visible final-state particles. As the signal contains no genuine source of missing energy, apart from detector effects and soft radiation, $E_{\rm miss}$ provides an efficient discriminator against background processes involving missing particles in the final state (see Fig.~\ref{fig:dists2}).
\end{itemize}
The cuts employed for the $bs/bd$ channels at the different FCC-ee stages, together with the corresponding $\epsilon_{S}$ and $\epsilon_{B}$, are presented below.

\subsubsection*{Stage 1: $\boldsymbol{(\sqrt{s} = 91.2\;{\rm GeV}\,, \mathfrak{L}_{\rm int} = 205\;{\rm ab}^{-1})}$}
At this stage, the dominant SM background arises from $jj$ and $\tau^{+}\tau^{-}$ process. To suppress this background, the following kinematic requirements are imposed:
\begin{equation}
M_{bj} > 75~\mathrm{GeV}\,, \qquad E_{\rm miss} < 20~\mathrm{GeV}\,.
\end{equation}
The resulting signal and background efficiencies are
\begin{equation}
\epsilon_{S}:
\begin{cases}
bs &= 0.34\,,\\
bd &= 0.35\,,
\end{cases}
\qquad
\epsilon_{B} :
\begin{cases}
jj &= 7.5\times10^{-2}\,,\\
\tau^+\tau^- &= 2.0\times10^{-3}\,.
\end{cases}
\end{equation}

\subsubsection*{Stage 2: $\boldsymbol{(\sqrt{s} = 161\;{\rm GeV}\,, \mathfrak{L}_{\rm int} = 19.2\;{\rm ab}^{-1})}$}
At this stage, $W^+W^-$ emerges as an additional background, alongside $jj$ and $\tau^{+}\tau^{-}$. The following kinematic cuts are imposed:
\begin{equation}
M_{bj} > 135~\mathrm{GeV}\,, \qquad E_{\rm miss} < 25~\mathrm{GeV}\,.
\end{equation}
The signal and background efficiencies following the cuts are
\begin{equation}
\epsilon_{S}:
\begin{cases}
bs &= 0.37\,,\\
bd &= 0.39\,,
\end{cases}
\qquad
\epsilon_{B} :
\begin{cases}
jj &= 8.2\times10^{-2}\,,\\
\tau^+\tau^- &= 1.8\times10^{-3}\,,\\
W^+W^- &= 1.6\times10^{-3}\,.
\end{cases}
\end{equation}

\subsubsection*{Stage 3: $\boldsymbol{(\sqrt{s} = 240\;{\rm GeV}\,, \mathfrak{L}_{\rm int} = 10.8\;{\rm ab}^{-1})}$}
At this stage, the $ZZ$ background emerges. We impose following kinematic cuts:
\begin{equation}
M_{bj} > 200~\mathrm{GeV}\,, \qquad E_{\rm miss} < 50~\mathrm{GeV}\,.
\end{equation}
The cut efficiencies are
\begin{equation}
\epsilon_{S}:
\begin{cases}
bs &= 0.37\,,\\
bd &= 0.41\,,
\end{cases}
\qquad
\epsilon_{B} :
\begin{cases}
jj &= 8.7\times10^{-2}\,,\\
\tau^+\tau^- &= 1.8\times10^{-3}\,,\\
W^+W^- &= 8.0\times10^{-4}\,,\\
ZZ &= 6.7\times10^{-4}\,.
\end{cases}
\end{equation}

\subsubsection*{Stage 4: $\boldsymbol{(\sqrt{s} = 365\;{\rm GeV}\,, \mathfrak{L}_{\rm int} = 2.70\;{\rm ab}^{-1})}$}
The backgrounds are same as Stage 3. The following kinematic cuts are applied:
\begin{equation}
M_{bj} > 250~\mathrm{GeV}\,, \qquad E_{\rm miss} < 75~\mathrm{GeV}\,.
\end{equation}
The signal and background efficiencies are
\begin{equation}
\epsilon_{S}:
\begin{cases}
bs &= 0.39\,,\\
bd &= 0.42\,,
\end{cases}
\qquad
\epsilon_{B} :
\begin{cases}
jj &= 9.4\times10^{-2}\,,\\
\tau^+\tau^- &= 1.1\times10^{-3}\,,\\
W^+W^- &= 2.2\times10^{-3}\,,\\
ZZ &= 4.5\times10^{-3}\,.
\end{cases}
\end{equation}

For optimal sensitivity estimation, we employ the observable $\cos\theta_j$, where $\theta_j$ denotes the polar angle of the light jet with respect to the beam axis. The 95\% C.L. sensitivities to the WCs (in TeV$^{-2}$) for the processes $e^{+}e^{-}\to bs$ and $e^{+}e^{-}\to bd$ are presented in Tab.~\ref{tab:sensitivity2}.

\begin{table}[htb!]
    \centering
    \renewcommand{\arraystretch}{1.25}{
    \begin{tabular}{|
    >{\centering\arraybackslash}p{0.075\textwidth}|
    >{\centering\arraybackslash}p{0.195\textwidth}|
    >{\centering\arraybackslash}p{0.195\textwidth}|
    >{\centering\arraybackslash}p{0.195\textwidth}|
    >{\centering\arraybackslash}p{0.195\textwidth}|}
    \hline \hline
    \multirow{3}*{\textbf{WCs}} & \textbf{Stage 1} & \textbf{Stage 2} & \textbf{Stage 3} & \textbf{Stage 4} \\
    & $\boldsymbol{\sqrt{s}=91.2\,{\rm GeV}}$ & $\boldsymbol{\sqrt{s}=161\,{\rm GeV}}$ & $\boldsymbol{\sqrt{s}=240\,{\rm GeV}}$ & $\boldsymbol{\sqrt{s}=365\,{\rm GeV}}$ \\
     & $\boldsymbol{\mathfrak{L}_{\rm int}= 205\,{\rm ab}^{-1}}$ & $\boldsymbol{\mathfrak{L}_{\rm int}= 19.2\,{\rm ab}^{-1}}$ & $\boldsymbol{\mathfrak{L}_{\rm int}= 10.8\,{\rm ab}^{-1}}$ & $\boldsymbol{\mathfrak{L}_{\rm int}= 2.70\,{\rm ab}^{-1}}$ \\
    \hline \hline
    $[C_{dW}]_{bd}$ & $\pm5.42\times10^{-2}$ & $\pm7.43\times10^{-2}$ & $\pm7.83\times10^{-2}$ & $\pm9.44\times10^{-2}$ \\
    $[C_{dB}]_{bd}$ & $\pm1.02\times10^{-1}$ & $7.03\pm\times10^{-2}$ & $\pm6.63\times10^{-2}$ & $\pm7.83\times10^{-2}$ \\ \hline
    $[C^{(+)}_{Hq}]_{bd}$ & $\pm1.40\times10^{-2}$ & $\pm2.23\times10^{-1}$ & $\pm3.63\times10^{-1}$ & $\pm2.15\times10^{-1}$ \\
    $[C_{Hd}]_{bd}$ & $\pm1.40\times10^{-2}$ & $\pm2.23\times10^{-1}$ & $\pm3.67\times10^{-1}$ & $\pm2.15\times10^{-1}$ \\
    \hline \hline
    $[C_{dW}]_{bs}$ & $\pm4.62\times10^{-2}$ & $\pm7.43\times10^{-2}$ & $\pm8.23\times10^{-2}$ & $\pm9.84\times10^{-2}$ \\
    $[C_{dB}]_{bs}$ & $\pm9.04\times10^{-2}$ & $\pm7.43\times10^{-2}$ & $\pm7.03\times10^{-2}$ & $\pm8,23\times10^{-2}$ \\ \hline
    $[C^{(+)}_{Hq}]_{bs}$ & $\pm1.40\times10^{-2}$ & $\pm2.27\times10^{-1}$ & $\pm3.76\times10^{-1}$ & $\pm2.19\times10^{-1}$ \\
    $[C_{Hd}]_{bs}$ & $\pm1.40\times10^{-2}$ & $\pm2.31\times10^{-1}$ & $\pm3.80\times10^{-1}$ & $\pm2.19\times10^{-1}$ \\
    \hline \hline
    \end{tabular}}
    \caption{Projected sensitivities on WCs (in TeV$^{-2}$) from $e^{+}e^{-} \to bs\,/\,bd$ at the FCC-ee.}
    \label{tab:sensitivity2}
\end{table}

Similar to the $\tau \mu/\tau e$ case, at the $Z$ pole, the sensitivities are significantly enhanced owing to the resonant intermediate $Z$ boson production, resulting in the strongest constraints for both the dipole and Higgs-current operators. For the dipole operators, the sensitivities remain relatively stable across all the stages, whereas the Higgs-current operators exhibit comparatively weaker sensitivities away from the $Z$ pole. Similar sensitivities are obtained for the $bd$ and $bs$ flavor structures, reflecting their analogous production mechanisms and event topologies. This similarity arises because jets initiated by $d$ and $s$ quarks exhibit nearly identical detector signatures and kinematic characteristics, making them difficult to distinguish experimentally.

\begin{figure}[htb!]
    \centering
    \includegraphics[width=0.475\linewidth]{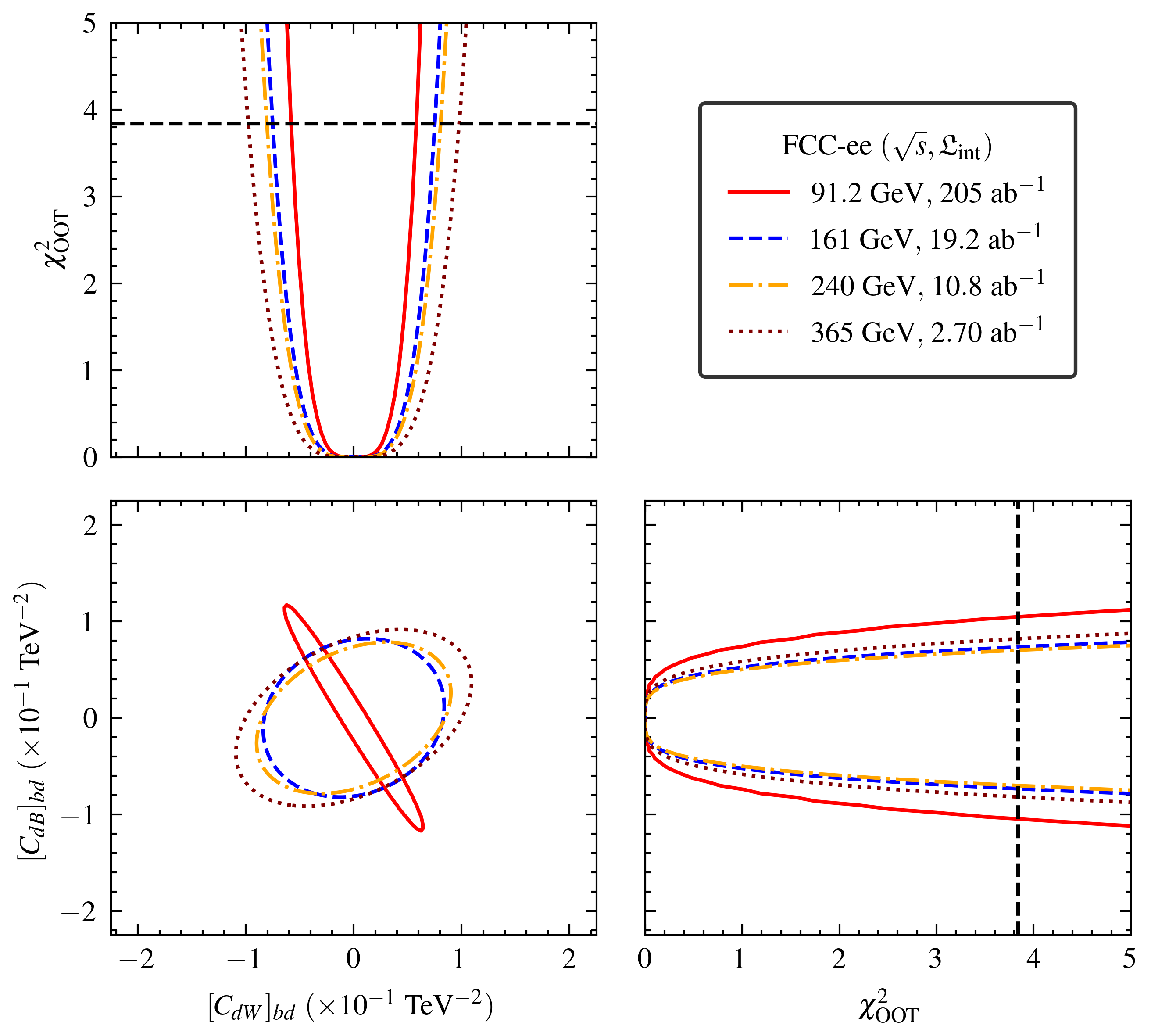}
    \includegraphics[width=0.475\linewidth]{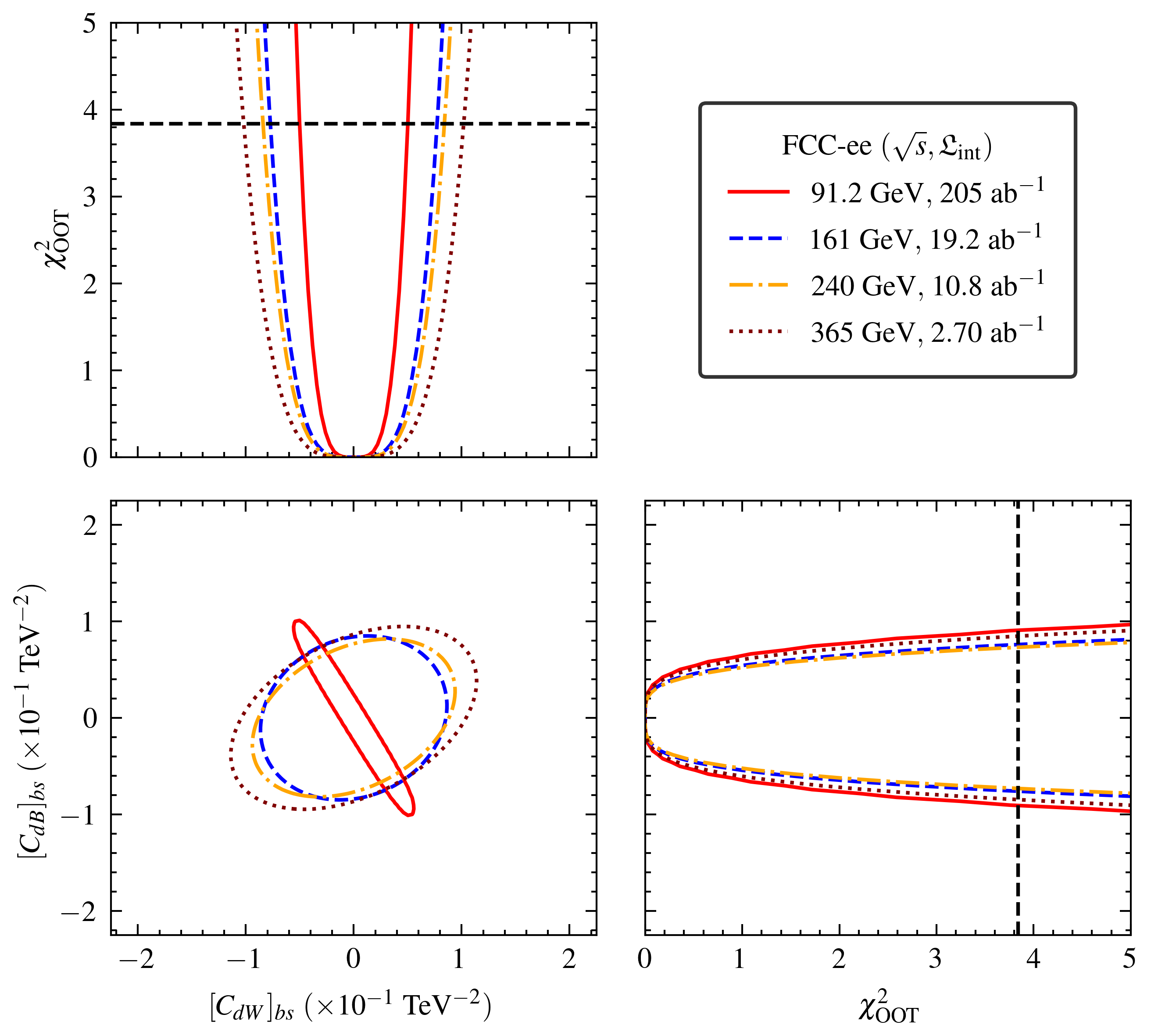} \\
    \includegraphics[width=0.475\linewidth]{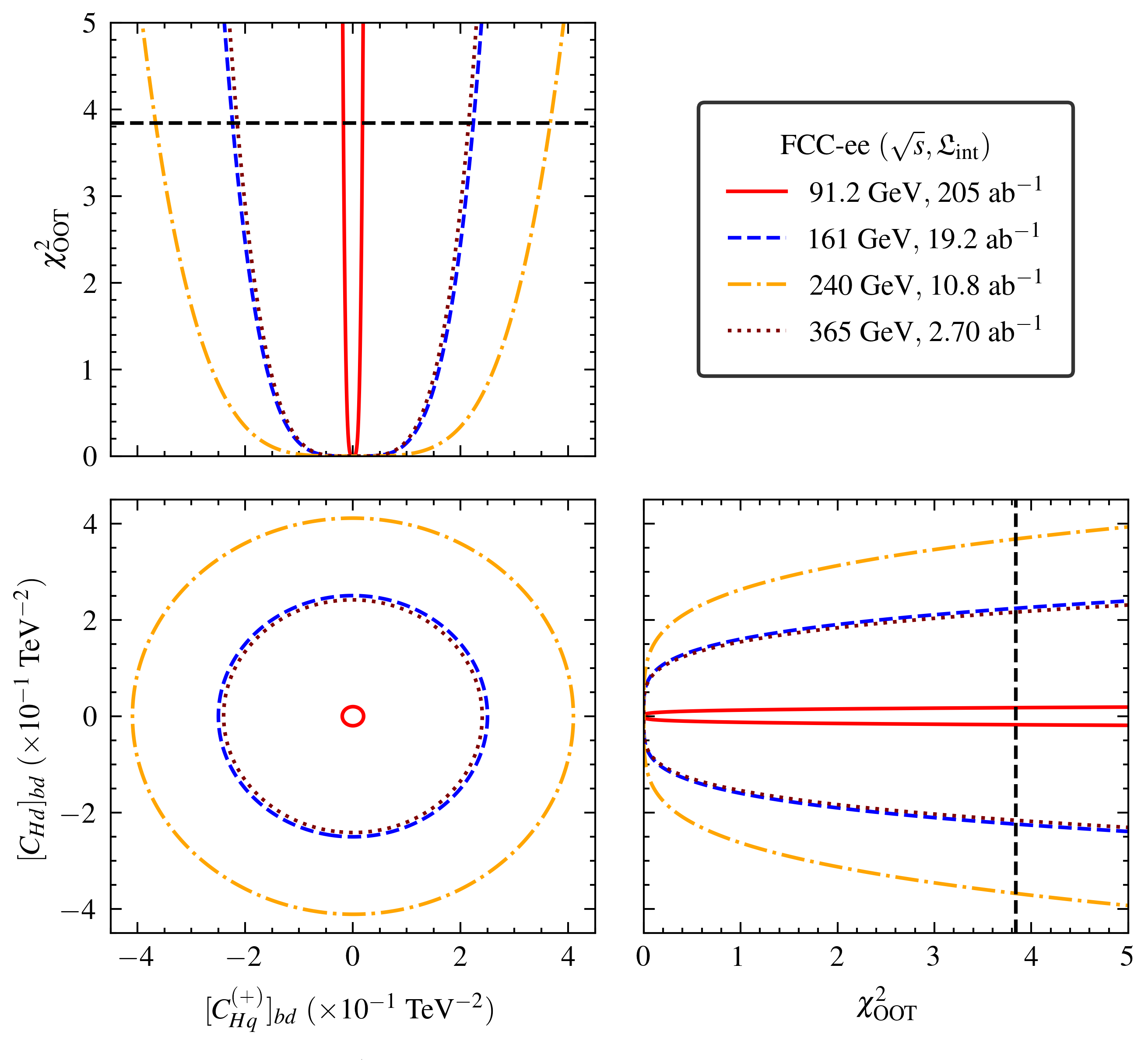}
    \includegraphics[width=0.475\linewidth]{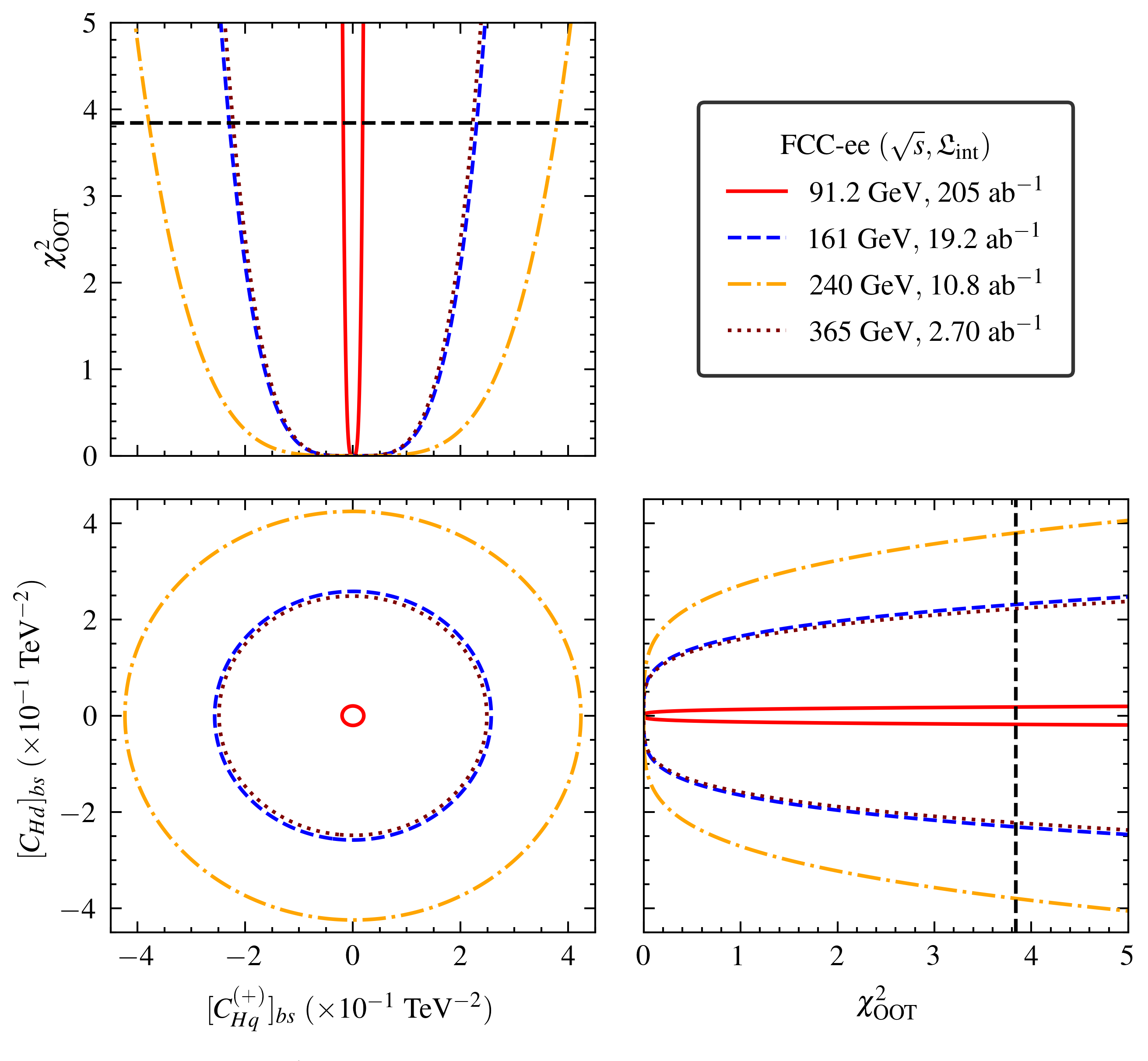}
    \caption{Projected 95\% C.L. plots/sensitivity contours for the WCs (in TeV$^{-2}$) obtained from the processes: $e^{+}e^{-} \to bs\,/\,bd$ at the FCC-ee.}
    \label{fig:contour2}
\end{figure}

In Fig.~\ref{fig:contour2}, we present the projected 95\% C.L. sensitivity contours for the WCs (in TeV$^{-2}$) obtained from the processes $e^{+}e^{-}\to bs/bd$ at the different FCC-ee energy stages. The overall behavior closely resembles that observed for the $\tau\mu/\tau e$ channels. In particular, the dipole operators exhibit constructive interference at the $Z$ pole, while the interference becomes destructive at higher CM energies, leading to a corresponding change in the orientation of the sensitivity contours. This behavior follows from the effective couplings entering the photon and $Z$-mediated amplitudes,
\begin{equation}
\begin{aligned}
C^{d}_\gamma &= -\sin\theta_W\,C_{dW}+\cos\theta_W\,C_{dB}\,,\\
C^{d}_Z &= -\cos\theta_W\,C_{dW}-\sin\theta_W\,C_{dB}\,.
\end{aligned}
\end{equation}
Since these combinations are identical to those in the charged-lepton sector, the interference pattern and, consequently, the orientation of the sensitivity contours closely resemble those obtained for the $\tau\mu/\tau e$ channels. Also similar to the charged lepton case, the Higgs-current operators do not interfere, resulting in contours that are nearly circular. However, the sensitivity contours for the $bs$ and $bd$ flavor structures are almost completely degenerate, reflecting the nearly indistinguishable detector signatures and kinematic properties of jets initiated by $d$ and $s$ quarks.

\subsection{Channel: $e^{+}e^{-} \to tc\,/\,tu$}
\label{sec:qfv2}
For the $tc/tu$ signal, we consider the leptonic decay mode of the top quark, $t\to b\ell\nu$. The signal is therefore characterized by one charged lepton ($N_e/N_\mu=1$), one $b$-tagged jet ($N_j^b=1$), one light jet ($N_j=1$), and missing energy arising from the neutrino. Although the hadronic decay mode of the top quark has a larger branching fraction, it results in a four-jet final state, making the signal considerably more difficult to distinguish from the SM backgrounds. Since the top quark mass is $m_t \simeq 173$ GeV, the $tc/tu$ channels are kinematically accessible only from Stage 3 onwards. The dominant SM background processes are $W^+W^-$, $ZZ$, and $t\overline{t}$ (the latter only at Stage 4). To discriminate the signal from the backgrounds, we employ the following observables:
\begin{itemize}
    \itemsep-0em
    \item The invariant mass of the system comprising the (b)-tagged jet, the charged lepton, and the reconstructed missing particle, defined as
    \begin{equation}
    M^{\rm miss}_{bl}= \sqrt{\left(p_b + p_l + p_{\rm miss}\right)^2} = \sqrt{\left(E_b+E_l+E_{\rm miss}\right)^2 -\left|\vec{p_b}+\vec{p_l}+\vec{p_{\rm miss}}\right|^2}\,,
    \end{equation}
    where $p_b\equiv(E_b,\vec{p_b})$, $p_l\equiv(E_l,\vec{p_l})$, and $p_{\rm miss}\equiv(E_{\rm miss},\vec{p_{\rm miss}})$ are the four-momenta of the $b$-tagged jet, the charged lepton, and the reconstructed missing particle, respectively. This observable reconstructs the top-quark candidate and is therefore expected to peak around the top-quark mass, for the signal. The missing 4-momentum is reconstructed from the event energy-momentum imbalance.
    \item The energy of the light jet, denoted by $E_j$, is another powerful discriminating observable. For the signal, the light jet is produced directly in the underlying $2\to2$ hard scattering process. Consequently, for a fixed center-of-mass energy, its energy is kinematically constrained, resulting in a sharp peak at
    \begin{equation}
        E_j=\frac{s-m_t^2}{2\sqrt{s}}\,.
    \end{equation}
    In contrast, for the background processes, the light jet typically originates from the decays of intermediate particles, resulting in a much broader energy distribution.
\end{itemize}
The distributions are shown in Fig.~\ref{fig:dists3}. The cuts employed for the $tc/tu$ channels at the different stages, together with the corresponding $\epsilon_{S}$ and $\epsilon_{B}$ efficiencies, are detailed below.

\subsubsection*{Stage 3: $\boldsymbol{(\sqrt{s} = 240\;{\rm GeV}\,, \mathfrak{L}_{\rm int} = 10.8\;{\rm ab}^{-1})}$}
The background processes are $W^+W^-$ and $ZZ$. We impose following kinematic cuts:
\begin{equation}
M^{\rm miss}_{bj} \in (160, 190)~\mathrm{GeV}\,, \qquad E_j \in (40,80)~\mathrm{GeV}\,.
\end{equation}
The cut efficiencies are
\begin{equation}
\epsilon_{S}:
\begin{cases}
tc &= 5.6 \times 10^{-2}\,,\\
tu &= 6.6 \times 10^{-2}\,,
\end{cases}
\qquad
\epsilon_{B} :
\begin{cases}
W^+W^- &= 4.1\times10^{-3}\,,\\
ZZ &= 1.1\times10^{-3}\,.
\end{cases}
\end{equation}

\subsubsection*{Stage 4: $\boldsymbol{(\sqrt{s} = 365\;{\rm GeV}\,, \mathfrak{L}_{\rm int} = 2.70\;{\rm ab}^{-1})}$}
In addition to $W^+W^-$ and $ZZ$, $t\overline{t}$ also emerges as a background, at this stage. We impose following kinematic cuts:
\begin{equation}
M^{\rm miss}_{bj} \in (150, 200)~\mathrm{GeV}\,, \qquad E_j \in (100,180)~\mathrm{GeV}\,.
\end{equation}
The cut efficiencies are
\begin{equation}
\epsilon_{S}:
\begin{cases}
tc &= 3.4 \times 10^{-2}\,,\\
tu &= 4.8 \times 10^{-2}\,,
\end{cases}
\qquad
\epsilon_{B} :
\begin{cases}
W^+W^- &= 1.1\times10^{-3}\,,\\
ZZ &= 3.7\times10^{-4}\,,\\
t\overline{t} &= 4.7\times10^{-4}\,.
\end{cases}
\end{equation}

For optimal sensitivity, we employ the observable $\cos\theta_j$, where $\theta_j$ denotes the polar angle of the light jet with respect to the beam axis. The 95\% C.L. sensitivities to the WCs (in TeV$^{-2}$) for the processes $e^{+}e^{-}\to tc$ and $e^{+}e^{-}\to tu$ are presented in Tab.~\ref{tab:sensitivity3}.

\begin{table}[htb!]
    \centering
    \renewcommand{\arraystretch}{1.25}{
    \begin{tabular}{|
    >{\centering\arraybackslash}p{0.075\textwidth}|
    >{\centering\arraybackslash}p{0.195\textwidth}|
    >{\centering\arraybackslash}p{0.195\textwidth}|
    >{\centering\arraybackslash}p{0.195\textwidth}|
    >{\centering\arraybackslash}p{0.195\textwidth}|}
    \hline \hline
    \multirow{3}*{\textbf{WCs}} & \textbf{Stage 1} & \textbf{Stage 2} & \textbf{Stage 3} & \textbf{Stage 4} \\
    & $\boldsymbol{\sqrt{s}=91.2\,{\rm GeV}}$ & $\boldsymbol{\sqrt{s}=161\,{\rm GeV}}$ & $\boldsymbol{\sqrt{s}=240\,{\rm GeV}}$ & $\boldsymbol{\sqrt{s}=365\,{\rm GeV}}$ \\
     & $\boldsymbol{\mathfrak{L}_{\rm int}= 205\,{\rm ab}^{-1}}$ & $\boldsymbol{\mathfrak{L}_{\rm int}= 19.2\,{\rm ab}^{-1}}$ & $\boldsymbol{\mathfrak{L}_{\rm int}= 10.8\,{\rm ab}^{-1}}$ & $\boldsymbol{\mathfrak{L}_{\rm int}= 2.70\,{\rm ab}^{-1}}$ \\
    \hline \hline
    $[C_{uW}]_{tu}$ & \multirow{2}*{$-$} & \multirow{2}*{$-$} & $\pm1.46\times10^{-1}$ & $\pm8.63\times10^{-2}$ \\
    $[C_{uB}]_{tu}$ &  &  & $\pm1.26\times10^{-1}$ & $\pm7.03\times10^{-2}$ \\ \hline
    $[C^{(-)}_{Hq}]_{tu}$ & \multirow{2}*{$-$} & \multirow{2}*{$-$} & $\pm7.83\times10^{-1}$ & $\pm7.43\times10^{-1}$ \\
    $[C_{Hu}]_{tu}$ &  &  & $\pm7.83\times10^{-1}$ & $\pm7.03\times10^{-1}$ \\
    \hline \hline
    $[C_{uW}]_{tc}$ & \multirow{2}*{$-$} & \multirow{2}*{$-$} & $\pm1.59\times10^{-1}$ & $\pm1.06\times10^{-1}$ \\
    $[C_{uB}]_{tc}$ &  &  & $\pm1.38\times10^{-1}$ & $\pm8.63\times10^{-2}$ \\ \hline
    $[C^{(-)}_{Hq}]_{tc}$ & \multirow{2}*{$-$} & \multirow{2}*{$-$} & $\pm8.63\times10^{-1}$ & $\pm8.63\times10^{-1}$ \\
    $[C_{Hu}]_{tc}$ &  &  & $\pm8.63\times10^{-1}$ & $\pm8.63\times10^{-1}$ \\
    \hline \hline
    \end{tabular}}
    \caption{Projected sensitivities on WCs (in TeV$^{-2}$) from $e^{+}e^{-} \to tc\,/\,tu$ at the FCC-ee.}
    \label{tab:sensitivity3}
\end{table}

The dipole operators are constrained more stringently than the Higgs-current operators at both stages, since the momentum dependence of the dipole interactions partially compensates for the suppression arising from the off-shell $s$-channel propagator, thereby enhancing their sensitivity. Furthermore, the projected sensitivities for the $tu$ and $tc$ flavor structures are broadly comparable, reflecting their similar production mechanisms and event topologies. The $tu$ channel, however, exhibits a slightly better sensitivity than the $tc$ channel. This is primarily because a non-negligible fraction of $c$-jets are misidentified as $b$-jets owing to their similar characteristics, causing a loss of signal events under the requirement of exactly one $b$-tagged jet and one light jet in the final state.

\begin{figure}[htb!]
    \centering
    \includegraphics[width=0.475\linewidth]{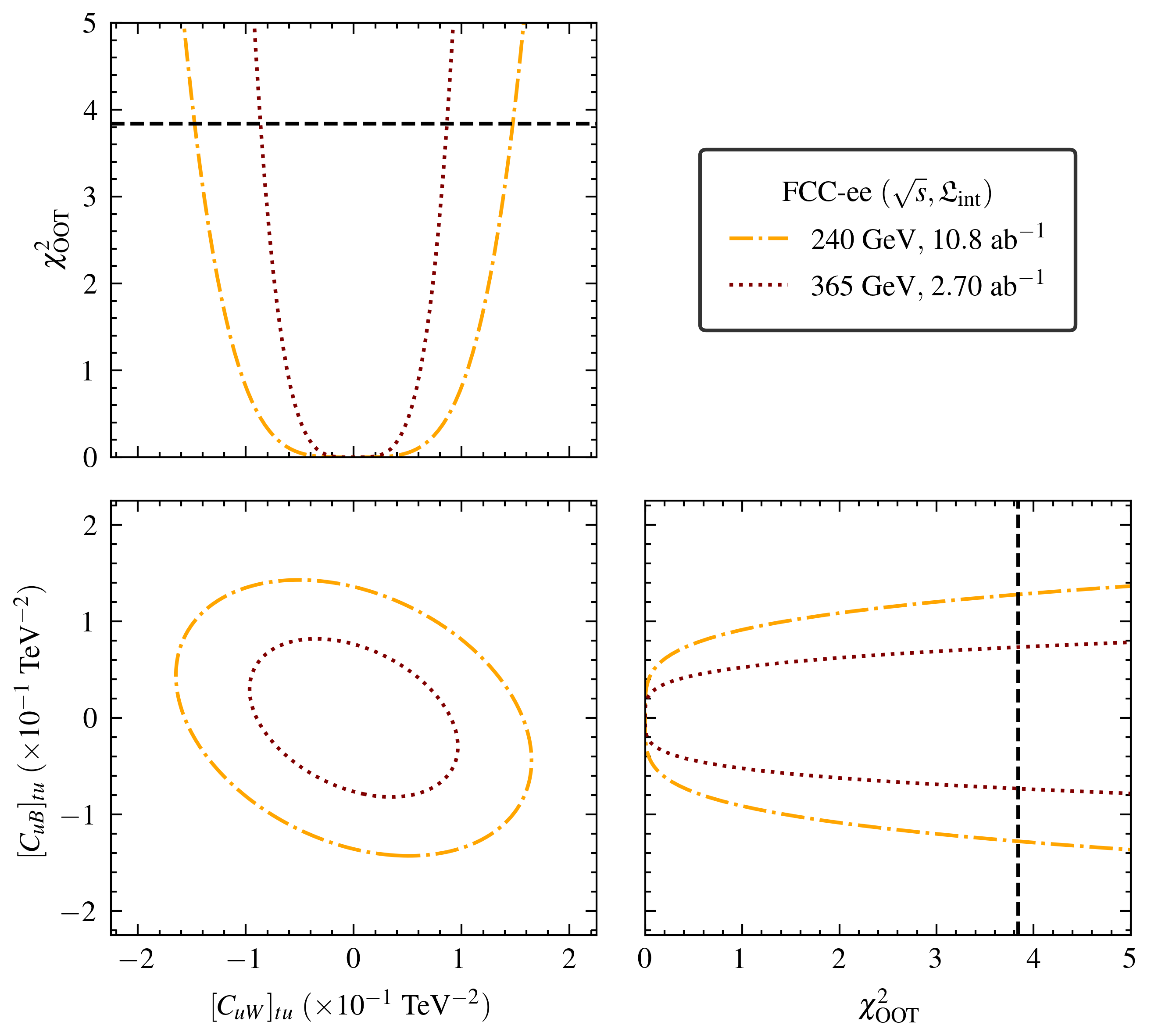}
    \includegraphics[width=0.475\linewidth]{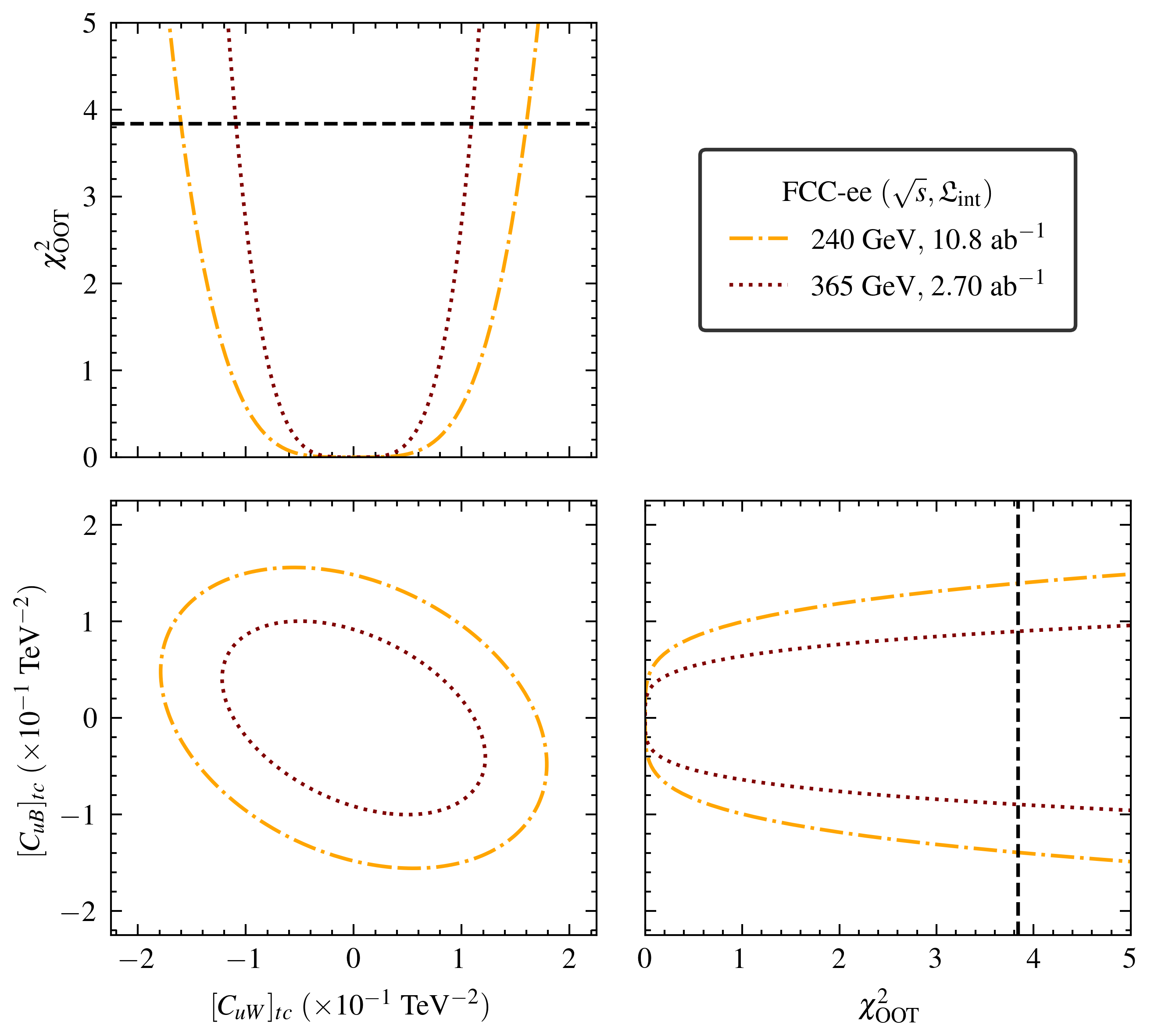} \\
    \includegraphics[width=0.475\linewidth]{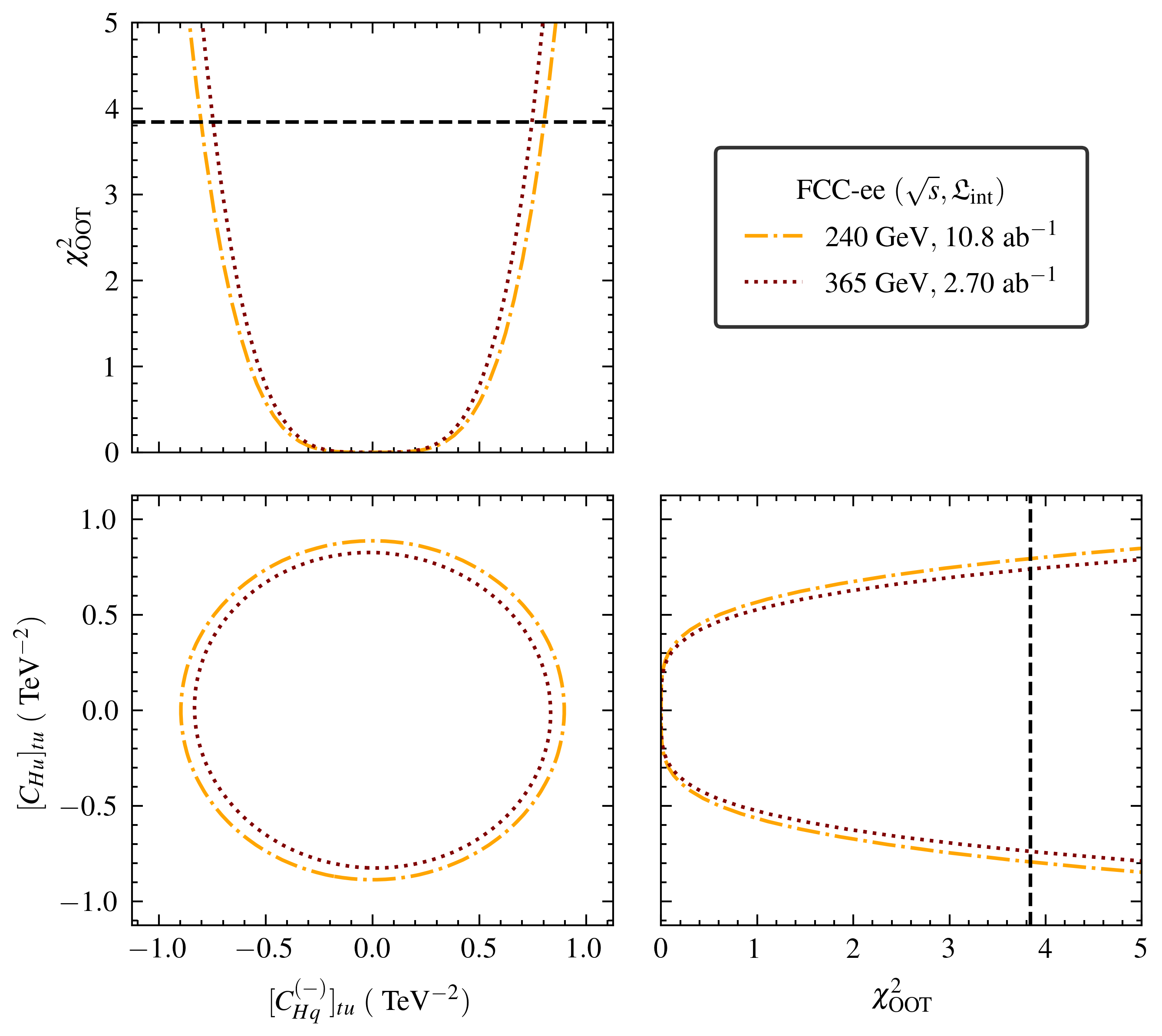}
    \includegraphics[width=0.475\linewidth]{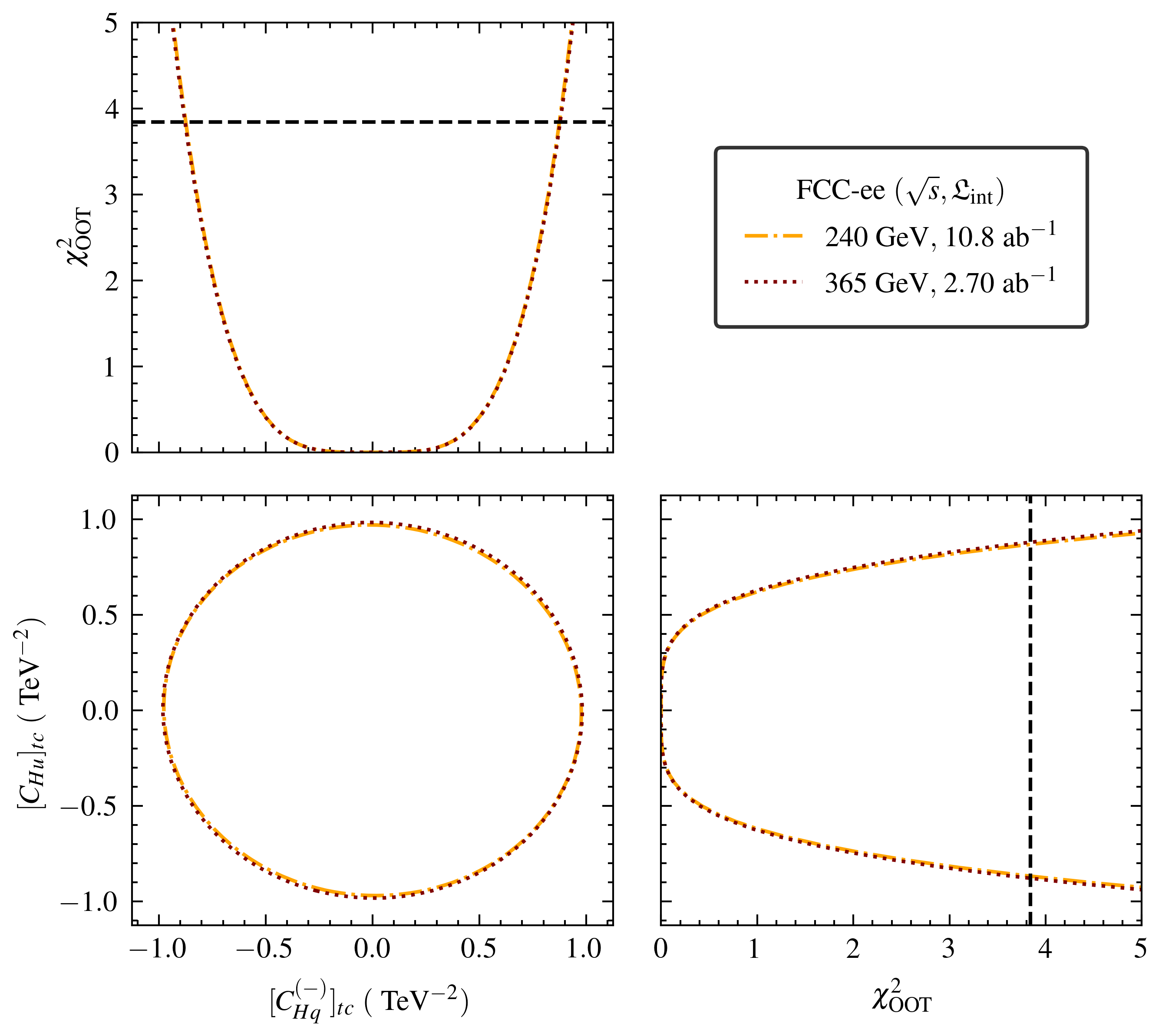}
    \caption{Projected 95\% C.L. plots/sensitivity contours for the WCs (in TeV$^{-2}$) obtained from the processes: $e^{+}e^{-} \to tc\,/\,tu$ at the FCC-ee.}
    \label{fig:contour3}
\end{figure}

In Fig.~\ref{fig:contour3}, we present the projected 95\% C.L. sensitivity contours for the WCs (in TeV$^{-2}$) obtained from the processes $e^{+}e^{-}\to tc\,/\,tu$ at the two FCC-ee energy stages. The behavior of the dipole operators differs from that observed for the $\tau\mu\,/\,\tau e$ and $bs\,/\,bd$ channels. In particular, at both Stages 3 and 4, the dipole operators exhibit constructive interference, in contrast to the other two channels, where the interference is destructive at higher center-of-mass energies. This behavior follows from the effective couplings entering the photo and $Z$-mediated amplitudes,
\begin{equation}
\begin{aligned}
C^{u}_\gamma &= \sin\theta_W\,C_{uW}+\cos\theta_W\,C_{uB}\,,\\
C^{u}_Z &= \cos\theta_W\,C_{uW}-\sin\theta_W\,C_{uB}\,.
\end{aligned}
\end{equation}
The difference in the relative signs originates from the presence of $\widetilde{H}=i\sigma^{2}H^{*}$, instead of $H$, in the definition of the up-type dipole operators. Consequently, the photon-mediated contribution exhibits constructive interference, while the $Z$-mediated contribution exhibits destructive interference. The Higgs-current operators do not interfere, resulting in nearly circular sensitivity contours. Furthermore, the projected sensitivities for the $tu$ flavor structure are slightly stronger than those for the $tc$ flavor structure, owing to the better retention of signal events.

\section{Conclusion}
\label{sec:conc}
In this study, we have presented a comprehensive study of flavor-violating interactions within the SMEFT framework at the FCC-ee. We considered a unified set of dimension-6 dipole and Higgs-current operators that induce FCNCs through $Z/\gamma$ mediation, covering both the charged lepton and quark sectors. In the lepton sector, we investigated the $\tau\mu$ and $\tau e$ flavor structures, while in the quark sector we considered the $bs/bd$ and $tc/tu$ transitions. For each flavor structure, we performed a detailed collider analysis at the relevant FCC-ee energy stages, employing kinematic selections to separate the signal from the corresponding SM backgrounds and deriving projected 95\% C.L. sensitivities using the OOT to the relevant WCs. We further studied the interference patterns between the different operator contributions and identified the characteristic energy dependence of the dipole and Higgs-current operators. To place the projected FCC-ee sensitivities in the broader context of existing constraints, we also derived bounds from direct and indirect flavor-transition processes, including bottom-quark FCNCs, rare top decays, charged LFV decays, and low-energy observables sensitive to the same flavor-violating interactions. This allows us to provide a systematic comparison between direct collider probes and existing low-energy constraints across the different flavor sectors.

\begin{figure}[h!]
    \centering
    \includegraphics[width=1.0\linewidth]{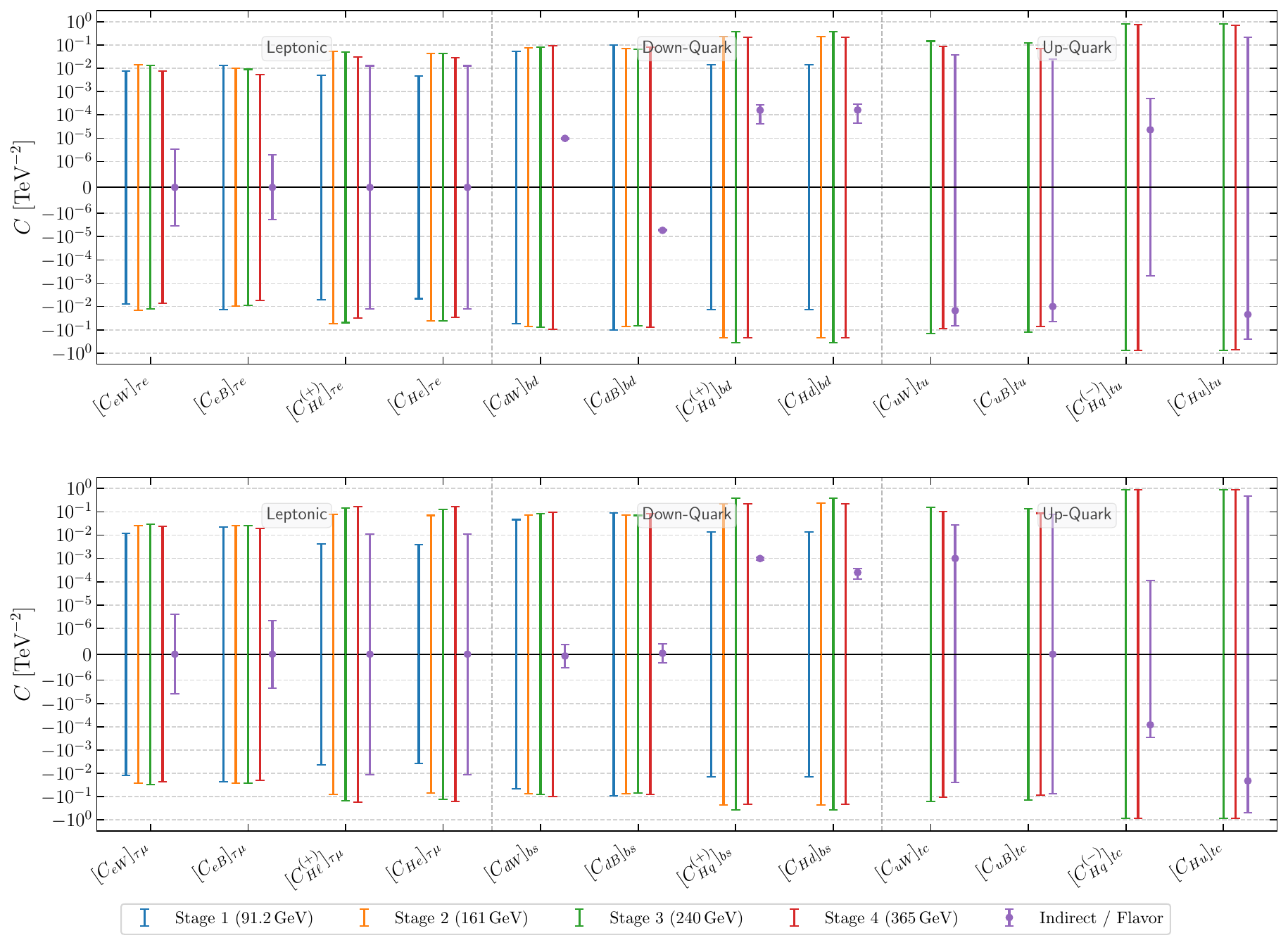}
    \caption{Projected $95\%$ C.L. sensitivities on flavor-violating SMEFT WCs in the leptonic, down-quark, and up-quark sectors. The top and bottom panels display bounds on $3 \to 1$ and $3 \to 2$ generation transitions, respectively, comparing the four operating stages of the FCC-ee with the most stringent flavor (direct/indirect) constraints.}
    \label{fig:summary_plot}
\end{figure}

A summary of the projected FCC-ee sensitivities and the corresponding most stringent flavor (direct/indirect) constraints is presented in Fig.~\ref{fig:summary_plot}. The figure provides a unified comparison across the leptonic, down-quark, and up-quark sectors, covering both $3 \to 1$ and $3 \to 2$ generation flavor transitions. We begin the summary with the leptonic operators. For the Higgs-current operators, the FCC-ee, particularly the $Z$ pole run, provides marginally better sensitivity compared to the existing constraints obtained from $\tau$ decay processes. This improvement is observed for both the $\tau\mu$ and $\tau e$ flavor transitions and highlights the advantage of operating at the $Z$ resonance, where the large event yield allows the FCC-ee to probe small deviations in the flavor-violating $Z$ couplings with high precision. The situation is markedly different for the dipole operators. The non-observation of radiative $\tau \to l\gamma$ decays imposes very stringent constraints of $\sim \mathcal{O}(10^{-5})\,\text{TeV}^{-2}$, which are substantially stronger than the corresponding FCC-ee sensitivities. Even with the enhanced energy dependence of the dipole interactions at higher CM energies, the collider reach remains well above these low-energy limits. Thus, while the FCC-ee offers a competitive and, in some cases, improved probe of the Higgs-current interactions, the radiative $\tau$ decays provide the dominant constraints on the leptonic dipole operators.

In the context of down-type quark operators, the existing flavor constraints are several orders of magnitude stronger than the projected FCC-ee sensitivities. This is largely a consequence of the exceptional precision achieved at dedicated $B$-physics experiments, where the primary objective is to study flavor-changing processes involving $B$ hadrons with minimal contamination from the complexities of hadronic final states. In contrast, at high-energy $e^+e^-$ colliders, the production of quarks is followed by hadronization and jet formation, which inevitably leads to a loss of information on the underlying partonic flavor transition. Nevertheless, the FCC-ee provides a complementary probe of these interactions at the electroweak scale, allowing the same flavor-violating couplings to be tested through a fundamentally different environment.

In the context of up-type operators inducing top FCNC interactions, the comparison between the FCC-ee projections and existing indirect constraints reveals a more competitive landscape. For both the $tc$ and $tu$ flavor structures, the dipole operators $\mathcal{O}_{uW}$ and $\mathcal{O}_{uB}$, together with the Higgs-current operator $\mathcal{O}_{Hu}$, exhibit sensitivities at the FCC-ee that are broadly comparable to those obtained from indirect low-energy measurements. This indicates that the electroweak-scale environment of the FCC-ee can provide a complementary probe of these interactions, particularly for operators whose effects are not overwhelmingly constrained by low-energy flavor observables. In contrast, the $\mathcal{O}^{(+)}_{Hq}$ operator is subject to substantially stronger indirect constraints, with the corresponding limits being approximately $2-3$ orders of magnitude stronger than the projected FCC-ee sensitivities. Nevertheless, the FCC-ee projections show a clear improvement over the existing direct constraints derived from top FCNC decays at the LHC.

Overall, this work provides a comprehensive assessment of the prospects for probing heavy-flavor-violating interactions at the FCC-ee. We have systematically studied flavor-violating transitions across the charged-lepton, down-type quark, and up-type quark sectors, covering all four proposed FCC-ee operating stages and thereby accounting for the complementary sensitivity offered by different CM energies and integrated luminosities. A key aspect of this study is the simultaneous comparison of the projected collider sensitivities with both direct and indirect constraints, allowing the relative strengths and complementarity of the FCC-ee and existing flavor experiments to be assessed on a common footing. Furthermore, the analysis provides insight into the interference structure of different SMEFT operators, including the energy-dependent interplay between photon and $Z$-mediated contributions, which can lead to characteristic sensitivity patterns that are directly accessible at the FCC-ee. Beyond presenting projected bounds, our results therefore provide a useful guideline for identifying the flavor structures and operators for which FCC-ee measurements can be particularly informative, as well as for determining which channels merit further detailed investigation. In this sense, the study serves as a broad roadmap for exploring flavor violation at the FCC-ee and for identifying the directions in which future dedicated analyses can yield the better physics reach.

\section*{Acknowledgements}
Amir Subba and Yu Shi acknowledge the support of National Natural Science Foundation of China under Grant Nos. T2241005 and 12075059. Subhajit Kala and Abhik Sarkar gratefully acknowledge Prof. Soumitra Nandi for useful discussions and valuable insights regarding the derivation of indirect bounds from flavor sectors.

\appendix
\section{Kinematic Distributions}
\begin{figure}[htb!]
    \centering
    \includegraphics[width=0.4\linewidth]{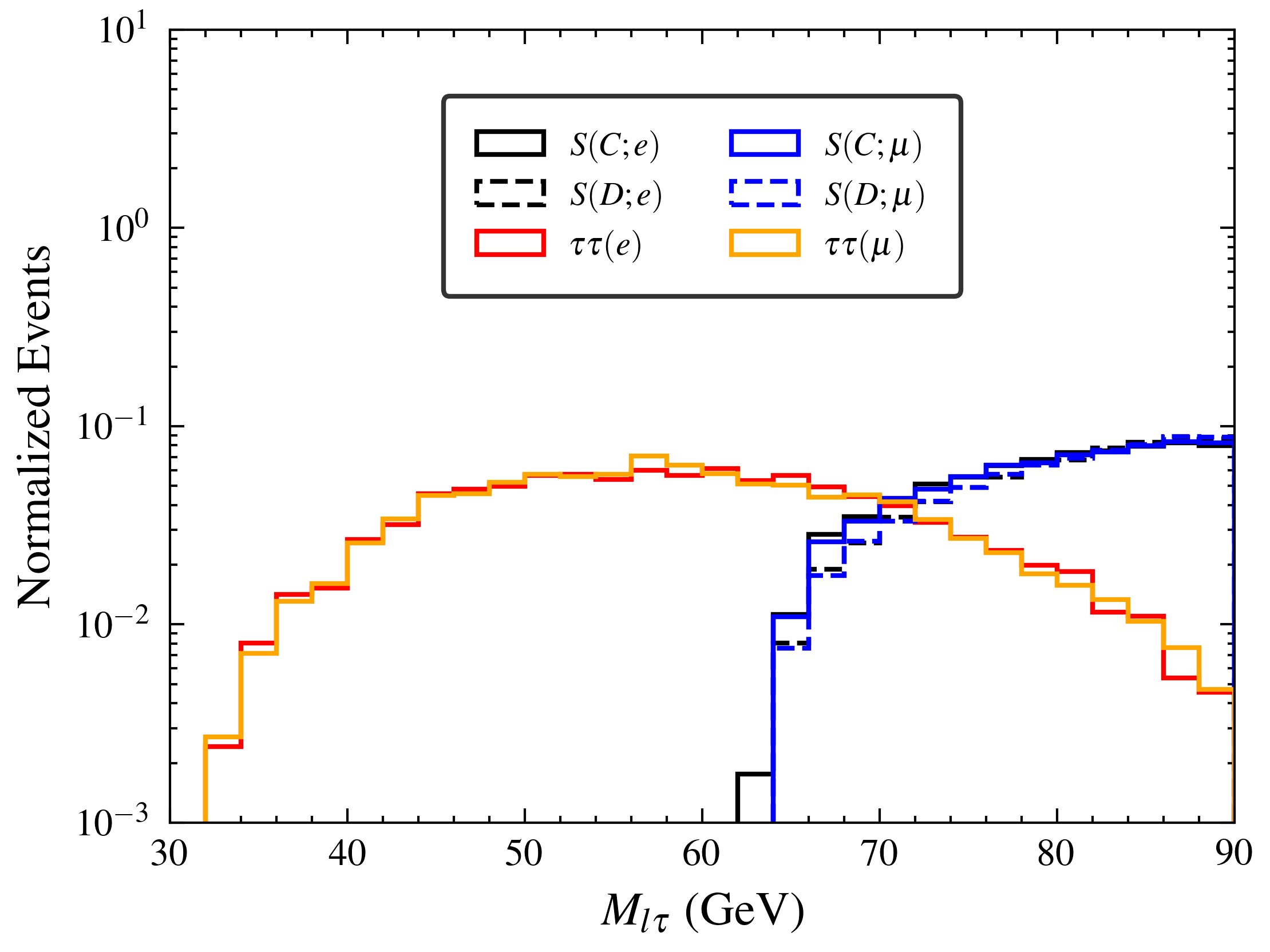} \qquad
    \includegraphics[width=0.4\linewidth]{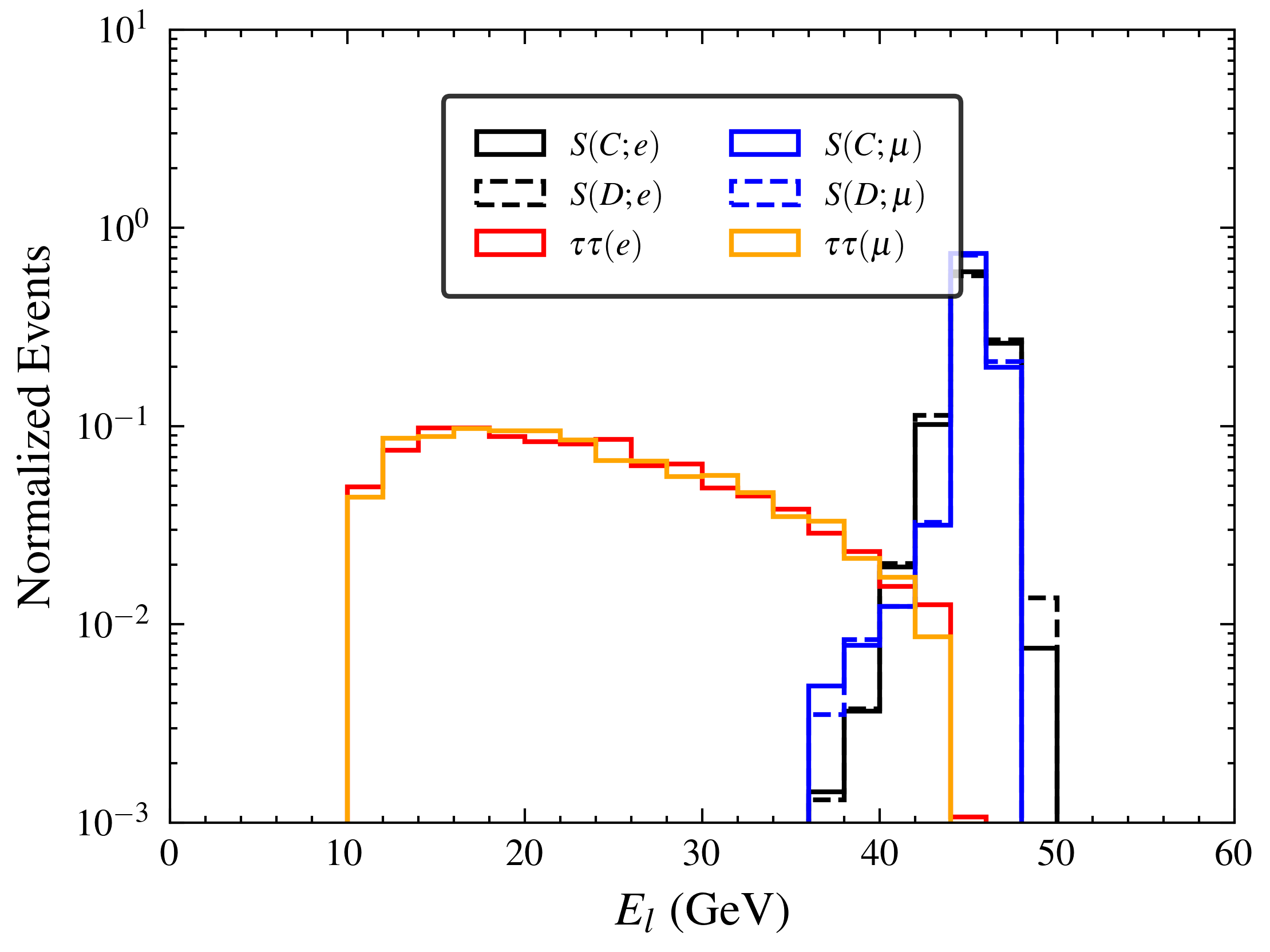} \\
    \includegraphics[width=0.4\linewidth]{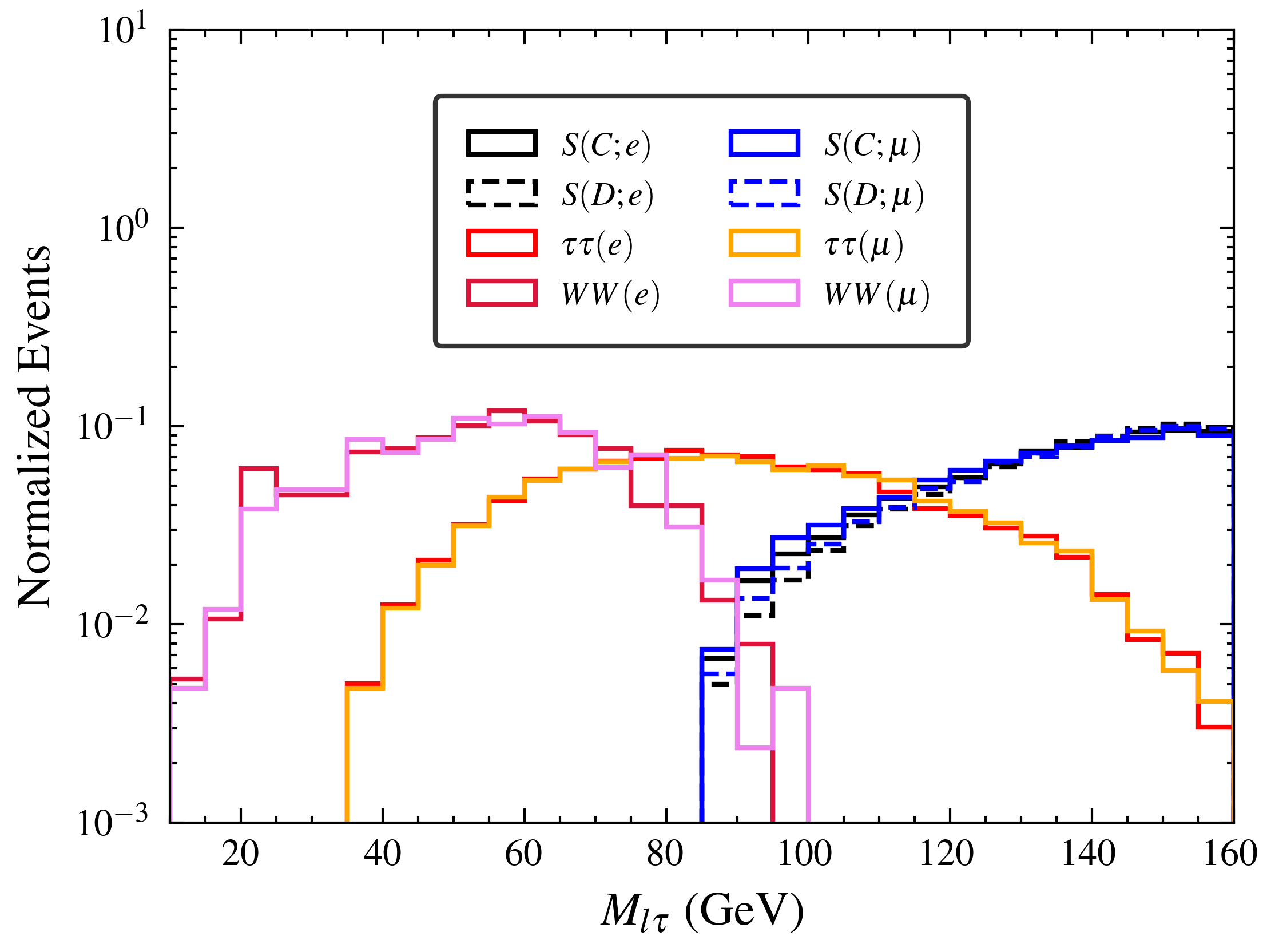} \qquad
    \includegraphics[width=0.4\linewidth]{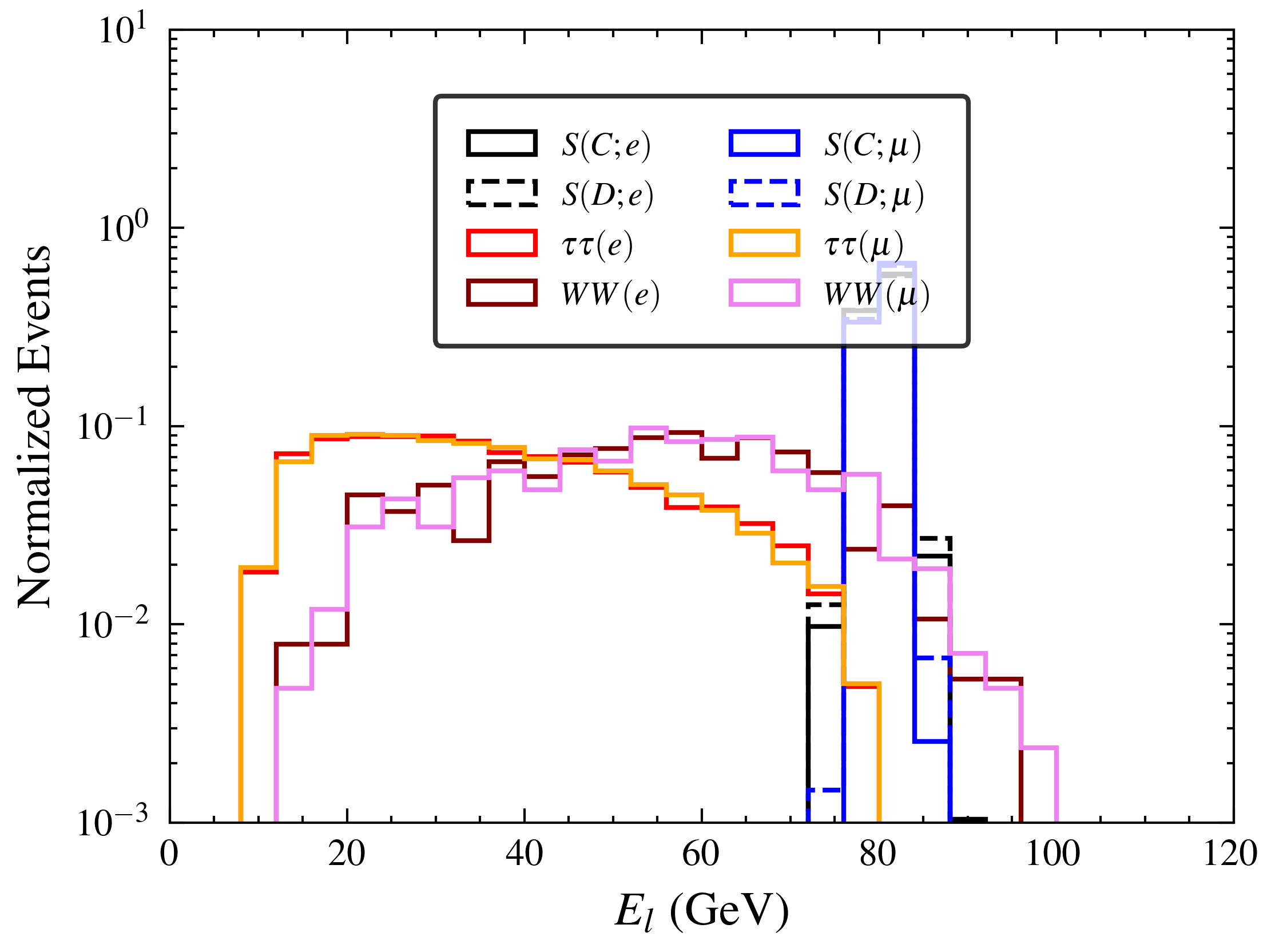} \\
    \includegraphics[width=0.4\linewidth]{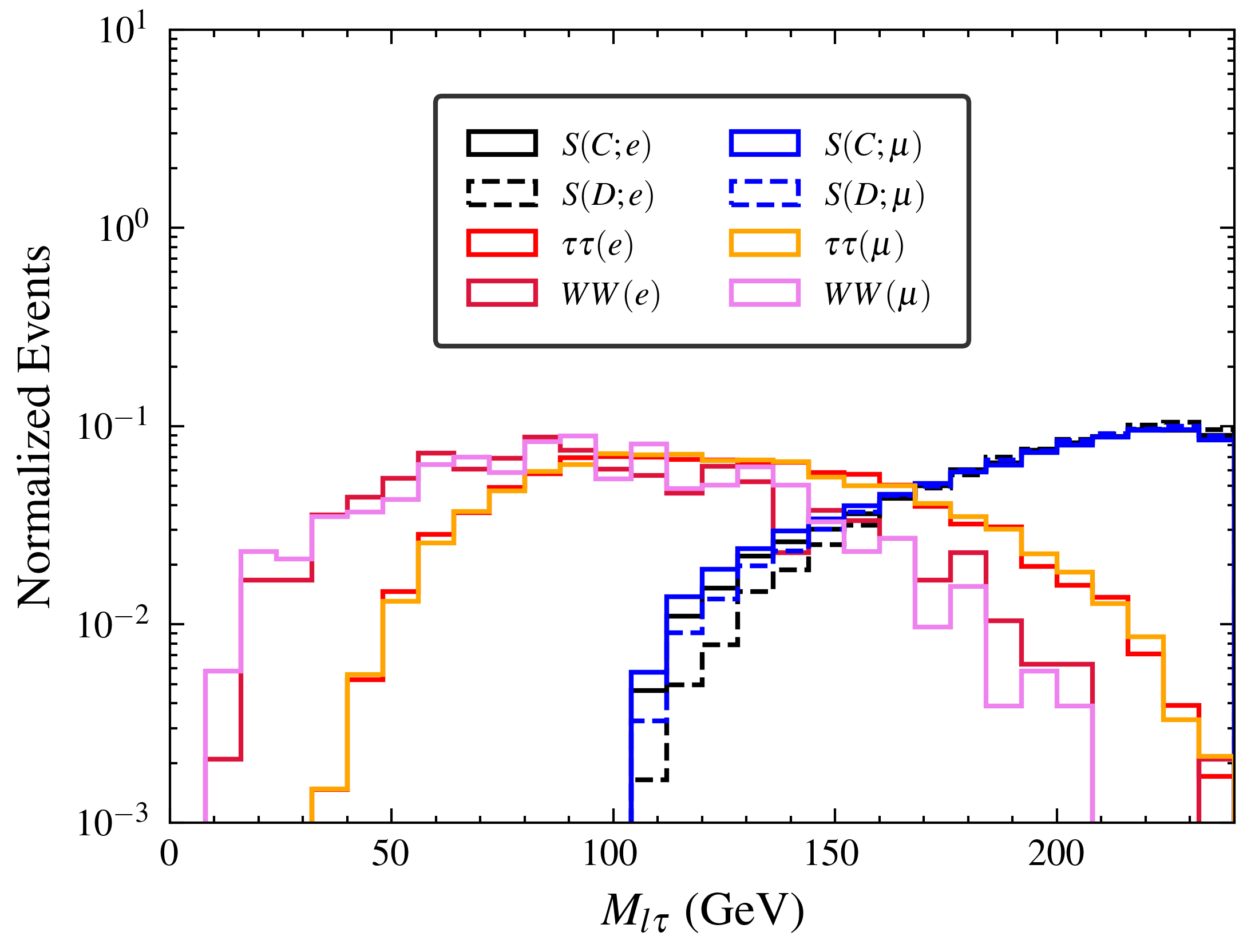} \qquad
    \includegraphics[width=0.4\linewidth]{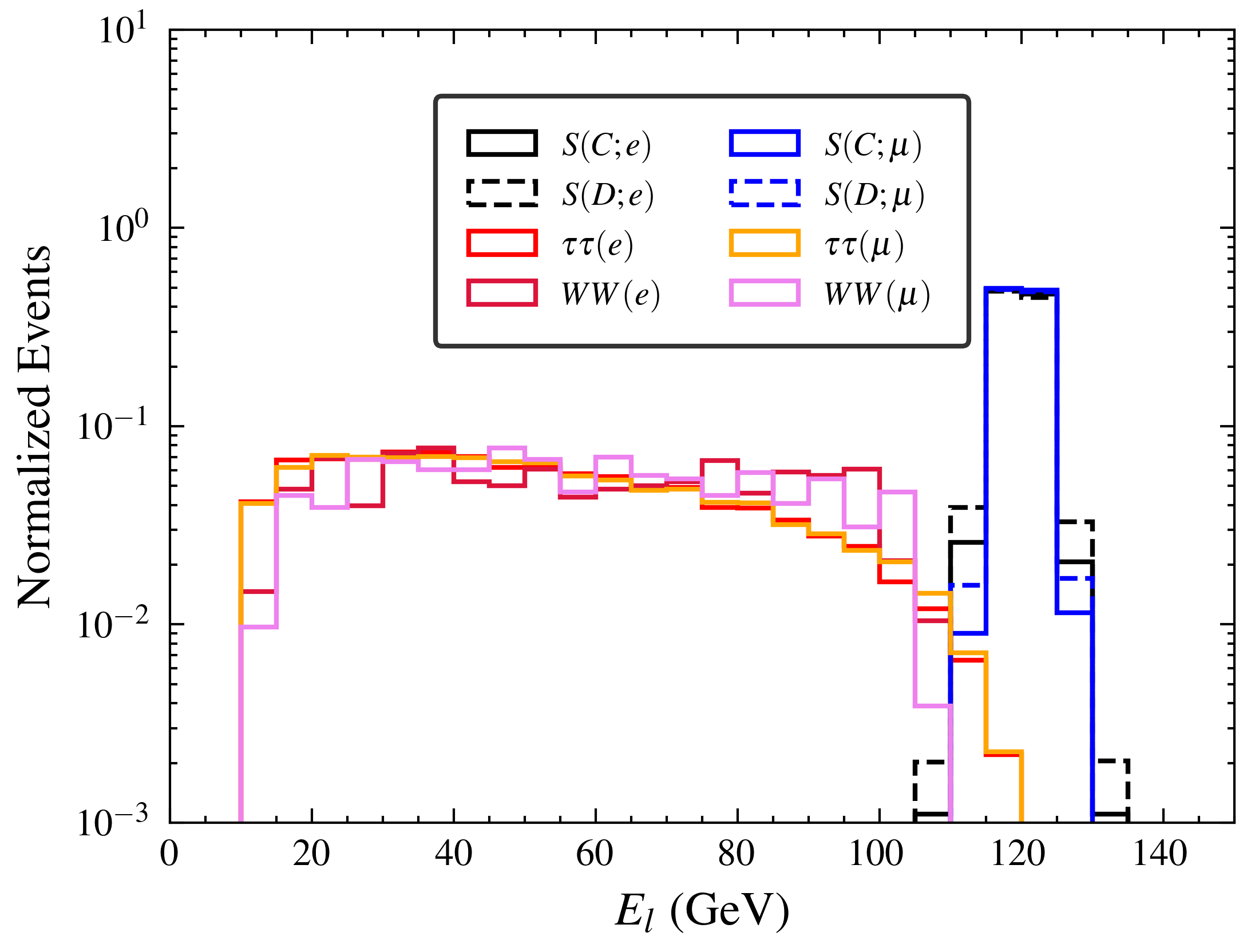} \\
    \includegraphics[width=0.4\linewidth]{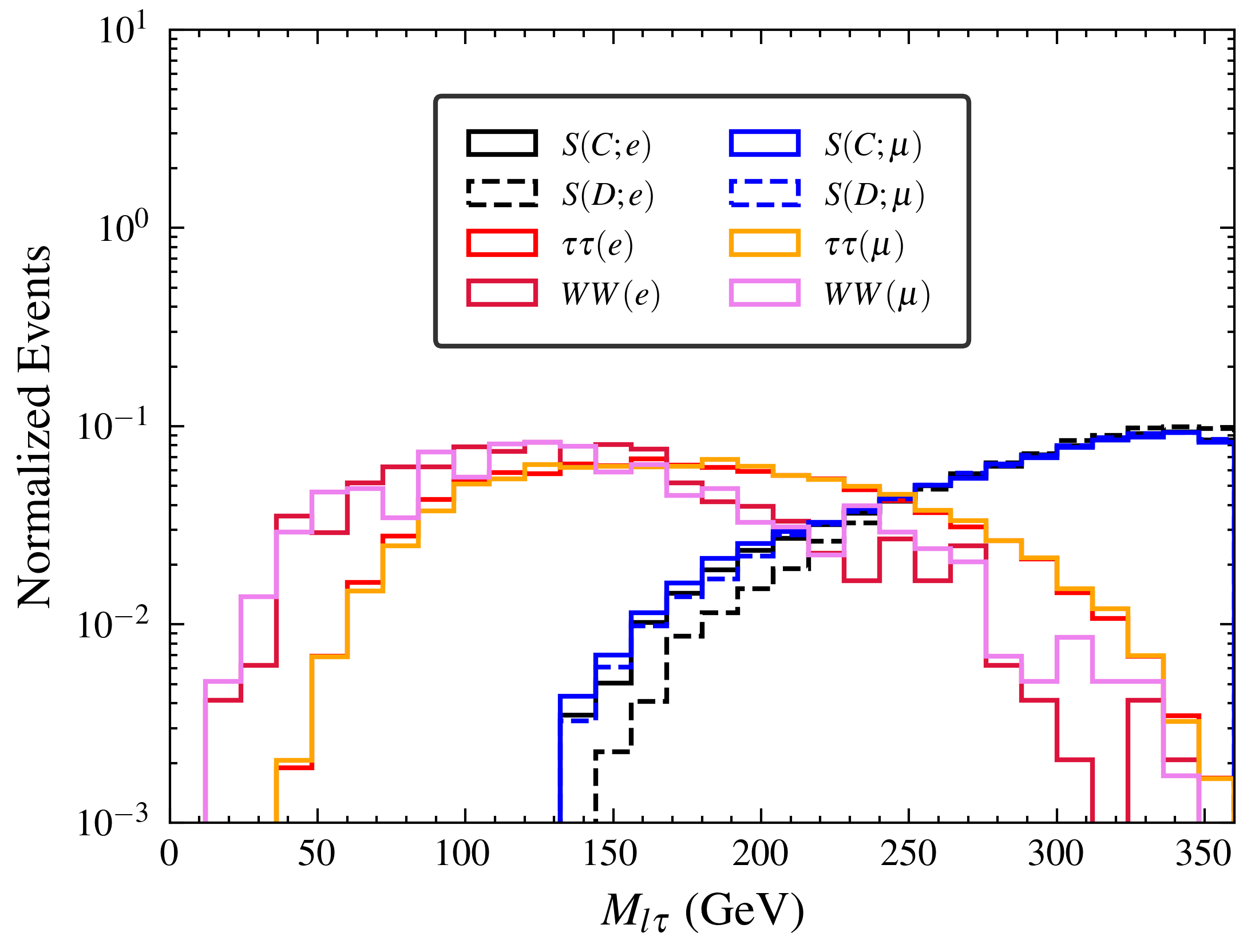} \qquad
    \includegraphics[width=0.4\linewidth]{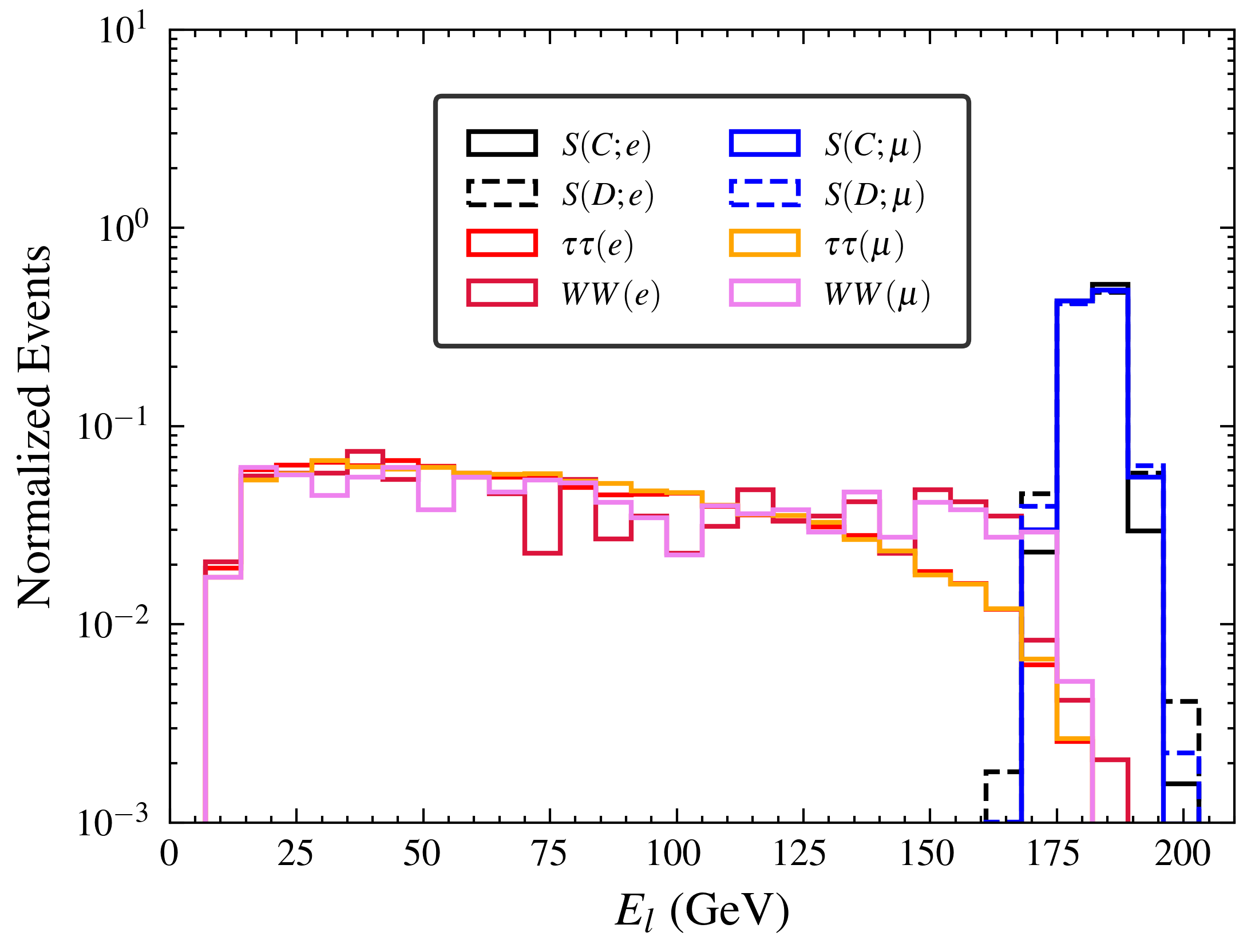}
    \caption{Kinematic distributions corresponding to signal and background processes for $\tau\mu/\tau e$ signal at different stages: Stage 1 (\textit{top-most}), Stage 2 (\textit{upper-middle}), Stage 3 (\textit{lower-middle}), and Stage 4 (\textit{bottom-most}).}
    \label{fig:dists1}
\end{figure}

\begin{figure}[htb!]
    \centering
    \includegraphics[width=0.4\linewidth]{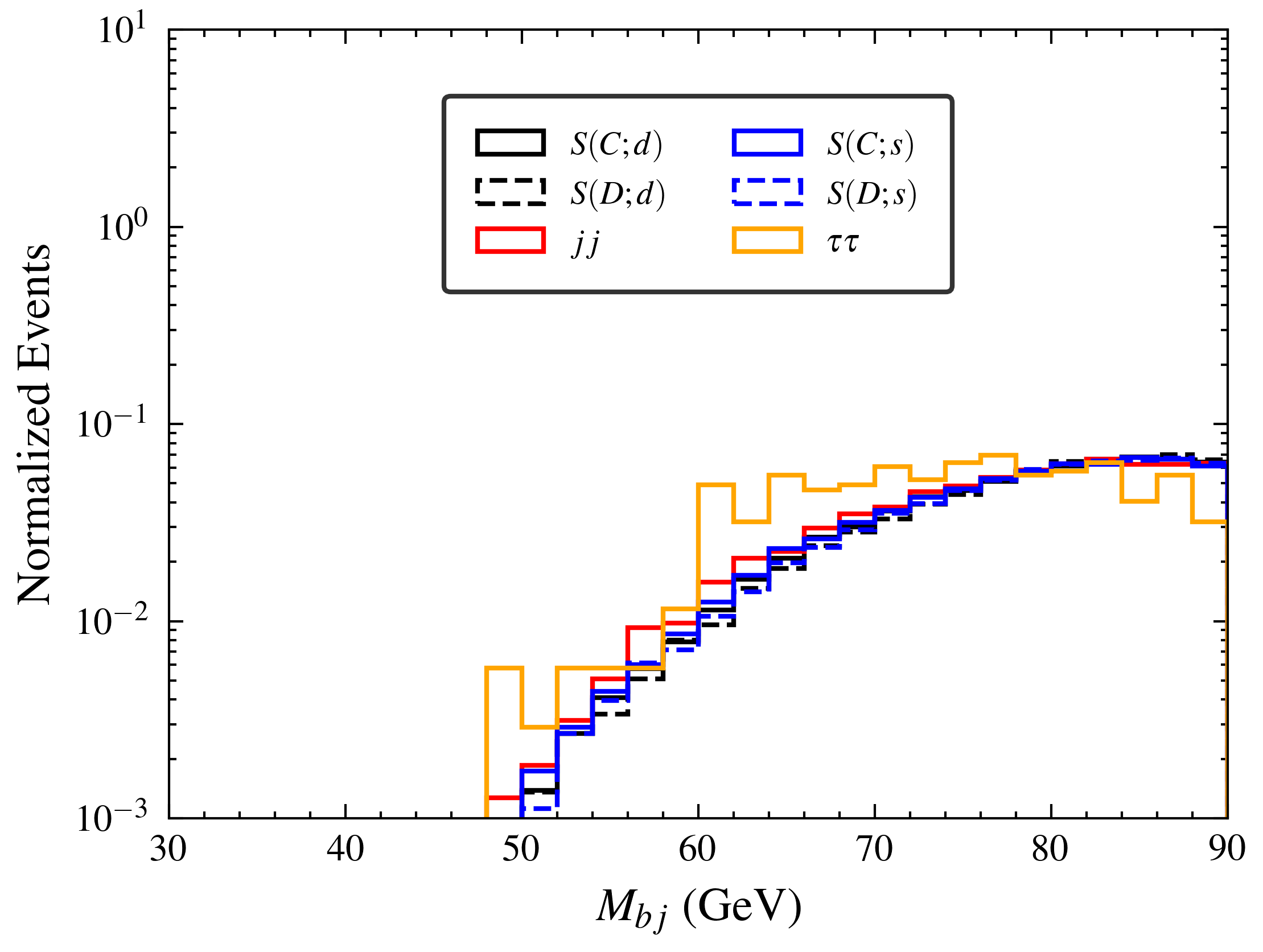} \qquad
    \includegraphics[width=0.4\linewidth]{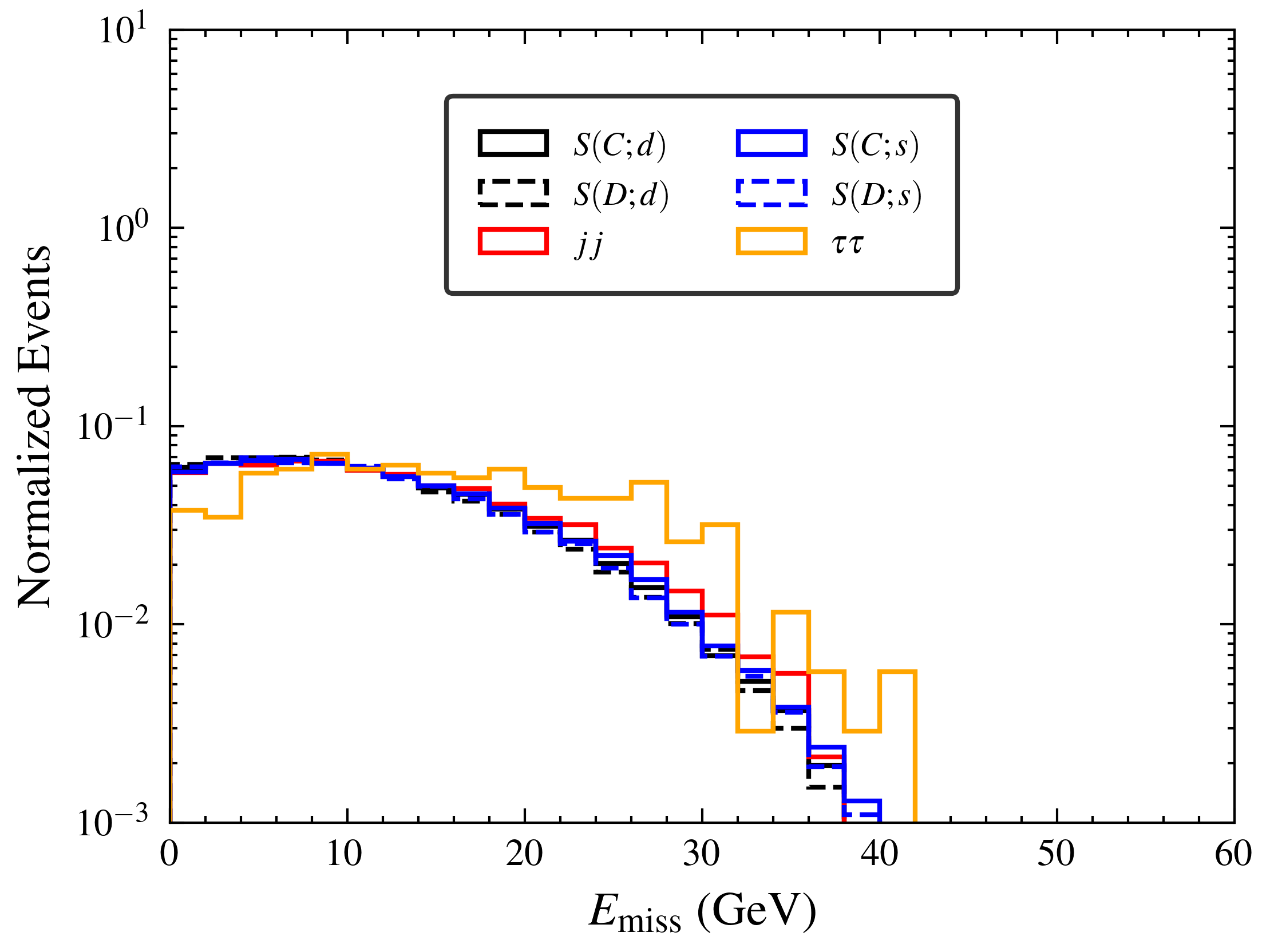} \\
    \includegraphics[width=0.4\linewidth]{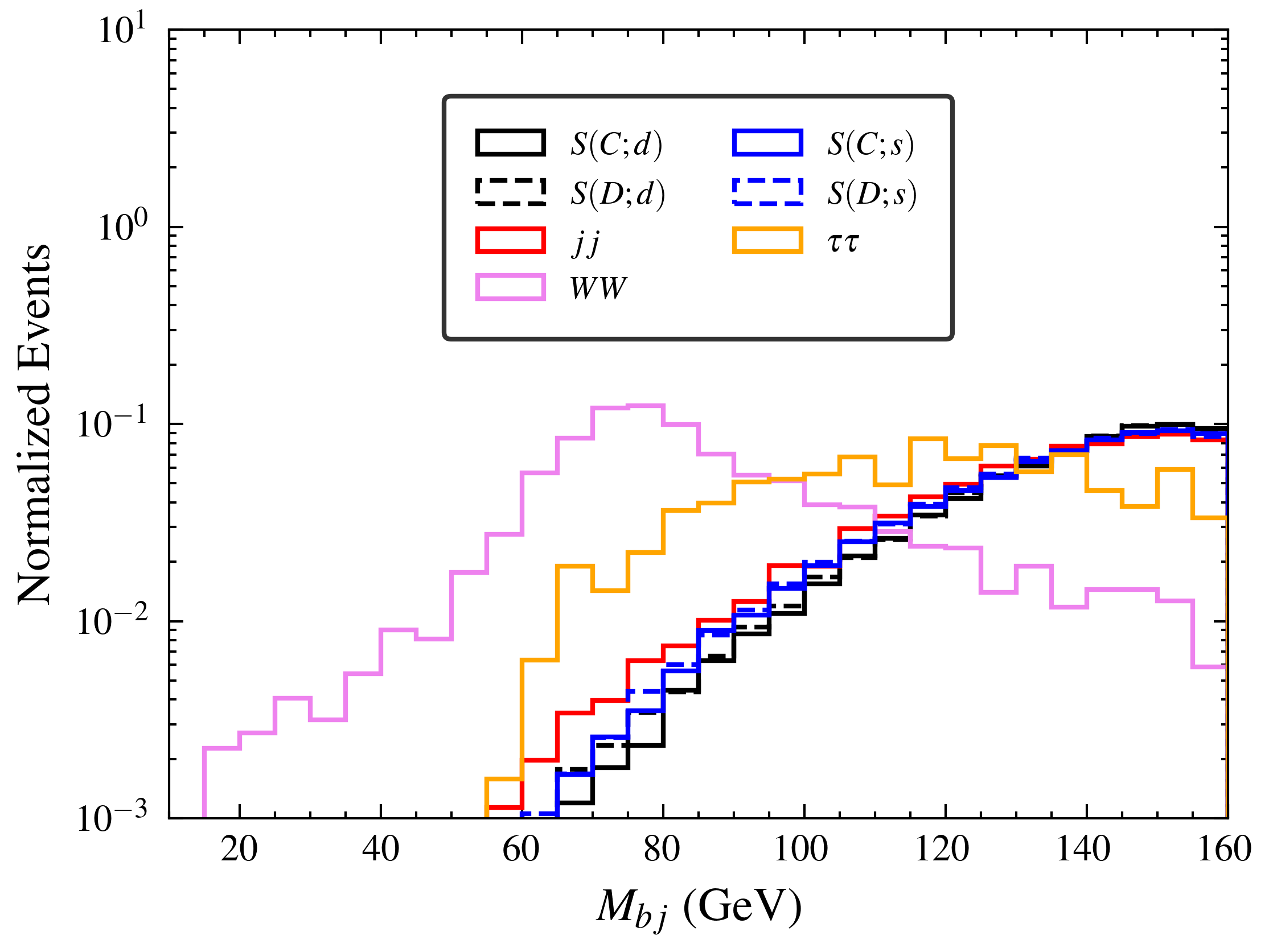} \qquad
    \includegraphics[width=0.4\linewidth]{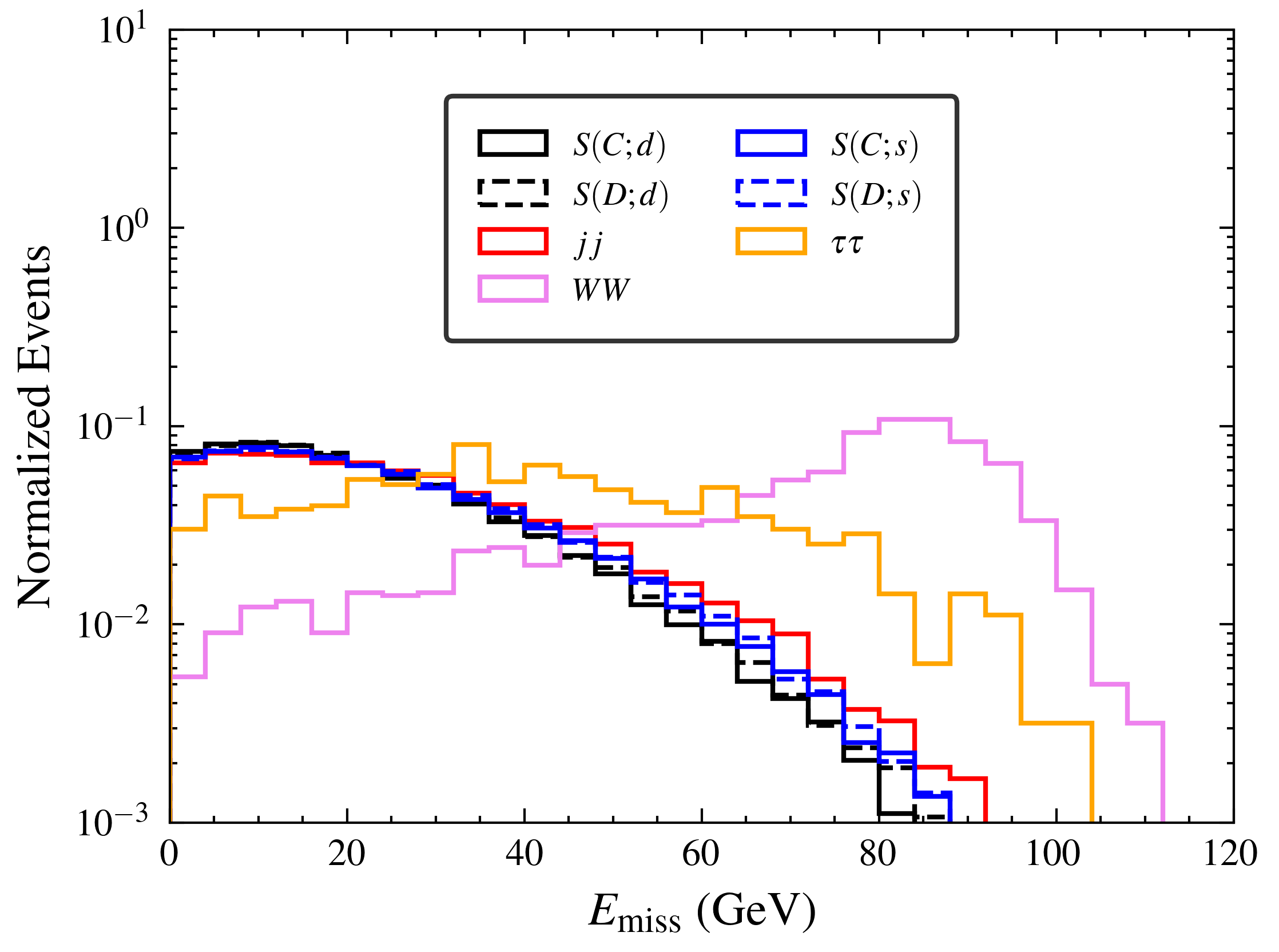} \\
    \includegraphics[width=0.4\linewidth]{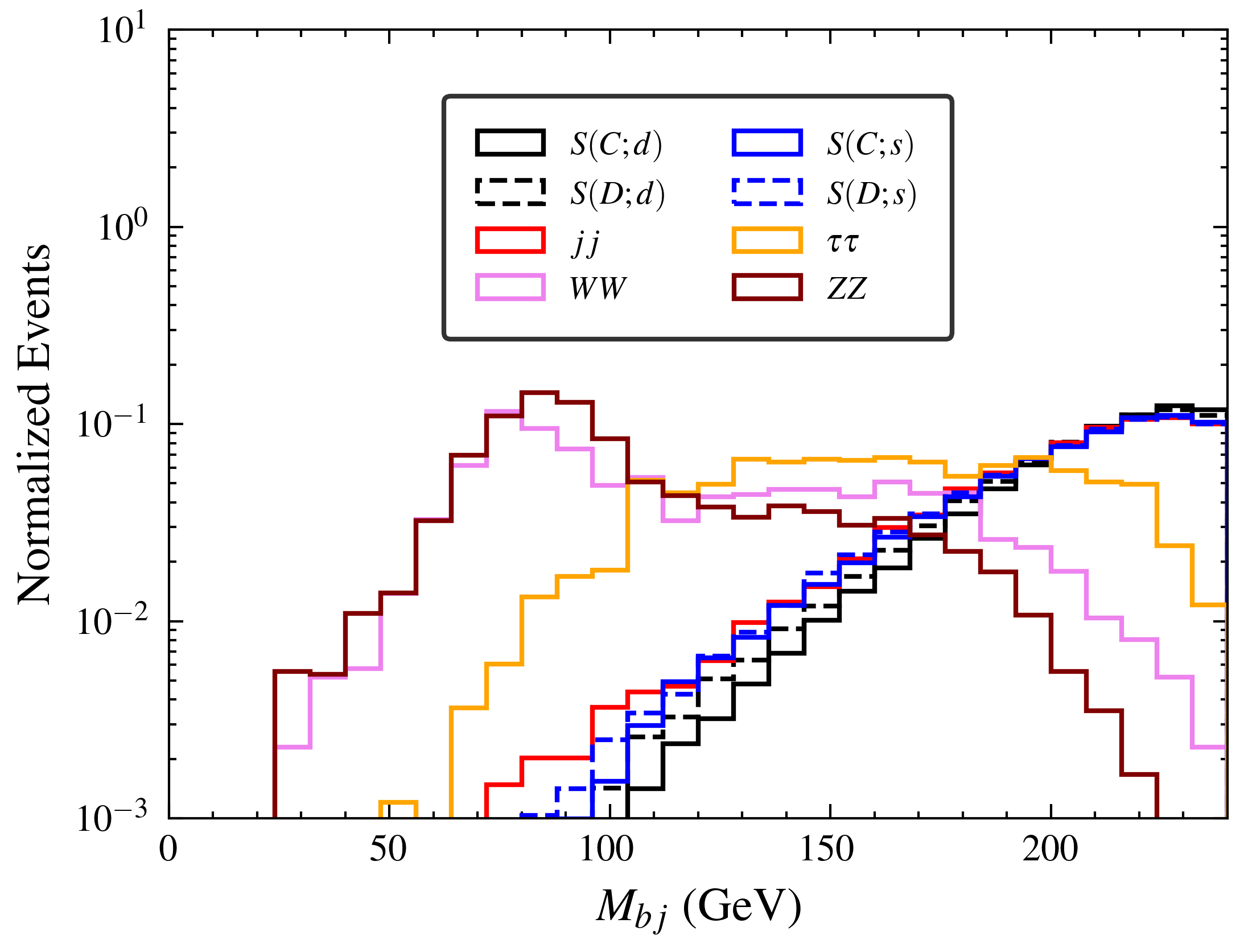} \qquad
    \includegraphics[width=0.4\linewidth]{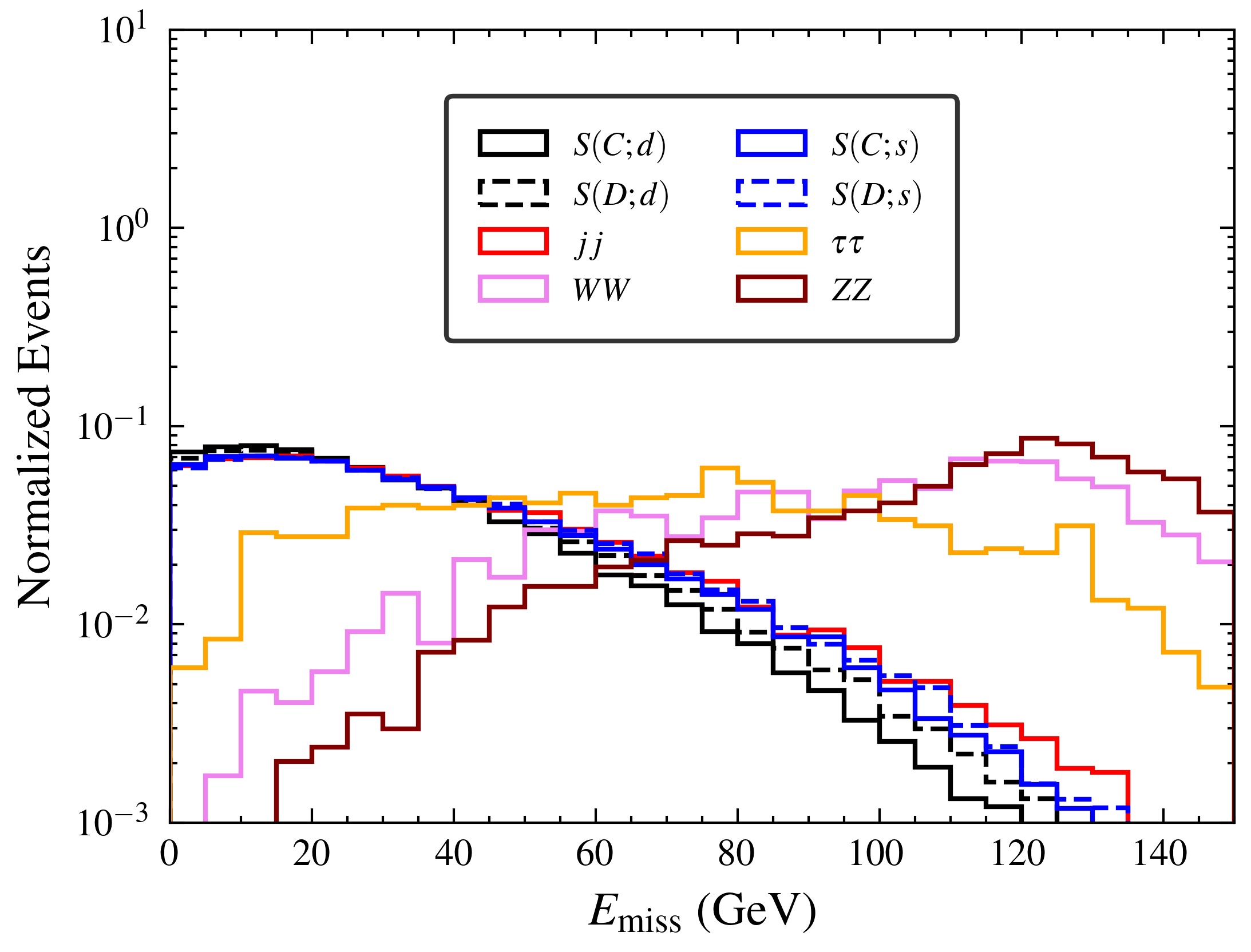} \\
    \includegraphics[width=0.4\linewidth]{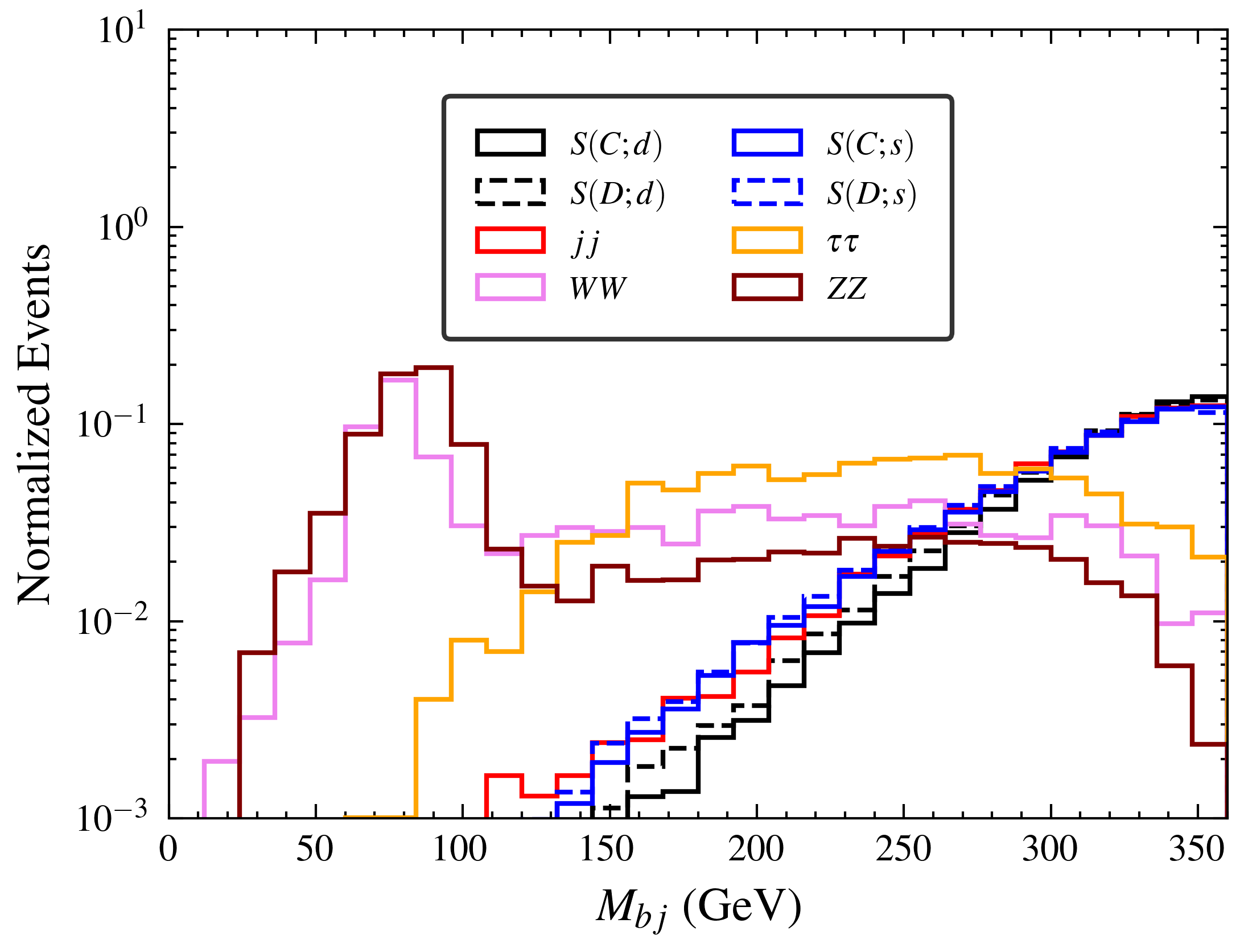} \qquad
    \includegraphics[width=0.4\linewidth]{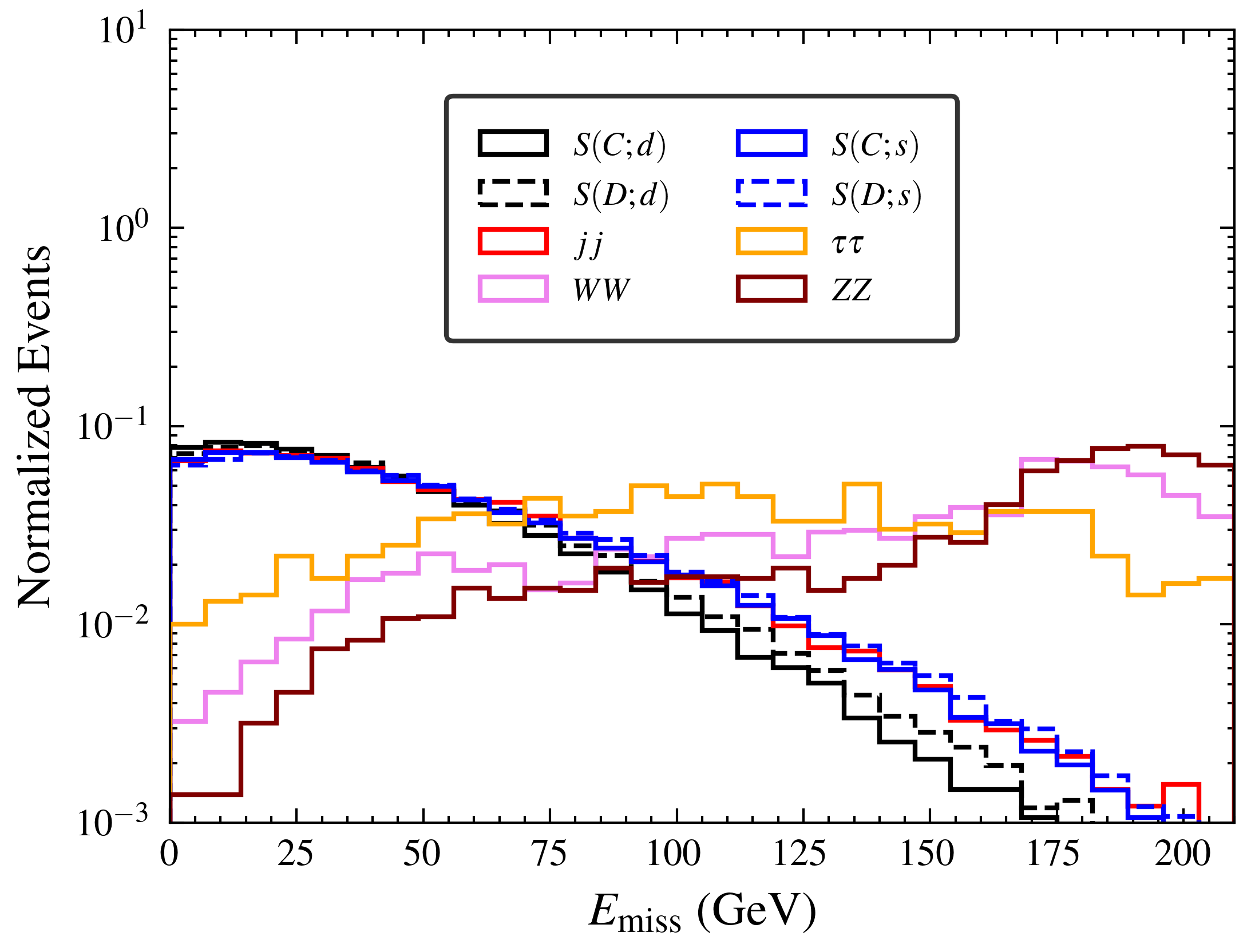}
    \caption{Kinematic distributions corresponding to signal and background processes for $bs/bd$ signal for different stages: Stage 1 (\textit{top-most}), Stage 2 (\textit{upper-center}), Stage 3 (\textit{lower-center}), and Stage 4 (\textit{bottom-most}).}
    \label{fig:dists2}
\end{figure}

\begin{figure}[htb!]
    \centering
    \includegraphics[width=0.4\linewidth]{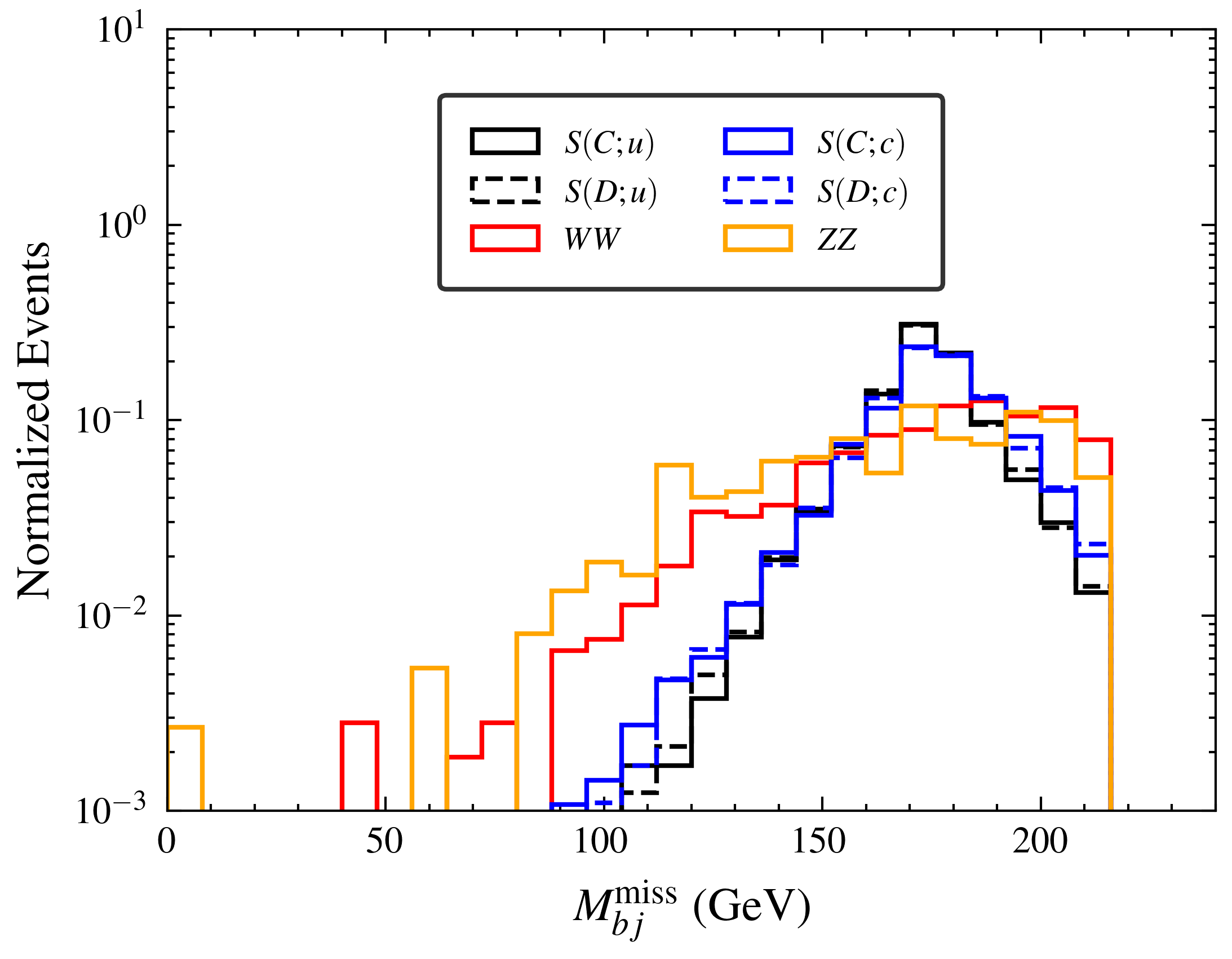} \qquad
    \includegraphics[width=0.4\linewidth]{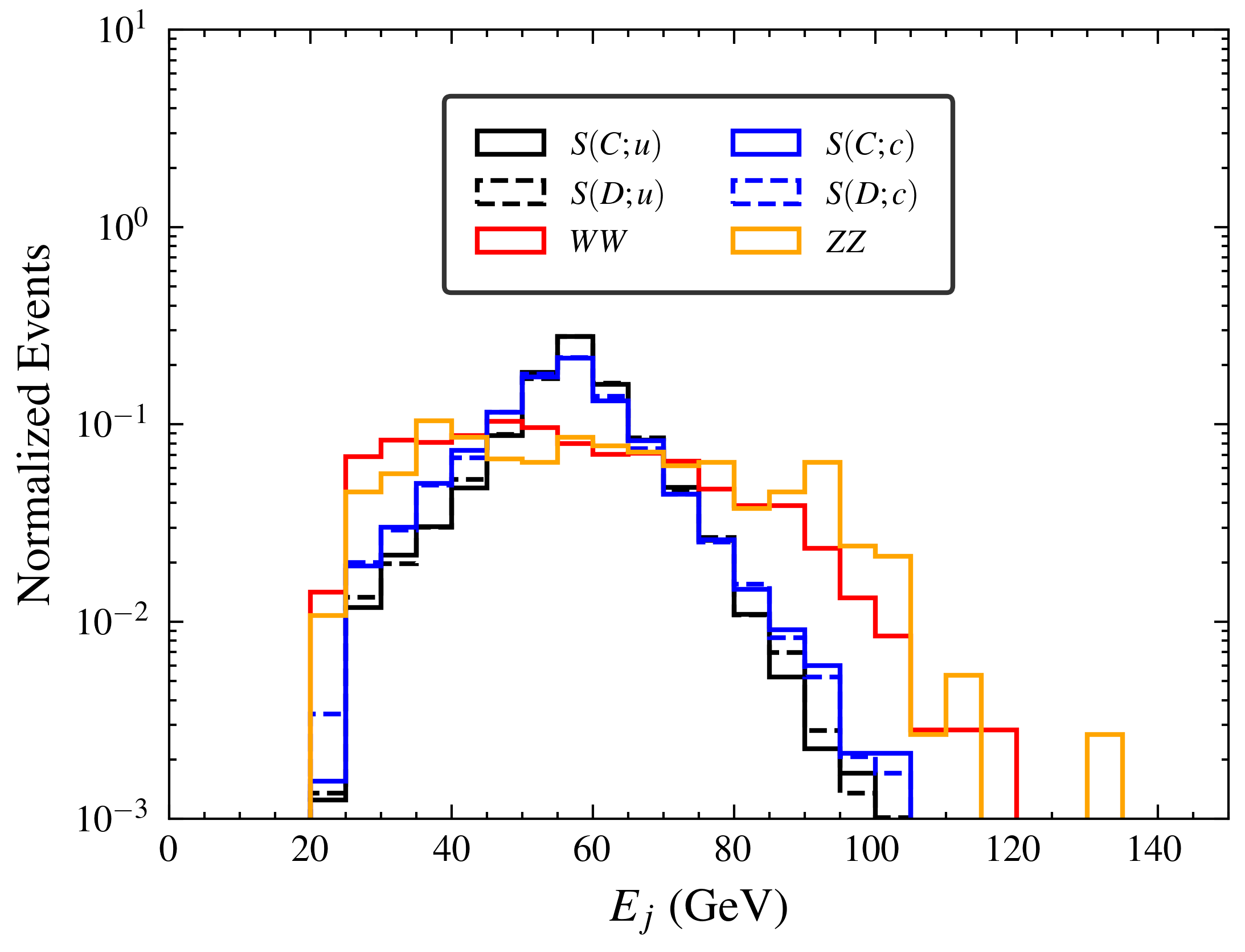} \\
    \includegraphics[width=0.4\linewidth]{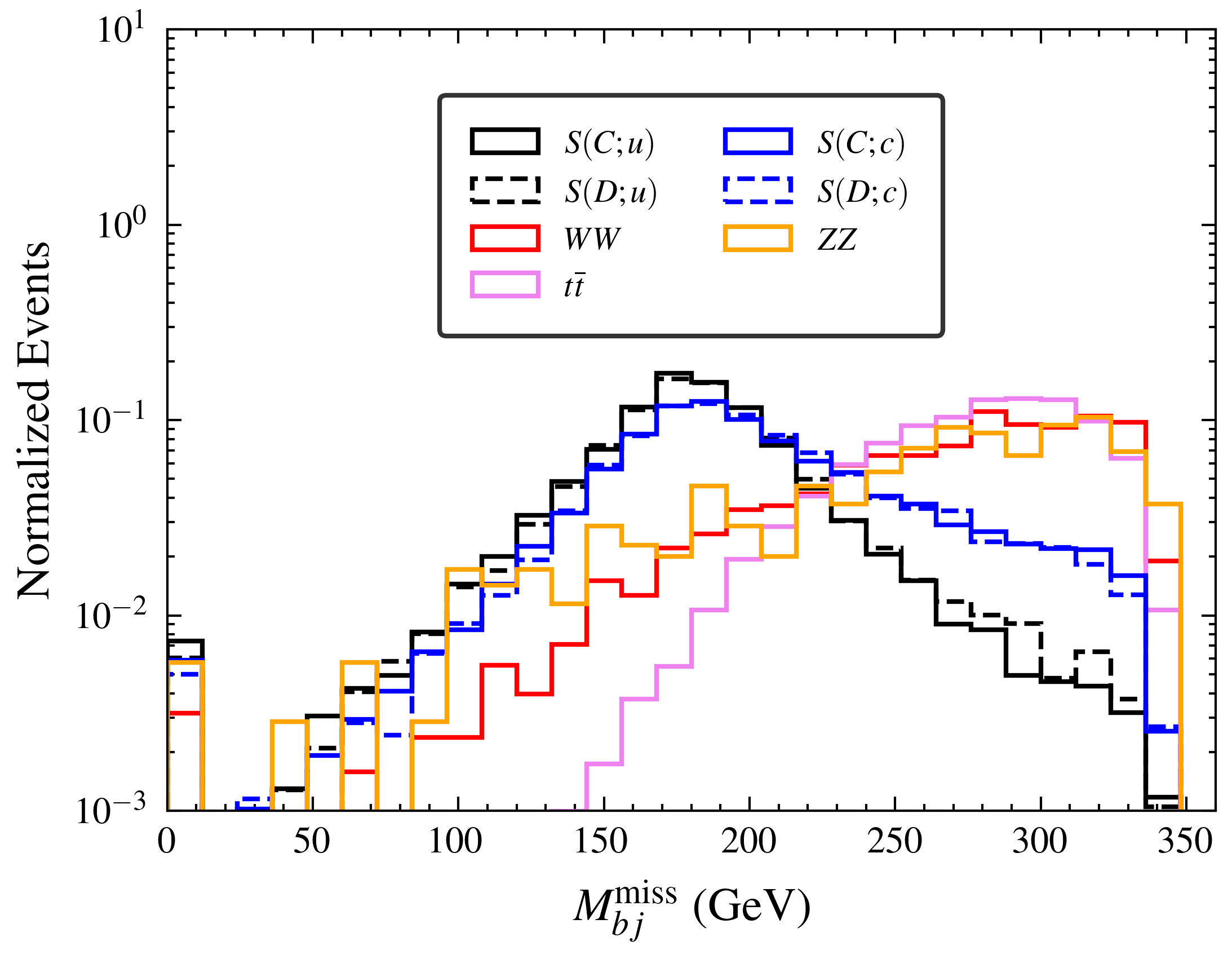} \qquad
    \includegraphics[width=0.4\linewidth]{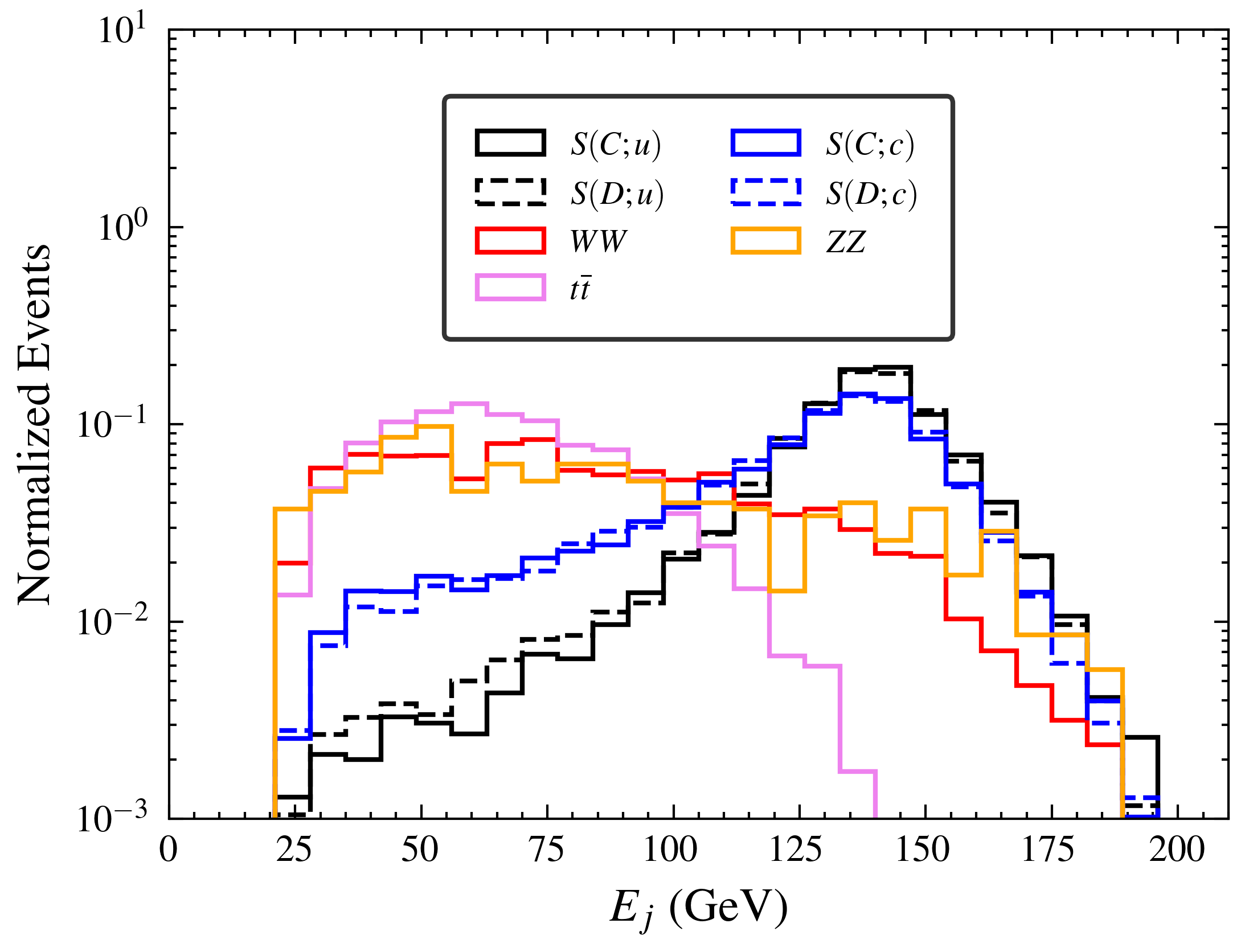}
    \caption{Kinematic distributions corresponding to signal and background processes for $tc/tu$ signal for different stages: Stage 3 (\textit{upper}), and Stage 4 (\textit{lower}).}
    \label{fig:dists3}
\end{figure}

In this section, we provide the kinematic distributions for the signal and background processes. The signal distributions correspond to the following signal processes and parameter benchmarks for $\tau\mu\,/\,\tau e$ production:
\begin{equation}
\begin{split}
    S(C;e)&: \quad \tau e\quad,\quad\left([C^{(+)}_{H\ell}]_{\tau e}\,,[C_{He}]_{\tau e}\right) = (1.0, 1.0)\text{ TeV}^{-2}\,,\\
    S(D;e)&: \quad \tau e\quad,\quad \left([C_{eW}]_{\tau e}\,,[C_{eB}]_{\tau e}\right) = (1.0, 1.0)\text{ TeV}^{-2}\,,\\
    S(C;\mu)&: \quad \tau \mu\quad,\quad \left([C^{(+)}_{H\ell}]_{\tau \mu}\,,[C_{He}]_{\tau \mu}\right) = (1.0, 1.0)\text{ TeV}^{-2}\,,\\
    S(C;\mu)&: \quad \tau \mu\quad,\quad \left([C_{eW}]_{\tau \mu}\,,[C_{eB}]_{\tau \mu}\right) = (1.0, 1.0)\text{ TeV}^{-2}\,,\\
\end{split}
\end{equation}
The kinematic distributions corresponding to signal and background processes for $\tau\mu/\tau e$ signal for different stages are shown in Fig.~\ref{fig:dists1}. Here, $\tau\tau(e/\mu)$, and $WW(e/\mu)$ correspond to backgrounds with signal selection criteria $N_{e/\mu} = 1$, respectively.
Similarly, the signal benchmarks for the $bs/bd$ signal are as follows:
\begin{equation}
\begin{split}
    S(C;d)&: \quad bd\quad,\quad\left([C^{(+)}_{Hq}]_{bd}\,,[C_{Hd}]_{bd}\right) = (1.0, 1.0)\text{ TeV}^{-2}\,,\\
    S(D;d)&: \quad bd\quad,\quad \left([C_{dW}]_{bd}\,,[C_{dB}]_{bd}\right) = (1.0, 1.0)\text{ TeV}^{-2}\,,\\
    S(C;s)&: \quad bs \quad,\quad \left([C^{(+)}_{Hq}]_{bs}\,,[C_{Hd}]_{bs}\right) = (1.0, 1.0)\text{ TeV}^{-2}\,,\\
    S(C;s)&: \quad bs \quad,\quad \left([C_{dW}]_{bs}\,,[C_{dB}]_{bs}\right) = (1.0, 1.0)\text{ TeV}^{-2}\,,\\
\end{split}
\end{equation}
The kinematic distributions corresponding to signal and background processes for $bs/bd$ signal for different stages are shown in Fig.~\ref{fig:dists2}.
Likewise, the signal benchmarks for the $tc/tu$ signal are as follows:
\begin{equation}
\begin{split}
    S(C;u)&: \quad tu\quad,\quad\left([C^{(-)}_{Hq}]_{tu}\,,[C_{Hu}]_{tu}\right) = (1.0, 1.0)\text{ TeV}^{-2}\,,\\
    S(D;u)&: \quad tu\quad,\quad \left([C_{uW}]_{tu}\,,[C_{uB}]_{tu}\right) = (1.0, 1.0)\text{ TeV}^{-2}\,,\\
    S(C;c)&: \quad tc\quad,\quad \left([C^{(-)}_{Hq}]_{tc}\,,[C_{Hu}]_{tc}\right) = (1.0, 1.0)\text{ TeV}^{-2}\,,\\
    S(C;c)&: \quad tc\quad,\quad \left([C_{uW}]_{tc}\,,[C_{uB}]_{tc}\right) = (1.0, 1.0)\text{ TeV}^{-2}\,,\\
\end{split}
\end{equation}
The kinematic distributions corresponding to signal and background processes for $tc/tu$ signal for different stages are shown in Fig.~\ref{fig:dists3}.

\section{Complex SMEFT Couplings} \label{Append:complex_fit}
In this section, we present the fit results for the non-Hermitian, dipole-type WCs, which are summarized in Tab.~\ref{tab:complexbounds}. Unlike our primary analysis, here we fully retain the complex nature of these couplings by simultaneously fitting both their real and imaginary components. The simultaneous extraction allows for correlations among the coefficients, leading to a broader allowed parameter space and, consequently, comparatively weaker constraints than those obtained in the single-parameter or purely real scenarios.
\begin{table}[htb!]
    \centering
    \renewcommand{\arraystretch}{1.25}
    \begin{tabular}{|
    >{\centering\arraybackslash}p{0.25\textwidth}|
    >{\centering\arraybackslash}p{0.55\textwidth}|
    }
    \hline \hline
    \textbf{WCs} & \textbf{Values $\boldsymbol{(\mu_{\rm EW})\,[\mathrm{TeV}^{-2}]}$} \\
    \hline \hline
    $\mathfrak{Re}[C_{dW}]_{bs},\mathfrak{Im}[C_{dW}]_{bs}$ & $(-0.11 \pm 0.80)\times 10^{-6},~(0.11 \pm 1.53)\times 10^{-6}$ \\
    $\mathfrak{Re}[C_{dW}]_{bd},\mathfrak{Im}[C_{dW}]_{bd}$ & $(-0.07 \pm 2.65)\times 10^{-5},~(0.01 \pm 1.66)\times 10^{-4}$ \\
    $\mathfrak{Re}[C_{dB}]_{bs},\mathfrak{Im}[C_{dB}]_{bs}$ & $(0.58\pm 4.08)\times 10^{-7},~(0.0\pm 4.93)\times 10^{-7}$ \\
    $\mathfrak{Re}[C_{dB}]_{bd},\mathfrak{Im}[C_{dB}]_{bd}$ & $(0.0 \pm 1.66)\times 10^{-5}, (0.15 \pm 2.64)\times 10^{-5}$ \\
    \hline \hline
    $\mathfrak{Re}[C_{uW}]_{tc},\mathfrak{Im}[C_{uW}]_{tc}$ & $(0.10 \pm 2.62)\times 10^{-2},~(0.03 \pm 4.50)\times 10^{-2}$ \\
    $\mathfrak{Re}[C_{uW}]_{tu},\mathfrak{Im}[C_{uW}]_{tu}$ & $(-1.52 \pm 5.26)\times 10^{-2},~(0.01 \pm 8.03)\times 10^{-2}$ \\
    $\mathfrak{Re}[C_{uB}]_{tc},\mathfrak{Im}[C_{uB}]_{tc}$ & $(0.0\pm 7.66)\times 10^{-2},~(0.04\pm 1.38)\times 10^{-2}$ \\
    $\mathfrak{Re}[C_{uB}]_{tu},\mathfrak{Im}[C_{uB}]_{tu}$ & $(-1.03 \pm 3.39)\times 10^{-2}, (-0.89 \pm 3.71)\times 10^{-2}$ \\
    \hline \hline
    \end{tabular}
    \caption{Best-fit values and allowed ranges for the complex top- and down-type dipole WCs, evaluated at the electroweak scale $\mu_{\rm EW}$.}
    \label{tab:complexbounds}
\end{table}

\paragraph{Electric Dipole Moment:}
To constrain the imaginary parts of the leptonic dipole-type operators, EDMs provide an excellent avenue, as these observables are measured with high precision. In Tab.~\ref{tab:edm_bounds}, we present a comparative analysis detailing the upper bounds on the imaginary components of these WCs derived from the electron ($d_e$) and muon ($d_\mu$) EDMs. It is important to note that within the theoretical calculation, connecting these specific WCs to the low-energy EDM observables requires two insertions of the relevant operator, so we get bounds on the products of WCs. 

\begin{table}[htb!]
    \centering
    \renewcommand{\arraystretch}{1.25}
    \begin{tabular}{|
    >{\centering\arraybackslash}p{0.22\textwidth}|
    >{\centering\arraybackslash}p{0.22\textwidth}|
    >{\centering\arraybackslash}p{0.22\textwidth}|
    >{\centering\arraybackslash}p{0.22\textwidth}|
    }
    \hline \hline
    \textbf{WCs} & \textbf{$d_e$} & \textbf{WCs} & \textbf{$d_\mu$} \\
    \hline \hline
    $\mathfrak{Im}\left|[C_{eB}]_{e\tau}[C_{eB}]_{\tau e}\right|$ & $< 3.30 \times 10^{-10}$ & $\mathfrak{Im}\left|[C_{eB}]_{\mu\tau}[C_{eB}]_{\tau \mu}\right|$ & $< 4.82\times 10^{-3}$ \\
    $\mathfrak{Im}\left|[C_{eW}]_{e\tau}[C_{eW}]_{\tau e}\right|$ & $< 1.10 \times 10^{-9}$ & $\mathfrak{Im}\left|[C_{eW}]_{\mu\tau}[C_{eW}]_{\tau \mu}\right|$ & $< 1.60 \times 10^{-2}$ \\
    \hline \hline
    \end{tabular}
     \caption{Comparative bounds on the imaginary parts of the leptonic dipole-type WCs derived from the electron and muon EDMs.}
    \label{tab:edm_bounds}
\end{table}
The current experimental upper limits on the electron and muon EDMs are given by
\begin{equation}
    |d_e| < 4.1 \times 10^{-30}~e\,\text{cm}~~(90\%~\text{CL})~\text{\cite{Roussy:2022cmp}}\,, \qquad
    |d_\mu| < 1.8 \times 10^{-19}~e\,\text{cm}~~(95\%~\text{CL})~\text{\cite{Muong-2:2008ebm}}\,.
\end{equation}
However, the upcoming High-Intensity Muon Beams (HIMB) project at PSI is expected to reach a significantly improved future sensitivity of $|d_\mu| < 6 \times 10^{-23}~e\,\text{cm}$ (Phase II). In our analysis, we have utilized this projected sensitivity to forecast the future constraints on the corresponding WCs.

\bibliographystyle{JHEP}
\bibliography{biblio.bib}

\end{document}